\documentclass[10pt,onecolumn]{IEEEtran}
\usepackage{epsfig}
\usepackage{amssymb}
\usepackage[usenames]{color}
\usepackage{umoline}
\usepackage{setspace}
\usepackage[ruled,boxed]{algorithm2e}
\usepackage{url}
\usepackage{enumerate}
\usepackage{cite}
\usepackage{mathtools}
\usepackage{mathabx}
\usepackage{mathrsfs}
\usepackage{multirow}
\usepackage{datetime}
\usepackage{cleveref}
\usepackage{pdflscape}

\usepackage{epstopdf}

\newcommand{\norm}[1]{\Vert#1\Vert}

\newcommand{\rank}{\mathrm{rank}}

\newcommand{\supp}[1]{\mathrm{supp}\left(#1\right)}

\newcommand{\Set}[1]{\mathcal{{#1}}}

\newtheorem{theorem}{Theorem}[section]
\newtheorem{lemma}[theorem]{Lemma}
\newtheorem{proposition}[theorem]{Proposition}
\newtheorem{corollary}[theorem]{Corollary}

\numberwithin{equation}{section}
\renewcommand{\theequation}{\arabic{section}.\arabic{equation}}

\newtheorem{definition}[theorem]{Definition}
\newtheorem{remark}[theorem]{Remark}
\newtheorem{example}[theorem]{Example}

\newcommand{\qed}{\nobreak \ifvmode \relax \else
      \ifdim\lastskip<1.5em \hskip-\lastskip
      \hskip1.5em plus0em minus0.5em \fi \nobreak
      \vrule height0.75em width0.5em depth0.25em\fi}

\title{Oblique Pursuits for Compressed Sensing}
\author{Kiryung Lee, Yoram Bresler~\IEEEmembership{Fellow,~IEEE}, and Marius Junge
\thanks{K. Lee and Y. Bresler are with Coordinated Science Laboratory and Department of Electrical and Computer Engineering, University of Illinois at Urbana-Champaign, Urbana, IL 61801, USA (e-mail: \{klee81,ybresler\}@illinois.edu).}
\thanks{M. Junge is with Department of Mathematics, University of Illinois at Urbana-Champaign, Urbana, IL 61801, USA (e-mail: junge@math.uiuc.edu).}}

\begin{document}
\doublespacing


\vspace{0.5in}

\maketitle

\begin{abstract}
Compressed sensing is a new data acquisition paradigm
enabling universal, simple, and reduced-cost acquisition, by exploiting a sparse signal model.
Most notably, recovery of the signal by computationally efficient algorithms is guaranteed for certain randomized acquisition systems.
However, there is a discrepancy between the theoretical guarantees and practical applications.
In applications, including Fourier imaging in various modalities,
the measurements are acquired by inner products with vectors selected randomly (sampled) from a frame.
Currently available guarantees are derived using a so-called restricted isometry property (RIP),
which has only been shown to hold under ideal assumptions.
For example, the sampling from the frame needs to be
independent and identically distributed with the uniform distribution, and the frame must be tight.
In practice though, one or more of the ideal assumptions is typically violated
and none of the existing guarantees applies.

Motivated by this discrepancy, we propose two related changes in the existing framework:
(i) a generalized RIP called the restricted biorthogonality property (RBOP);
and (ii) correspondingly modified versions of existing greedy pursuit algorithms, which we call oblique pursuits.
Oblique pursuits are guaranteed using the RBOP without requiring ideal assumptions;
hence, the guarantees apply to practical acquisition schemes.
Numerical results show that oblique pursuits also perform competitively with, or sometimes better than their conventional counterparts.
\end{abstract}


\section{Introduction}
\label{sec:intro}

\subsection{Compressed Sensing}

Many natural and man-made signals admit sparse representations \cite{mallat08}.
Compressed sensing is a new paradigm of data acquisition that takes advantage of this property
to reduce the amount of data that needs to be acquired to recover the signal of interest.
Unlike the conventional paradigm, in which large quantities of data are acquired, often followed by compression,
compressed sensing acquires minimally redundant data directly in a universal way that does not depend on the data \cite{BreGV99,Don06,CanRT06}.

The model for the acquisition is formally stated as the following linear system:
Let $f \in \mathbb{K}^d$ (where $\mathbb{K} = \mathbb{R}$ or $\mathbb{K} = \mathbb{C}$) be the unknown signal.
The measurement vector $y \in \mathbb{K}^m$ obtained by sensing matrix $A \in \mathbb{K}^{m \times d}$ is
\[
y = A f + w
\]
where $w \in \mathbb{K}^m$ denotes additive noise.
In the conventional paradigm, arbitrary signal $f \in \mathbb{K}^d$ is stably reconstructed when the rows of $A$ constitute a frame for $\mathbb{K}^d$,
which requires redundant measurements ($m \geq d$).
In contrast, compressed sensing aims to reconstruct signals that
are (approximately) $s$-sparse over a dictionary $D \in \mathbb{K}^{d \times n}$ (cf. \cite{Don06,CanRom07}) from compressive measurements ($m < d$).
Let $x \in \mathbb{K}^n$ be the coefficient vector of $f$ over $D$ such that $f \approx D x$ with $x$ being $s$-sparse.\footnote{When $w$ is assumed arbitrary, the model error term $A (f - D x)$ can be absorbed into $w$. Alternatively, $x$ can be assumed approximately sparse. We consider the former case in this paper.}
Then, the composition $\Psi = A D$ can be viewed as a sensing matrix for $x$ that produces the measurement vector $y$.
Once an estimate $\widehat{x}$ of $x$ is computed, $D \widehat{x}$ provides an estimate of the unknown signal $f$.
Hence, we may focus on the recovery of sparse $x$.

In an ideal case with exact sparse signal model and noise-free measurements,
if any $2s$ columns of $\Psi$ are linearly independent,
the unknown $s$-sparse $x$ is recovered as the unique solution to the linear system $\Psi x = y$ \cite{DonEla03,Feng97,VenBre98}.
In typical examples of compressed sensing (e.g., $\Psi$ is a matrix with independently and identically distributed (i.i.d.) Gaussian entries),
this is often achieved with $m = 2s$.
However, this algebraic guarantee only shows the uniqueness of the solution.
Furthermore, it is only valid in the absence of measurement noise and no error in the sparse signal model.

In practice, both computational cost of signal recovery, and its robustness against noise and model error are of interest.
For certain matrices $\Psi$, the unknown $x$ is stably recovered using efficient algorithms from compressive measurements.
The required number of measurements for a stable recovery is quantified through a property of $\Psi$ called
the \textit{restricted isometry property} (RIP) \cite{CanTao05}.
\begin{definition}
The \textit{$s$-restricted isometry constant} $\delta_s(\Psi)$ of $\Psi$ is defined as the smallest $\delta$ that satisfies
\begin{equation}
(1 - \delta) \norm{x}_2^2 \leq \norm{\Psi x}_2^2 \leq (1 + \delta) \norm{x}_2^2, \quad \text{$\forall$$s$-sparse $x$}.
\label{eq:rip}
\end{equation}
\end{definition}
Matrix $\Psi$ satisfies the RIP of order $s$ if $\delta_s(\Psi) < c$ for some constant $c \in (0,1)$.
Intuitively, smaller $\delta_s(\Psi)$ implies that $\Psi^* \Psi x$ is closer to $x$ for all $s$-sparse $x$.
Although, in general, the recovery of $s$-sparse $x$ from compressive measurements is NP hard even in the noiseless case,
the recovery can be accomplished efficiently (in polynomial time) and with guaranteed accuracy,
when $\Psi$ satisfies the RIP with certain parameters (order and threshold).
Such results are among the major achievements  of compressed sensing theory.
For example, when $\delta_{2s}(\Psi) < \sqrt{2} - 1$,
the solution to an $\ell_1$-norm-based convex optimization formulation provides a good approximation of the unknown $s$-sparse $x$ \cite{Can08rip}.
The approximation error in this result is guaranteed to be small, and vanishes in the noiseless case.
A computationally efficient alternative is provided by iterative greedy algorithms \cite{NeeTro09,DaiMil09,BluDav09,Fou11HTP},
which exploit the RIP of $\Psi$ to compute an approximation of $x$.
These iterative greedy algorithms provide similar approximation guarantees when $\delta_{ks}(\Psi) < c$,
where $k \in \{2,3,4\}$ and $c \in (0,1)$ are constants specified by the algorithms.
Different applications of the RIP require different values for the parameters $k$ and $c$.
Henceforth, we assume that $k$ and $c$ are arbitrarily fixed constants as above.

The question of feasibility of compressed sensing then reduces to determining whether,
and with how many measurements, $\Psi$ satisfies the RIP.\footnote{
There also exist analyses not in terms of the RIP (e.g., \cite{DonTan10}, \cite{Wai09lasso}).
However, these analyses only apply to certain ideal random matrices such as an i.i.d. Gaussian matrix,
which although reasonable in models for regression problems in statistics,
is rarely used in practical acquisition systems.
}
Certain random matrices $\Psi \in \mathbb{K}^{m \times n}$ satisfy $\delta_s(\Psi) < \delta$ with high probability
when the number of measurements $m$ satisfies $m = O(\delta^{-2} s \ln^\alpha n)$ for some small integer $\alpha$ \cite{BDDVW08,RudVer08,Rau08,Rau10}.
This result, when combined with the aforementioned RIP-based guarantees of the recovery algorithms, enables ``compressive sensing'' ($m < d$).
For example, if $\Psi$ satisfies the \emph{strong concentration property},
that is, $\norm{\Psi x}_2^2$ is highly concentrated around its expectation for all $x$,
then $\delta_s(\Psi) < \delta$ holds with $m = O(\delta^{-2} s \ln(n/s))$ \cite{BDDVW08}.
In words, a number $m$ of measurements that is proportional to the number $s$ of nonzeros,
and only logarithmic in the number $n$ of unknowns, suffices for stable and computationally efficient recovery.
This celebrated result of compressed sensing has been extended to the case
where $A$ satisfies the strong concentration property with $\delta_s(A) < \delta$ and $D$ satisfies the RIP,
stating that $\delta_s(A D) < \delta_s(D) + \delta + \delta \cdot \delta_s(D)$ holds with $m = O(\delta^{-2} s \ln(n/s))$ \cite{RauSV08}.
Now, the RIP of $D$ is often relatively easy to satisfy.
Recall that the role of $D$ is to provide a sparse representation of $f$.
Although redundant $D$ (with $n \geq d$) performs better in this respect,
it is often the case that $f$ is sparse over a $D$ that is a basis (e.g., a piecewise smooth signal $f$ over a wavelet basis $D$).
In this case, $\delta_s(D)$ is easily bounded using the condition number of $D$.
Furthermore, if $D$ is an orthonormal basis, then $\delta_s(D)=0$ for any $s \leq n$.
As for the strong concentration property of $A$, it is satisfied by an i.i.d. Gaussian or Bernoulli matrix \cite{BDDVW08}.
This has been extended recently to any matrix satisfying the RIP with certain parameters,
when postmultiplied by a random diagonal matrix of $\pm1$ \cite{KraWar11}.
When implementing such a sensing system is technically feasible, it would provide a sensing matrix $A$ that admits efficient computation \cite{CENR11}.

However, although the aforementioned random matrix models are interesting in theory, they are rarely used in practice.
In most practical signal acquisition systems,
the linear functionals used for acquiring the measurements (rows of $A$) are determined by
the physics of the specific modality and by design constraints of the sensor.
In compressed sensing applied to these systems \cite{BreGV99,LusDP07},
the sensing matrix $A$ does not follow the aforementioned random matrix models;
instead its rows are i.i.d. samples from the uniform distribution on a set that constitutes a frame in $\mathbb{K}^d$.\footnote{
The use of the i.i.d. sampling may end up with a repetition of the same row.
However, repeating one row of $A$ as an additional row does not increase the RIC of $A$.
A similar construction of $A$, where the rows are selected from a frame using the Bernoulli sampling,
has also been studied \cite{CanRT06,RudVer08}.
While the Bernoulli sampling does not cause the repetition, the size of selection is no longer deterministic, i.e.,
it is concentrated around $m$ with high probability.
The imperfection with these two sampling schemes becomes negligible as the size of $A$ increases.
We focus on the i.i.d. sampling scheme in this paper.}

To describe the sensing matrix more precisely, we recall the definition of a frame \cite{Chr03frame}.
We denote by $L_2(\Omega, \nu)$ the Hilbert space of functions defined on a compact set $\Omega$
that are square integrable with respect to a probability measure $\nu$ on $\Omega$,
and by $\ell_2^d$ the $d$-dimensional Euclidean space.

\begin{definition}
Let $\mu$ denote the uniform probability measure on a compact set $\Omega$.
Let $(\phi_\omega)_{\omega \in \Omega}$ be a set of vectors in $\mathbb{K}^d$.
Let $\Phi: L_2(\Omega,\mu) \to \ell_2^d$ be the synthesis operator associated with $(\phi_\omega)_{\omega \in \Omega}$ defined as
\begin{equation}
\Phi h = \int_{\Omega} \phi_\omega h(\omega) d\mu(\omega), \quad \forall h \in L_2(\Omega,\mu),
\label{eq:defsynthesis}
\end{equation}
with its adjoint $\Phi^*: \ell_2^d \to L_2(\Omega,\mu)$, which is the corresponding analysis operator given by
\begin{equation}
(\Phi^* f)(\omega) = \langle \phi_\omega, f \rangle, \quad \forall \omega \in \Omega, ~ \forall f \in \ell_2^d.
\label{eq:defanalysis}
\end{equation}
Then, $(\phi_\omega)_{\omega \in \Omega}$ is a \textit{frame}, if the frame operator $\Phi \Phi^*$ satisfies
$\alpha \leq \lambda_{\min}(\Phi \Phi^*) \leq \lambda_{\max}(\Phi \Phi^*) \leq \beta$ for some positive real numbers $\alpha$ and $\beta$.
In particular, if the frame operator $\Phi \Phi^*$ is a scaled identity, then $(\phi_\omega)_{\omega \in \Omega}$ is a \textit{tight frame}.
\label{def:frame}
\end{definition}

Let $\nu$ be a probability measure on $\Omega$.
Let $\bar{z}$ denote the complex conjugate of $z \in \mathbb{C}$ and $[m]$ denote the set $\{1, \ldots, m\}$.
The sensing matrix $A \in \mathbb{K}^{m \times d}$ is constructed from a frame $(\phi_{\omega})_{\omega \in \Omega}$ as
\begin{equation}
A_{k,\ell} = \frac{1}{\sqrt{m}} \overline{(\phi_{\omega_k})_\ell}, \quad \forall k \in [m],~ \ell \in [d]
\label{eq:contrA}
\end{equation}
for random indices $(\omega_k)_{k=1}^m$ in $\Omega$ chosen i.i.d. with respect to $\nu$.
We call this type of matrix a \emph{random frame matrix}.
It is the model for a sensing matrix of primary interest in this paper,
and we will assume henceforth that $A$ is defined by (\ref{eq:contrA}).

Random frame matrices arise in numerous applications of compressed sensing.
We list a few below. For simplicity, they are described for the 1D case.
\begin{example}
An important example of a random frame matrix is a random partial discrete Fourier transform (DFT) matrix.
Let $\phi_\omega \triangleq [1, e^{-j2\pi\omega}, \ldots, e^{-j2\pi(d-1)\omega}]^T$ be defined
for $\omega \in \Omega \triangleq \{1/d,\ldots,(d-1)/d,1\}$.
In this setup, $\nu: \Omega \to [0,1]$ is a cumulative density function on $\Omega$
and $\frac{d\nu}{d\mu}(\omega)$ denotes the probability that $\omega$ will be chosen, multiplied by $d$.
Then, an $m \times d$ random partial DFT matrix is constructed from $(\phi_\omega)_{\omega \in \Omega}$ using $\nu$ by (\ref{eq:contrA}).
The frame $(\phi_\omega)_{\omega \in \Omega}$ in this example is a tight frame,
and $\sup_\omega \norm{\phi_\omega}_{\ell_\infty^d} / \norm{\phi_\omega}_{\ell_2^d}$, which will play a role in our subsequent discussion,
achieves its minimum $\frac{1}{\sqrt{d}}$.
Sensing matrix of this kind arise in practical applications of compressed sensing such as the multi-coset sampling and spectrum-blind recovery of multiband signals at sub-Nyquist rates \cite{FenBre96,Feng97,MisEld09}.\footnote{This was the invention of compressed sensing of analog signals. See \cite{Bre08} for a survey of this early work.}
Similar random matrices also arise in more recent studies on compressed sensing of analog signals \cite{Eld09,MisEld10mwc,TLDRB10,MisEE11}.
\label{example:dft}
\end{example}

\begin{example}
One author of this paper proposed the compressive acquisition of signals in Fourier imaging systems \cite{BreGV99,BreFen96,VenBre98},
which is one of the works that invented the notion of compressed sensing.
This idea has been applied with refinements to various modalities such as magnetic resonance imaging (MRI) \cite{YeBM02,LusDP07},
photo-acoustic tomography \cite{ProLes09}, radar \cite{HerStr09}, radar imaging \cite{BarSte07,PEPC10}, and astronomical imaging \cite{BobSO08}, etc.
The sensing matrix $A$ for compressed sensing in Fourier imaging systems is a random partial Fourier transform matrix with continuous-valued frequencies (\textit{continuous random partial Fourier matrix}, henceforth), which is obtained similarly to the previous example.
Let $\phi_\omega \triangleq [1, e^{-j2\pi\omega}, \ldots, e^{-j2\pi(d-1)\omega}]^T$ be defined
for $\omega \in \Omega \triangleq [-\frac{1}{2},\frac{1}{2})$.
The frame $(\phi_\omega)_{\omega \in \Omega}$ in this example is a continuous tight frame,
and the quantity $\sup_\omega \norm{\phi_\omega}_{\ell_\infty^d} / \norm{\phi_\omega}_{\ell_2^d}$ achieves its minimum $\frac{1}{\sqrt{d}}$.
\label{example:cft}
\end{example}

\begin{example}
In MRI, the Fourier measurements are usually modeled as obtained from the input signal
modified by pointwise multiplication with a mask $\lambda \in \mathbb{K}^d$, representing the receiving coil sensitivity profile.
Let $\Lambda = \mathrm{diag}(\lambda)$ denote the diagonal matrix with the elements of $\lambda$ on the diagonal.
Let $\phi_\omega \triangleq \Lambda^* [1, e^{-j2\pi\omega}, \ldots, e^{-j2\pi(d-1)\omega}]^T$ be defined
for $\omega \in \Omega \triangleq [-\frac{1}{2},\frac{1}{2})$.
If $\lambda$ has no zero element, then $(\phi_\omega)_{\omega \in \Omega}$ is a frame that spans $\mathbb{K}^d$.
Otherwise, $(\phi_\omega)_{\omega \in \Omega}$ is a frame for the subspace $\Set{S}$
of $\mathbb{K}^d$ spanned by the standard basis vectors corresponding to the nonzero elements of $\lambda$.
In the latter case, letting the signal space be $\Set{S}$ instead of $\mathbb{K}^d$,
we modify the inverse problem so that $A$ constructed by (\ref{eq:contrA}) is a map from $\Set{S}$ to $\mathbb{K}^m$.
Note that each vector in the frame is multiplied from the right by $\Lambda$ compared to that in Example~\ref{example:cft}.
In this example, unless the nonzero elements of $\lambda$ have the same magnitudes,
$(\phi_\omega)_{\omega \in \Omega}$ does not satisfy the two properties coming from a Fourier system (tightness and minimal $\sup_\omega \norm{\phi_\omega}_{\ell_\infty^d} / \norm{\phi_\omega}_{\ell_2^d}$).
Therefore, we do not restrict our interest to the Fourier case and consider a general frame $(\phi_\omega)_{\omega \in \Omega}$.
\label{example:cft2}
\end{example}

Because random frame matrices are so ubiquitous in compressed sensing, the analysis of their RIP is of major interest.
Although random frame matrices do not satisfy the strong concentration property, other tools are available for the analysis of their RIP.
In particular, the RIP of a partial Fourier matrix has been studied using noncommutative probability theory \cite{RudVer08,Rau08}.
The extension of this analysis to the RIP of a random frame matrix \cite{Rau10} enables handling a more general class of sensing matrices.
Notably, all known analyses \cite{RudVer08,Rau08,Rau10,CanPla11ripless} focused on the case where $D$ corresponds to an orthonormal basis.
These analyses also assumed either the exact isotropy property, $\mathbb{E} A^* A = I_d$ \cite{RudVer08,Rau08,Rau10},
or the so-called near isotropy property, $\norm{\mathbb{E} A^* A - I_d} = O(\frac{1}{\sqrt{n}})$ \cite{CanPla11ripless}.
There is no alternative sufficient condition that does not require these properties.
In fact though, these RIP analyses further extend to the following Theorem~\ref{thm:uripsimple} (proved in Section~\ref{sec:rbp}),
which addresses the case of $\Psi = A D$, where $A$ is a random frame matrix and $D$ is not necessarily an orthonormal basis,
and furthermore, allows a non-vanishing deviation from isotropy.

\begin{theorem}
Let $A \in \mathbb{K}^{m \times d}$ be a random matrix constructed from a frame $(\phi_\omega)_{\omega \in \Omega}$ by (\ref{eq:contrA})
and let $D = [d_1,\ldots,d_n] \in \mathbb{K}^{d \times n}$ satisfy $\delta_s(D) < 1$.
Suppose that $\sup_\omega \max_j |\langle \phi_\omega, d_j \rangle| \leq K$.
Let $\Psi = A D$.
Then, $\delta_s(\Psi) < \delta + \norm{\mathbb{E} A^* A - I_d} + \delta_s(D) + \norm{\mathbb{E} A^* A - I_d} \delta_s(D)$ holds with high probability for $m = O(\delta^{-2} s \ln^4 n)$.
\label{thm:uripsimple}
\end{theorem}

The inequality in Theorem~\ref{thm:uripsimple} indicates that $\delta_{ks}(\Psi) < c$ holds with high probability for $m = O(s \ln^4 n)$
if $D$ satisfies $\delta_{ks}(D) \leq \frac{c}{4}$, and $A$ satisfies $\norm{\mathbb{E} A^* A - I_d} \leq \frac{c}{4}$.
Combined with the aforementioned RIP-based guarantees,
this result again enables compressive sensing, when the conditions given in Theorem~\ref{thm:uripsimple} are satisfied.

\subsection{Motivation: Failure of Guarantees in Practical Applications}
\label{subsec:motivation}

While the RIP is essential for all existing performance guarantees for compressed sensing with random frame sensing matrices,
it turns out that this property is satisfied only under certain nonrealistic assumptions.
Most notably, although compressed sensing has been proposed to accelerate the acquisition in imaging systems \cite{BreGV99,CanRT06,LusDP07}
and some of the most widely studied applications of compressed sensing to date are in such systems,
the RIP has not been shown to hold for the associated sensing matrices in a realistic setup.
More specifically, $\norm{\mathbb{E} A^* A - I_d}$ is not negligible,
which makes even the upper bound on $\delta_s(\Psi)$ given by Theorem~\ref{thm:uripsimple},
which is the most relaxed condition on deviation from isotropy known to date,
too conservative to be used for RIP-based recovery guarantees.

One reason for the increase $\norm{\mathbb{E} A^* A - I_d}$ from the ideal case is the use of a nonuniform distribution in the construction of $A$.
In Examples~\ref{example:dft} and \ref{example:cft},
the sensing matrix $A$ were constructed from i.i.d. samples from a tight frame $(\phi_\omega)_{\omega \in \Omega}$.
In this case, if the i.i.d. sampling is done in accordance to the uniform distribution, then $\mathbb{E} A^* A = I_d$.
However, in practice, i.i.d. sampling using a nonuniform distribution is often preferred for natural signals:
it is desirable to take more measurements of lower frequency components, which contain more of the signal energy.
Therefore, acquisition at frequencies sampled non-uniformly with a variable density is preferred \cite{LusDP07}.
As a consequence, the exact isotropy property is violated.
Depending on the probability distribution, $\norm{\mathbb{E} A^* A - I_d}$ is often not negligible, and even larger than 1,
which renders the upper bound on $\delta_s(\Psi)$ in Theorem~\ref{thm:uripsimple} useless.
Therefore, no known RIP analysis applies to Fourier imaging applications.

Another reason for the increase $\norm{\mathbb{E} A^* A - I_d}$  from the ideal case is that $(\phi_\omega)_{\omega \in \Omega}$ is a not tight frame.
As shown in Example~\ref{example:cft2}, even in a Fourier imaging system,
$(\phi_\omega)_{\omega \in \Omega}$ can be a non-tight frame due to the presence of a mask.
Furthermore, the application of compressed sensing is not restricted to Fourier imaging systems.
The idea of compressed sensing and recovery using sparsity also applies to other inverse problems in imaging described by non-orthogonal operators (e.g., a Fredholm integral equations of the first kind).
Optical diffusion tomography \cite{LeeKBY11} is a concrete example of compressed sensing with such a scenario.
As another example, the sensing matrix that arises in compressed sensing in shift-invariant spaces \cite{Eld09} is not necessarily obtained from a tight frame.

Yet another reason for the failure of the upper bound on $\delta_s(\Psi)$ in Theorem~\ref{thm:uripsimple} has to do with the dictionary $D$.
Indeed, to achieve $\delta_s(\Psi) < c$ with $m = O(s \ln^4 n)$,
it is necessary that both $\norm{\mathbb{E} A^* A - I_d}$ and $\delta_s(D)$ are less than a certain threshold.
However, verification of this condition for $\delta_s(D)$ is usually computationally expensive.
For the special case where $D$ has full column rank (hence, $d \geq n$), $\delta_s(D)$ is easily bounded from above by $\norm{D^* D - I_n}$.
In particular, if $D$ corresponds to an orthonormal basis, then $D^* D = I_n$, which implies $\delta_s(D) = 0$.
Otherwise, $\delta_s(D)$ vanishes as $D$ approaches an orthonormal basis.
However, it is often too restrictive to make $\norm{D^* D - I_n}$ less than a small threshold below 1.
Moreover, without this constraint, $D$ can provide a better sparse representation, which is also desired for stable recovery.
In particular, for a data-adaptive $D$, the property that $\norm{D^* D - I_n}$ is less than a given threshold is not guaranteed.
In this all too common situation, all known RIP analyses break down:
they only provide a conservative upper bound on $\delta_s(\Psi)$, which does enable the RIP-based recovery-guarantees.

In summary, in most practical compressed sensing applications,
the effective sensing matrix $\Psi = A D$ may fail to satisfy the RIP for one or more of the following reasons:
the i.i.d. sampling in the construction of $A$ does not use the uniform distribution;
the frame used in the construction of $A$ is not tight;
or the dictionary $D$ does not have a sufficiently small restricted isometry constant.
From these observations, we conclude that none of the existing performance guarantees for recovery algorithms applies to the aforementioned applications of compressed sensing.

\subsection{Contributions}

Recall that unlike the $\ell_1$-norm-based recovery,
greedy recovery algorithms were designed to exploit the property that $\Psi^* \Psi x \approx x$ for sparse $x$, explicitly.
For example, in the derivation of the CoSaMP algorithm \cite{NeeTro09},
the procedure of applying $\Psi^*$ to $y = \Psi x$ for $x$ sparse was called the computation of a ``proxy'' signal,
which reveals the information about the locations of nonzero elements of $x$.
The same idea was also used for deriving other iterative greedy algorithms \cite{DaiMil09,BluDav09,Fou11HTP}.
Indeed, if $\Psi$ satisfies the RIP, then the use of the (transpose of) the same matrix $\Psi$ to compute a proxy is a promising approach.
Otherwise, one can employ a different matrix $\widetilde{\Psi}$ to get a better proxy $\widetilde{\Psi}^* y$.
The required property is that $\widetilde{\Psi}^* \Psi x \approx x$ for sparse $x$.
To improve the recovery algorithms in this direction,
we first extend the RIP to a property of a pair of matrices $\Psi,\widetilde{\Psi} \in \mathbb{K}^{m \times n}$ called
the \textit{restricted biorthogonality property} (RBOP).
\begin{definition}
The \textit{$s$-restricted biorthogonality constant} $\theta_s(M)$ of $M \in \mathbb{K}^{n \times n}$ is defined as the smallest $\delta$ that satisfies
\begin{align}
|\langle y, M x \rangle - \langle y, x \rangle| \leq \delta \norm{x}_2 \norm{y}_2, \quad \text{$\forall$$s$-sparse $x,y$ with common support}.
\label{eq:rbp}
\end{align}
\end{definition}
The pair $(\Psi,\widetilde{\Psi})$ satisfies the RBOP of order $s$ if $\theta_s(\widetilde{\Psi}^* \Psi) < c$ for some constant $c \in (0,1)$.\footnote{As in the case of the RIP, the threshold value of $c$ for which the RBOP is said to be satisfied depends on the application.}
Intuitively, smaller $\theta_s(\widetilde{\Psi}^* \Psi)$ implies that $\widetilde{\Psi}^* \Psi x$ becomes closer to $x$ for all $s$-sparse $x$.
In other words, any $s$ columns of $\Psi$ and $\widetilde{\Psi}$ corresponding to the same indices behave like a biorthogonal basis.
If $\widetilde{\Psi} = \Psi$, then $\theta_s(\widetilde{\Psi}^* \Psi)$ reduces to $\delta_s(\Psi)$;
hence, the RBOP of $(\Psi,\widetilde{\Psi})$ reduces to the RIP of $\Psi$.

We then modify the greedy recovery algorithms so that the modified algorithms employ both $\Psi$ and $\widetilde{\Psi}$
and, in particular, exploit the RBOP of $(\Psi,\widetilde{\Psi})$ to provide an approximation guarantee.
In fact, modified thresholding and forward greedy algorithms using a different matrix $\widetilde{\Psi}$
have been already proposed by Schnass and Vandergheynst \cite{SchVan08}.
However, our work is different from theirs in several important respects.
Schnass and Vandergheynst \cite{SchVan08} propose to use $\widetilde{\Psi}$ numerically optimized to minimize a version of the Babel function.
However, although sufficient conditions given in terms of the Babel function are easily computable,
the resulting guarantees for the recovery performance are conservative.
Furthermore, their numerical algorithm to design $\widetilde{\Psi}$ is a heuristic,
and does not provide any guarantee on the value of the Babel function achieved.
In contrast, we propose an explicit construction of $\widetilde{\Psi}$ so that $\theta_s(\widetilde{\Psi}^* \Psi) \ll 1$ holds.
To show the construction, we recall the definition of a biorthogonal frame that extends the notion of a biorthogonal basis.
\begin{definition}
Let $(\phi_\omega)_{\omega \in \Omega}$ and $(\widetilde{\phi}_\omega)_{\omega \in \Omega}$ be sets of vectors in $\mathbb{K}^d$.
Let $L_2(\Omega,\mu)$ be as defined in Definition~\ref{def:frame}.
Let $\Phi^*: \ell_2^d \to L_2(\Omega,\mu)$ be the analysis operator associated to $(\phi_\omega)_{\omega \in \Omega}$ defined in (\ref{eq:defanalysis}).
Let $\widetilde{\Phi}: L_2(\Omega,\mu) \to \ell_2^d$ be the synthesis operator associated to $(\widetilde{\phi}_\omega)_{\omega \in \Omega}$ defined similarly to (\ref{eq:defsynthesis}).
Then, $(\phi_\omega,\widetilde{\phi}_\omega)_{\omega \in \Omega}$ is a \textit{biorthogonal frame} if $\widetilde{\Phi} \Phi^* = I_d$.
\end{definition}

Matrix $\widetilde{\Psi}$ is then constructed as the composition $\widetilde{\Psi} = \widetilde{A} \widetilde{D}$.
We construct $\widetilde{A} \in \mathbb{K}^{m \times d}$ from the dual frame $(\widetilde{\phi}_{\omega})_{\omega \in \Omega}$ by
\begin{align}
(\widetilde{A})_{k,\ell} {} & = \frac{1}{\sqrt{m}} \left[ \frac{d\nu}{d\mu}(\omega_k) \right]^{-1} \overline{(\widetilde{\phi}_{\omega_k})_\ell}, \quad \forall k \in [m],~ \ell \in [d] \label{eq:contrtildeA}
\end{align}
where $(\omega_k)_{k=1}^m$ are the same indices
as used to define the samples from $(\psi_\omega)_{\omega \in \Omega}$ in the construction of $A$ in (\ref{eq:contrA}).
Assuming $\frac{d\nu}{d\mu}(\omega) > 0$,
then, by the construction of $A$ and $\widetilde{A}$, it follows that the pair $(A,\widetilde{A})$ satisfies the dual isotropy property
\[
\mathbb{E} \widetilde{A}^* A = I_d.
\]

\begin{remark}
We proposed modified greedy pursuit algorithms in Section~\ref{sec:algorithms} that
use both $\Psi = A D$ and $\widetilde{\Psi} = \widetilde{A} \widetilde{D}$ and are guaranteed using the RBOP of $(\Psi,\widetilde{\Psi})$.
Therefore, it is important to check whether $\widetilde{\Psi}^* = \widetilde{D}^* \widetilde{A}^*$ can be efficiently implemented.
The discussion on $(D,\widetilde{D})$ is deferred to the next subsections and we only discuss the computational issue with $\widetilde{A}$ here.
In practice, $A^*$ is implemented using fast algorithms without forming a dense matrix explicitly.
For example, if $A$ is a partial DFT matrix,
then, $A^*$ is implemented as the fast Fourier transform (FFT) applied to the zero padded vector.
Likewise, if $A$ is a continuous partial Fourier matrix, then, the nonuniform FFT (NUFFT) \cite{FesSut03nufft} can be used for fast computation.
In this case, since our construction of $\widetilde{A}$ in (\ref{eq:contrtildeA}) only involves row-wise rescaling of $A$ by constant factors,
$\widetilde{A}^*$ is also implemented using the same fast algorithms.
In the more general biorthogonal case, once the synthesis operator $\widetilde{A}$ is implemented as a fast algorithm,
$\widetilde{A}^*$ is also computed efficiently using the same algorithm.
In fact, in many applications, the biorthogonal dual system is given analytically.
For example, if the frame $(\phi_\omega)_{\omega \in \Omega}$ is given as a filter bank system,
designing perfect reconstruction filters that provide the corresponding biorthogonal dual frame is well studied \cite{Vai93fb}.
Similar arguments apply to the analysis operator of analytic frames such as overcomplete DCT or wavelet packets.
\end{remark}

Regarding the construction of $\widetilde{D}$, we consider the following two cases:
(i) $D$ corresponds to a basis for $\mathbb{K}^n$ ($d = n$); (ii) $D$ satisfies the RIP with certain parameter.
We let $\widetilde{D} = D (D^*D)^{-1}$ for the former case and $\widetilde{D} = D$ for the latter case.
The RBOP of this construction is deferred to after the exposition of new recovery algorithms.

Now, we return to the discussion of the recovery algorithms.
While Schnass and Vandergheynst \cite{SchVan08} only replaced $\Psi$ by $\widetilde{\Psi}$
in the steps of computing a proxy in forward greedy algorithms (MP and OMP),
we also replace the orthogonal projection used in the update of the residual in OMP
by a corresponding oblique projection obtained from $\Psi$ and $\widetilde{\Psi}$.
Therefore, we propose a different variation of OMP called Oblique Matching Pursuit (ObMP),
which is guaranteed using the RBOP of $(\Psi,\widetilde{\Psi})$.
We also propose similar modifications of iterative greedy recovery algorithms and their RIP-based guarantees.
The modified algorithms are different from the original algorithms:
we assign them new names, with the modifier ``oblique''.
For example, SP is extended to \textit{oblique subspace pursuit} (ObSP).
CoSaMP, IHT, and HTP are likewise extended to ObCoSaMP, ObIHT, ObHTP, respectively.
We call these modified greedy algorithms based on the RBOP \textit{oblique pursuits}.
In the numerical experiments in this paper, in scenarios where one or more of the ideal assumptions
(i.i.d. sampling according to the uniform distribution, tight frame $(\phi_\omega)_{\omega \in \Omega}$, or orthonormal basis $D$) are violated,
the oblique pursuits perform better than, or at least competitively with their conventional counterparts.

Importantly, the oblique pursuits come with RBOP-based approximation guarantees.
In particular, similarly to its conventional counterpart,
each iterative oblique pursuit algorithm is guaranteed when $\theta_{ks}(\widetilde{\Psi}^* \Psi) < c$,
where $k \in \{2,3,4\}$ and $c \in (0,1)$ are constants specified by the algorithms.
The number of measurements required for the guarantees of oblique pursuits is also similar to
that required in the ideal scenario by their conventional counterparts.
When combined with the subsequent RBOP analysis of $(\Psi,\widetilde{\Psi})$ for random frame sensing matrices,
the recovery by the iterative oblique pursuit algorithms is guaranteed with $m = O(s \ln^4 n)$.
In particular, we show that it is no longer necessary to have $\norm{\mathbb{E} A^* A - I_n} \ll 1$.
Therefore, the obtained guarantees apply in realistic setups of the aforementioned CS applications.

The degrees of freedom added by the freedom to design $\widetilde{\Psi}$
make the RBOP easier to satisfy under milder assumptions than the RIP.
In particular, with the proposed construction of $\widetilde{\Psi}$,
the RBOP of $(\Psi,\widetilde{\Psi})$ holds without requiring the (near) isotropy property of $A$.
More specifically, depending on whether $D$ corresponds to a basis or satisfies the RIP,
the RIP analysis in Theorem~\ref{thm:uripsimple} is extended to the following theorems.
Recall that we proposed different constructions of $\widetilde{\Psi}$ for the two cases.
\begin{theorem}
Let $A,\widetilde{A} \in \mathbb{K}^{m \times d}$ be random matrices constructed from a biorthogonal frame $(\phi_\omega,\widetilde{\phi}_\omega)_{\omega \in \Omega}$ by (\ref{eq:contrA}) and (\ref{eq:contrhatA}), respectively.
Let $D = [d_1,\ldots,d_n]$ and $\widetilde{D} \in \mathbb{K}^{d \times n}$ ($d = n$) satisfy $\widetilde{D}^* D = I_d$.
Let $\Psi = A D$ and $\widetilde{\Psi} = \widetilde{A} \widetilde{D}$.
Suppose that $\sup_\omega \max_j |\langle \phi_\omega, d_j \rangle| \leq K$.
Then, $\theta_s(\widetilde{\Psi}^* \Psi) < \delta$ holds with high probability for $m = O(\delta^{-2} s \ln^4 n)$.
\label{thm:urbpsimpleBO}
\end{theorem}
\begin{theorem}
Let $A,\widetilde{A} \in \mathbb{K}^{m \times d}$ be random matrices constructed from
a biorthogonal frame $(\phi_\omega,\widetilde{\phi}_\omega)_{\omega \in \Omega}$ by (\ref{eq:contrA}) and (\ref{eq:contrhatA}), respectively.
Let $D = [d_1,\ldots,d_n] \in \mathbb{K}^{d \times n}$ satisfy $\delta_s(D) < 1$
Let $\Psi = A D$ and $\widetilde{\Psi} = \widetilde{A} D$.
Suppose that $\sup_\omega \max_j |\langle \phi_\omega, d_j \rangle| \leq K$.
Then, $\theta_s(\widetilde{\Psi}^* \Psi) < \delta + \delta_s(D)$ holds with high probability for $m = O(\delta^{-2} s \ln^4 n)$.
\label{thm:urbpsimpleOC}
\end{theorem}

Note that the upper bounds on $\theta_s(\widetilde{\Psi}^* \Psi)$ in Theorems~\ref{thm:urbpsimpleBO} and \ref{thm:urbpsimpleOC}
do not depend on $\norm{\mathbb{E} A^* A - I_d}$.
Therefore, unlike the RIP, which breaks down when the ideal assumptions,
such as i.i.d. sampling according to the uniform distribution and tight frame,
are violated, the RBOP continues to hold even with such violations.

In summary, we introduced a new tool for the design, analysis, and performance guarantees of sparse recovery algorithms,
and illustrate its application to derive new guaranteed versions of several of the most popular recovery algorithms.

\subsection{Organization of the Paper}

In Section~\ref{sec:algorithms}, we propose the oblique pursuit algorithms and their guarantees in terms of the RBOP.
In Section~\ref{sec:rbp}, we elaborate the RBOP analysis of random frame matrices in various scenarios.
The empirical performance of the oblique pursuit algorithms is studied in Section~\ref{sec:numres},
and we conclude the paper in Section~\ref{sec:conclusion}.

\subsection{Notation}
Symbol $\mathbb{N}$ is the set of natural numbers (excluding zero), and $[n]$ denotes the set $\{1, \ldots, n\}$ for $n \in \mathbb{N}$.
Symbol $\mathbb{K}$ denotes a scalar field, which is either the real field $\mathbb{R}$ or the complex field $\mathbb{C}$.
The vector space of $d$-tuples over $\mathbb{K}$ is denoted by $\mathbb{K}^d$ for $d \in \mathbb{N}$.
Similarly, for $m,n \in \mathbb{N}$, the vector space of $m \times n$ matrices over $\mathbb{K}$ is denoted by $\mathbb{K}^{m \times n}$.

We will use various notations on a matrix $A \in \mathbb{K}^{m \times n}$.
The range space spanned by the columns of $A$ will be denoted by $\mathcal{R}(A)$.
The adjoint operator of $A$ will be denoted by $A^*$.
This notation is also used for the adjoint of a linear operator that is not necessarily a finite matrix.
The $j$th column of $A$ is denoted by $a_j$ and the submatrix of $A$ with columns indexed by $J \subset [n]$ is denoted by $A_J$.
The $k$th row of $A$ is denoted by $a^k$,
and the submatrix of $A$ with rows indexed by $K \subset [m]$ is denoted by $A^K$.
Symbol $e_k$ will denote the $k$th standard basis vector of $\mathbb{K}^d$,
where $d$ is implicitly determined for compatibility.
The $k$th element of $d$-tuple $x \in \mathbb{K}^d$ is denoted by $(x)_j$.
The $k$th largest singular value of $A$ will be denoted by $\sigma_k(A)$.
For Hermitian symmetric $A$, $\lambda_k(A)$ will denote the $k$th largest eigenvalue of $A$.
The Frobenius norm and the spectral norm of $A$ are denoted by $\norm{A}_F$ and $\norm{A}$, respectively.
The inner product is denoted by $\langle \cdot, \cdot \rangle$.
The embedding Hilbert space, where the inner product is defined, is not explicitly mentioned when it is obvious from the context.
For a subspace $\Set{S}$ of $\mathbb{K}^d$,
matrices $P_{\Set{S}} \in \mathbb{K}^{d \times d}$ and $P_{\Set{S}}^\perp \in \mathbb{K}^{d \times d}$ denote
the orthogonal projectors onto $\Set{S}$ and its orthogonal complement $\Set{S}^\perp$, respectively.
For $J \subset [n]$, the coordinate projection $\Pi_J: \mathbb{K}^n \to \mathbb{K}^n$ is defined by
\begin{equation}
(\Pi_J x)_k =
\begin{cases}
(x)_k & \text{if $k \in J$} \\
0 & \text{else}.
\end{cases}
\label{eq:def:coorproj}
\end{equation}
Symbols $\mathbb{P}$ and $\mathbb{E}$ will denote the probability and the expectation with respect to a certain distribution.
Unless otherwise mentioned, the distribution shall be obvious from the context.

\section{Oblique Pursuit Algorithms}
\label{sec:algorithms}

In this section, we propose modified greedy pursuit algorithms that use both $\Psi$ and $\widetilde{\Psi}$,
and show that they are guaranteed by the RBOP of $(\Psi,\widetilde{\Psi})$ similarly to the way
that the corresponding conventional pursuit algorithms are guaranteed by the RIP of $\Psi$.
The modified greedy pursuit algorithms will be called \textit{oblique pursuit algorithms},
because they involve oblique projections instead of the orthogonal projections in the conventional algorithms.

Recall that greedy pursuit algorithms seek an approximation of signal $f$ that is exactly sparse over dictionary $D$.
Let $x^\star \in \mathbb{K}^n$ be an $s$-sparse vector such that
\[
x^\star = \arg\min_{x \in \mathbb{K}^n} \{ \norm{f - Dx}_2 : \norm{x}_0 \leq s\}.
\]
We assume that the approximation error $f - D x^\star$ is small compared to $\norm{f}_2$.

The measurement vector $y \in \mathbb{K}^m$ is then given by
\[
y = A (D x^\star) + z
\]
where the distortion term $z$ includes both the approximation error $A (f - D x^\star)$ in modeling $f$ as an $s$-sparse signal over $D$,
and additive noise $w$,
\[
z = A (f - D x^\star) + w.
\]

Let $D \widehat{x}$ be an estimate of $f$ given by a greedy pursuit algorithm such that $\widehat{x}$ is exactly $s$-sparse.
Then,
\begin{align*}
\norm{f - D \widehat{x}}_2 {} & \leq \norm{f - D x^\star}_2 + \norm{D(x^\star - \widehat{x})}_2 \\
{} & \leq \norm{f - D x^\star}_2 + \sqrt{1 + \delta_{2s}(D)} \norm{\widehat{x} - x^\star}_2.
\end{align*}
Since the first term $\norm{f - D x^\star}_2$ corresponds to a fundamental limit for any greedy algorithm,
we will focus in the remainder of this section on bounding $\norm{\widehat{x} - x^\star}_2$.

To describe both the original greedy pursuit algorithms and our modifications,
we recall the definition of the \textit{hard thresholding} operator
that makes a given vector exactly $s$-sparse by zeroing the elements except the $s$-largest.
Formally, $H_s : \mathbb{K}^n \to \mathbb{K}^n$ is defined by
\[
H_s(x) \triangleq \arg\min_w \{ \norm{x - w} : \norm{w}_0 \leq s \}.
\]

\begin{remark}
All algorithms that appear in this section extend straightforwardly to the versions that exploit the structure of the support,
a.k.a. recovery algorithms for model-based compressed sensing \cite{BCDH10}.
The only task required in this modification is to replace the hard thresholding operator by a projection onto
$s$-sparse vectors with supports satisfying certain structure (e.g., tree).
The extension to model-based CS explicitly depends on the support and is only available for the greedy algorithms.
To focus on the main contribution of this paper, we will not pursue the details in this direction here.
\end{remark}

\subsection{Oblique Thresholding}
\label{subsec:dthres}

We start with a modification of the simple thresholding algorithm.
The thresholding algorithm computes an estimate of the support $J$ as the indices of the $s$ largest entries of $\Psi^* y$,
which is the support of $H_s(\Psi^* y)$.

Let us consider a special case, where $\Psi$ has full column rank and $y = \Psi x^\star$ is noise free.
While exact support recovery by naive thresholding of $\Psi^* y$ is not guaranteed,
thresholding of $\widetilde{\Psi}^* y$ with the biorthogonal dual $\widetilde{\Psi} = (\Psi^\dagger)^*$ is guaranteed to provide exact support recovery.
This example leaves room to improve thresholding using another properly designed matrix $\widetilde{\Psi}$.
In compressed sensing, we are interested in an underdetermined system given by $\Psi$; hence, $\Psi$ cannot have full column rank.
In this setting, the use of the canonical dual $\widetilde{\Psi} = (\Psi^\dagger)^*$ is not necessarily a good choice of $\widetilde{\Psi}$.

Schnass and Vandergheynst \cite{SchVan08} proposed a version of the thresholding algorithm that uses another matrix $\widetilde{\Psi}$ different from $\Psi$.
We call this algorithm \textit{Oblique Thresholding (ObThres)}, as an example of the oblique pursuit algorithms that will appear in the sequel.

\begin{algorithm}
\LinesNumbered
\SetAlgoNoLine
\caption{Oblique Thresholding (ObThres)}
\label{alg:bthres}

\BlankLine
$\widehat{J} \leftarrow \supp{H_s(\widetilde{\Psi}^*y)}$\;
\BlankLine
\end{algorithm}

Schnass and Vandergheynst \cite[Theorem~3]{SchVan08} showed a sufficient condition for exact support recovery by ObThres
in the noiseless case ($z = 0$), given by
\begin{equation}
\frac{\widetilde{\mu}_1(s,\Psi,\widetilde{\Psi})}{\min_j |\widetilde{\psi}_j^* \psi_j|}
< \frac{\min_{j \in J^\star} |(x^\star)_j|}{2 \norm{x^\star}_\infty}
\label{eq:cond_thres_SV}
\end{equation}
where the cross Babel function $\widetilde{\mu}_1(s,\Psi,\widetilde{\Psi})$ is defined by
\[
\widetilde{\mu}_1(s,\Psi,\widetilde{\Psi}) \triangleq \max_k \max_{\begin{subarray}{c} |J| = s \\ k \not\in J \end{subarray}} \sum_{j \in J} |\widetilde{\psi}_j^*\psi_j|.
\]

Since the left-hand side of (\ref{eq:cond_thres_SV}) is easily computed for given $\Psi$ and $\widetilde{\Psi}$,
Schnass and Vandergheynst \cite{SchVan08} proposed a numerical algorithm that designs $\widetilde{\Psi}$ to minimize the left-hand side of (\ref{eq:cond_thres_SV}).
However, the minimization problem is not convex and there is no guarantee for the quality of the resulting $\widetilde{\Psi}$.
Moreover, their optimality criterion for $\widetilde{\Psi}$ is based on the sufficient condition in (\ref{eq:cond_thres_SV}),
which is conservative (see \cite[Fig.~1]{SchVan08}).
In particular, unlike the RBOP, there is no known analysis of the (cross) Babel function of random frame matrices.

Instead, we derive an alternative sufficient condition for exact support recovery by ObThres,
given in terms of the RBOP of $(\Psi,\widetilde{\Psi})$.

\begin{theorem}[ObThres]
Let $x^\star \in \mathbb{K}^n$ be $s$-sparse with support $J^\star \subset [n]$.
Let $y = \Psi x^\star + z$.
Suppose that $\Psi$ and $\widetilde{\Psi}$ satisfy
\begin{equation}
\min_{j \in J^\star} |(x^\star)_j| > 2 \theta_{s+1}(\widetilde{\Psi}^* \Psi) \norm{x^\star}_2 + 2 \max_j \norm{\widetilde{\psi}_j}_2 \norm{z}_2.
\label{eq:thm:dthres:cond}
\end{equation}
Then, ObThres will identify $J^\star$ exactly.
\label{thm:dthres}
\end{theorem}

Compared to the numerical construction of $\widetilde{\Psi}$ by Schnass and Vandergheynst \cite{SchVan08},
our construction of $\widetilde{\Psi}$ in (\ref{eq:contrtildeA}) for a random frame matrix $\Psi$ has two advantages:
it is analytic; and it guarantees the RBOP of $(\Psi,\widetilde{\Psi})$.
Therefore, with this construction, the computation of $\theta_{s+1}(\widetilde{\Psi}^* \Psi)$ for given $\Psi$ and $\widetilde{\Psi}$,
which involves a combinatorial search, is not needed.

For the noiseless case ($z = 0$), the sufficient condition in (\ref{eq:thm:dthres:cond}) reduces to
\begin{equation}
\theta_{s+1}(\widetilde{\Psi}^* \Psi) < \frac{\min_{j \in J^\star} |(x^\star)_j|}{2 \norm{x^\star}_2}.
\label{eq:thm:dthres:cond:noiseless}
\end{equation}
Even in this case though, the upper bound in (\ref{eq:thm:dthres:cond:noiseless}) depends on both the dynamic range of $x^\star$ and the sparsity level $s$.
Therefore, compared to the guarantees of the iterative greedy pursuit algorithms in Section~\ref{subsec:iterative_algs},
the guarantee of ObThres is rather weak.
In fact, the other algorithms in Section~\ref{subsec:iterative_algs} outperform ObThres empirically too.
However, ObThres will serve as a building block for the iterative greedy pursuit algorithms.

\subsection{Oblique Matching Pursuit}
\label{subsec:obmp}

Matching Pursuit (MP) and Orthogonal Matching Pursuit (OMP) are forward greedy pursuit algorithms.
Unlike thresholding, which selects the support elements by a single step of hard thresholding,
(O)MP increments an estimate $\widehat{J}$ of the support $J^\star$ by adding one element per step chosen by a greedy criterion:
\begin{equation}
k^\star = \arg\max_k \big|\big(\Psi^* (y - \Psi \widehat{x})\big)_k\big|
\label{eq:sel_mpomp}
\end{equation}
where $y - \Psi \widehat{x}$ is the residual vector computed with the estimate $\widehat{x}$ of $x^\star$ spanned by $\Psi_{\widehat{J}}$.

Given the estimated support $\widehat{J}$, OMP updates the estimate $\widehat{x}$ optimally in the sense that $\widehat{x}$ satisfies
\begin{equation}
\widehat{x} = \arg\min_x \{ \norm{y - \Psi x}_2 : \supp{x} \subset \widehat{J} \}.
\label{eq:ls_omp}
\end{equation}
Therefore, the criterion in (\ref{eq:sel_mpomp}) for OMP reduces to
\begin{align}
k^\star {} & = \arg\max_k \big|\big(\Psi^* P_{\mathcal{R}(\Psi_{\widehat{J}})}^\perp y\big)_k\big| \nonumber \\
{} & = \arg\max_k \big| \langle P_{\mathcal{R}(\Psi_{\widehat{J}})}^\perp \psi_k ,~ P_{\mathcal{R}(\Psi_{\widehat{J}})}^\perp y \rangle \big|,
\label{eq:sel_omp}
\end{align}
which clearly describes the idea of ``orthogonal matching''.

Schnass and Vandergheynst \cite{SchVan08} proposed variations of MP and OMP that, using $\widetilde{\Psi}$, replace (\ref{eq:sel_mpomp})  by
\begin{equation}
k^\star = \arg\max_k \big|\big(\widetilde{\Psi}^* (y - \Psi \widehat{x})\big)_k\big|
\label{eq:SVsel_mpomp}
\end{equation}
and provided the following sufficient condition \cite[Theorem~4]{SchVan08} for exact support recovery by the OMP using (\ref{eq:SVsel_mpomp})
\begin{equation}
\frac{\widetilde{\mu}_1(s,\Psi,\widetilde{\Psi})}{\min_j |\widetilde{\psi}_j^* \psi_j|}
< \frac{1}{2}. \label{eq:cond_mpomp_SV}
\end{equation}
As for ObThres, they proposed to use a numerically designed $\widetilde{\Psi}$ that minimizes the left-hand side of (\ref{eq:cond_mpomp_SV})
(the same criterion as in their analysis of ObThres).

As discussed in the previous subsection, while easily computable for given $\Psi$ and $\widetilde{\Psi}$,
this sufficient condition is conservative and is not likely to be satisfied
even when $\widetilde{\Psi}$ is numerically optimized.
Thus, the resulting algorithm will have no guarantee.
Another weakness of the sufficient condition in (\ref{eq:cond_mpomp_SV}) is
that it has been derived without considering the orthogonal matching in OMP,
and thus ignores the improvement of OMP over MP.
Indeed, the same condition provides a partial guarantee of MP
that each step of MP will select an element of the support $J^\star$, which is not necessarily different from the previously selected ones.

In view of the weaknesses of the approach based on coherence, we turn instead to the RIP.
Davies and Wakin \cite{DavWak10} provided a sufficient condition for exact support recovery by OMP in terms of the RIP,
which has been refined in the setting of joint sparsity by Lee \textit{et al.} \cite[Proposition~7.11]{LeeBJ11SAMUSICarxiv}.
These analyses explicitly reflect the ``orthogonal matching''.
In particular, one key property required for the RIP-based sufficient conditions is that
the RIP is preserved under the orthogonal projection with respect to a few columns of $\Psi$, i.e.,
for all $\widehat{J} \subset [n]$ satisfying $|\widehat{J}| < s$,
\begin{equation}
\delta_s(P_{\mathcal{R}(\Psi_{\widehat{J}})}^\perp \Psi_{[n] \setminus \widehat{J}}) \leq \delta_s(\Psi).
\label{eq:ripwproj}
\end{equation}
This condition is an improvement on \cite[Lemma~3.2]{DavWak10} and was shown \cite[Proof of Proposition~7.11]{LeeBJ11SAMUSICarxiv}
using the interlacing eigenvalues property of the Schur complement \cite[Lemma~A.2]{LeeBJ11SAMUSICarxiv}.

The objective function in the orthogonal matching in (\ref{eq:sel_omp}) can be rewritten as
\begin{align}
\big| \langle P_{\mathcal{R}(\Psi_{\widehat{J}})}^\perp \psi_k ,~ P_{\mathcal{R}(\Psi_{\widehat{J}})}^\perp y \rangle \big|
= \big| \sum_{j \in J^\star \setminus \widehat{J}} \langle P_{\mathcal{R}(\Psi_{\widehat{J}})}^\perp \psi_k ,~ P_{\mathcal{R}(\Psi_{\widehat{J}})}^\perp \psi_j \rangle (x^\star)_j + \langle P_{\mathcal{R}(\Psi_{\widehat{J}})}^\perp \psi_k , z \rangle \big|.
\label{eq:obj_om}
\end{align}
The RIP of $\Psi$ together with (\ref{eq:ripwproj}) imply that the left-hand side of (\ref{eq:obj_om}) is close to $|(\Pi_{J^\star \setminus \widehat{J}} x^\star)_k|$,
with the perturbation bounded as a function the RIC of $\Psi$.
Then, orthogonal matching will choose $k^\star$ as
\[
k^\star = \arg\max_{k \in J^\star \setminus \widehat{J}} |(x^\star)_k|.
\]
This explains why orthogonal matching is a good strategy when $\Psi$ satisfies the RIP.

The OMP using (\ref{eq:SVsel_mpomp}) by Schnass and Vandergheynst \cite{SchVan08} still employs the orthogonal matching.
However, we are interested in the scenario where $\Psi$ does not satisfy the RIP but instead satisfies the RBOP with a certain $\widetilde{\Psi}$.
Unfortunately, unlike the RIP of $\Psi$,
the RBOP of $(\Psi,\widetilde{\Psi})$ is no longer valid
when the orthogonal projection $P_{\mathcal{R}(\Psi_{\widehat{J}})}^\perp$ is applied to both matrices.
Instead, we show that the RBOP of $(\Psi,\widetilde{\Psi})$ is preserved under an oblique projection,
which is analogous to the RIP result in (\ref{eq:ripwproj}).
To this end, we recall the definition of an oblique projection.

\begin{definition}[Oblique projection]
Let $\Set{V},\Set{W} \subset \Set{H}$ be two subspaces such that $\Set{V} \oplus \Set{W}^\perp = \Set{H}$.
The oblique projection onto $\Set{V}$ along $\Set{W}^\perp$, denoted by $E_{\Set{V},\Set{W}^\perp}$, is defined
as a linear map $E_{\Set{V},\Set{W}^\perp} : \Set{H} \rightarrow \Set{H}$ that satisfies
\begin{enumerate}
\item $(E_{\Set{V},\Set{W}^\perp}) x = x, \quad \forall x \in \Set{V}$.
\item $(E_{\Set{V},\Set{W}^\perp}) x = 0, \quad \forall x \in \Set{W}^\perp$.
\end{enumerate}
\end{definition}

By the definition of the oblique projection, it follows that
\begin{align*}
I_\mathcal{H} - E_{\Set{V},\Set{W}^\perp} = E_{\Set{W}^\perp,\Set{V}}
\quad \text{and} \quad
E_{\Set{V},\Set{W}^\perp}^* = E_{\Set{W},\Set{V}^\perp}.
\end{align*}
When $\Set{V} = \Set{W}$, the oblique projection reduces to the orthogonal projection $P_\Set{V}$ onto $\Set{V}$.

\begin{lemma}
Suppose that $M,\widetilde{M} \in \mathbb{K}^{m \times k}$ for $k \leq m$ satisfy that $\widetilde{M}^* M$ has full rank.
Then, $\mathcal{R}(M)$ and $\mathcal{R}(\widetilde{M})^\perp$ are complementary,
i.e., $\mathcal{R}(M) \cap \mathcal{R}(\widetilde{M})^\perp = \{0\}$.
\label{lemma:compl}
\end{lemma}

\begin{IEEEproof}[Proof of Lemma~\ref{lemma:compl}]
Assume that there is a nonzero $x \in \mathcal{R}(M) \cap \mathcal{R}(\widetilde{M})^\perp$.
Then, $x = M y$ for some $y \in \mathbb{K}^k$ and
$\widetilde{M}^* M y = 0$ since $x \in \mathcal{R}(\widetilde{M})^\perp = \mathcal{N}(\widetilde{M}^*)$.
Since $\widetilde{M}^* M$ is invertible, it follows that $y = 0$, which is a contradiction.
\end{IEEEproof}

The RBOP of $(\Psi,\widetilde{\Psi})$ implies that $\widetilde{\Psi}_{\widehat{J}}^* \Psi_{\widehat{J}}$ is invertible.
Furthermore, $\mathcal{R}(\Psi_{\widehat{J}})$ and $\mathcal{R}(\widetilde{\Psi}_{\widehat{J}})$ are complementary by Lemma~\ref{lemma:compl}.
Therefore, $\Psi_{\widehat{J}} (\widetilde{\Psi}_{\widehat{J}}^* \Psi_{\widehat{J}})^{-1} \widetilde{\Psi}_{\widehat{J}}^*$
is an oblique projection onto $\mathcal{R}(\Psi_{\widehat{J}})$ along $\mathcal{R}(\widetilde{\Psi}_{\widehat{J}})^\perp$.
It follows that $E = I_{|\widehat{J}|} - \Psi_{\widehat{J}} (\widetilde{\Psi}_{\widehat{J}}^* \Psi_{\widehat{J}})^{-1} \widetilde{\Psi}_{\widehat{J}}^*$
is an oblique projection onto $\mathcal{R}(\widetilde{\Psi}_{\widehat{J}})^\perp$ along $\mathcal{R}(\Psi_{\widehat{J}})$.

\begin{lemma}
Suppose that $\Psi, \widetilde{\Psi} \in \mathbb{K}^{m \times n}$ satisfy
\[
\theta_s(\widetilde{\Psi}^* \Psi) < 1.
\]
Let $\widehat{J} \subset [n]$.
Let $E = I_{|\widehat{J}|} - \Psi_{\widehat{J}} (\widetilde{\Psi}_{\widehat{J}}^* \Psi_{\widehat{J}})^{-1} \widetilde{\Psi}_{\widehat{J}}^*$.
Then,
\[
\theta_s(\widetilde{\Psi}_{[n] \setminus \widehat{J}}^* E \Psi_{[n] \setminus \widehat{J}}) \leq \theta_s(\widetilde{\Psi}^* \Psi).
\]
\label{lemma:rbpwoblp}
\end{lemma}
\begin{remark}
When $\widetilde{\Psi} = \Psi$, Lemma~\ref{lemma:rbpwoblp} reduces to (\ref{eq:ripwproj}).
\end{remark}

\begin{IEEEproof}[Proof of Lemma~\ref{lemma:rbpwoblp}]
Follows directly from Lemma~\ref{lemma:biorwoblp} in the Appendix.
\end{IEEEproof}

Lemma~\ref{lemma:rbpwoblp} suggests that if $\Psi$ does not satisfy the RIP but $\Psi$ and $\widetilde{\Psi}$ satisfy the RBOP,
then it might better to replace the orthogonal matching by the ``oblique matching'' given by
\begin{equation}
k^\star = \arg\max_k \big| \langle E^* \widetilde{\psi}_k ,~ E y \rangle \big|,
\label{eq:sel_domp}
\end{equation}
where $E$ is an oblique projector defined as
\[
E = I_{|\widehat{J}|} - \Psi_{\widehat{J}} (\widetilde{\Psi}_{\widehat{J}}^* \Psi_{\widehat{J}})^{-1} \widetilde{\Psi}_{\widehat{J}}^*.
\]

To affect the appropriate modification in OMP,
recall that orthogonal matching in (\ref{eq:sel_omp}) corresponds to
matching each column of $\widetilde{\Psi}$ with the residual $y - \Psi \widehat{x}$
computed with a solution $\widehat{x}$ to the least square problem in (\ref{eq:ls_omp}).
Similarly, oblique matching is obtained by replacing
the least square problem in (\ref{eq:ls_omp}) by the following weighted least square problem:
\[
\widehat{x} = \arg\min_x \{ \norm{\widetilde{\Psi}_{\widehat{J}}^*(y - \Psi x)}_2 : \supp{x} \subset \widehat{J} \}.
\]
We call the resulting forward greedy pursuit algorithm with the oblique matching \textit{oblique matching pursuit (ObMP)}.
ObMP is summarized in Algorithm~\ref{alg:bomp}.
In particular, when $\widetilde{\Psi} = \Psi$, ObMP reduces to the conventional OMP.
Like OMP, ObMP does not select the same support element more than once.
This is guaranteed since the selected columns are within the null space of the oblique projection associated with the oblique matching.

\begin{algorithm}
\LinesNumbered
\SetAlgoNoLine
\caption{Oblique Matching Pursuit (ObMP)}
\label{alg:bomp}

\BlankLine
$\widehat{J} \leftarrow \emptyset$; $\widehat{x} \leftarrow 0$\;
\BlankLine
\While{$|\widehat{J}| < s$}{
    \BlankLine
    $\displaystyle k^\star \leftarrow \arg\max_{k \in [n] \setminus \widehat{J}} \big|\big(\widetilde{\Psi}^* (y - \Psi \widehat{x})\big)_k\big|$\;
    \BlankLine
    $\widehat{J} \leftarrow \widehat{J} \cup \{k^\star\}$\;
    \BlankLine
    $\widehat{x} \leftarrow \arg\min_x \{ \norm{\widetilde{\Psi}_{\widehat{J}}^*(y - \Psi x)}_2 : \supp{x} \subset \widehat{J} \}$\; \label{step:bomp:WLS}
    \BlankLine
}
\BlankLine
\end{algorithm}

Next, we present a guarantee of ObMP in terms of the RBOP.
\begin{proposition}[A Single Step of ObMP]
Let $x^\star \in \mathbb{K}^n$ be $s$-sparse with support $J^\star \subset [n]$.
Let $y = \Psi x^\star + z$ and $J \subsetneq J^\star$.
Suppose that $\Psi$ and $\widetilde{\Psi}$ satisfy
\begin{align}
\norm{\Pi_{J^\star \setminus J} x^\star}_\infty - 2 \theta_{s+1}(\widetilde{\Psi}^* \Psi) \norm{\Pi_{J^\star \setminus J} x^\star}_2
> \left(\frac{\norm{\Psi_{J^\star}} \norm{\widetilde{\Psi}_{J^\star}}}{1-\theta_{s+1}(\widetilde{\Psi}^* \Psi)}\right) 2 \max_j \norm{\widetilde{\psi}_j}_2 \norm{z}_2.
\label{eq:prop:doblmp:cond4}
\end{align}
where the coordinate projection $\Pi_{J^\star \setminus J}$ is defined in (\ref{eq:def:coorproj}).
Then, the next step of ObMP given $J$ will identify an element of $J^\star \setminus J$.
\label{prop:doblmp}
\end{proposition}

The following theorem is a direct consequence of Proposition~\ref{prop:doblmp}.
\begin{theorem}[ObMP]
Let $x^\star \in \mathbb{K}^n$ be $s$-sparse with support $J^\star \subset [n]$.
Let $y = \Psi x^\star + z$.
Suppose that $\Psi$ and $\widetilde{\Psi}$ satisfy
\begin{align}
\min_{j \in J^\star} |(x^\star)_j| \left( \min_{J \subset J^\star, J \neq \emptyset} \frac{\norm{\Pi_J x^\star}_\infty}{\norm{\Pi_J x^\star}_2} - 2 \theta_{s+1}(\widetilde{\Psi}^* \Psi) \right)
> \left(\frac{\norm{\Psi_{J^\star}} \norm{\widetilde{\Psi}_{J^\star}}}{1-\theta_{s+1}(\widetilde{\Psi}^* \Psi)}\right) 2 \max_j \norm{\widetilde{\psi}_j}_2 \norm{z}_2.
\label{eq:thm:doblmp:cond}
\end{align} %
Then, ObMP will identify $J^\star$ exactly.
\label{thm:doblmp}
\end{theorem}

If $\widetilde{\Psi} = \Psi$, then ObMP reduces to OMP; hence, Proposition~\ref{prop:doblmp} reduces to the single measurement vector case of \cite[Proposition~7.11]{LeeBJ11SAMUSICarxiv},
with the requirement on $\Psi$ in (\ref{eq:prop:doblmp:cond4}) reduced to
\begin{equation}
\norm{\Pi_{J^\star \setminus J} x^\star}_\infty - 2 \delta_{s+1}(\Psi) \norm{\Pi_{J^\star \setminus J} x^\star}_2
> 2 \max_j \norm{\widetilde{\psi}_j}_2 \norm{z}_2. \label{eq:prop:omp:cond2}
\end{equation}
In fact, the proof of Proposition~\ref{prop:doblmp} in the Appendix is carried out by modifying that of \cite[Proposition~7.11]{LeeBJ11SAMUSICarxiv}
so that the non-Hermitian case is appropriately managed.
Similarly, the guarantee of ObMP in Theorem~\ref{thm:doblmp} reduces to that of OMP given by
\begin{align}
\min_{j \in J^\star} |(x^\star)_j| \left( \min_{J \subset J^\star, J \neq \emptyset} \frac{\norm{\Pi_J x^\star}_\infty}{\norm{\Pi_J x^\star}_2} - 2 \delta_{s+1}(\Psi) \right)
> 2 \max_j \norm{\psi_j}_2 \norm{z}_2.
\label{eq:thm:omp:cond2}
\end{align}

To satisfy the condition in (\ref{eq:thm:omp:cond2}),
it is required that $\delta_{s+1}(\Psi) < c$ for some $c \in (0,1)$ that depends on $x^\star$.
As will be shown in Section~\ref{sec:rbp}, this RIP condition is often not satisfied in a typical scenario of practical applications.
In contrast, $\theta_{s+1}(\widetilde{\Psi}^* \Psi)$ is still satisfied with a properly designed $\widetilde{\Psi}$ in the same scenario.
Therefore, the guarantee of ObMP in Theorem~\ref{thm:doblmp} is less demanding than the corresponding guarantee of OMP.

We observe that the bound on the noise amplification in ObMP is larger by the factor
$\frac{\norm{\Psi_{J^\star}} \norm{\widetilde{\Psi}_{J^\star}}}{1-\delta_{s+1}(\Psi,\widetilde{\Psi})}$ than in OMP.
This factor is an upper bound on the spectral norm of the oblique projection onto $\mathcal{R}(\Psi_{\widehat{J}})$
along $\mathcal{R}(\widetilde{\Psi}_{\widehat{J}})^\perp$.
The analogous operator in OMP is an orthogonal projector and the spectral norm is trivially bounded from above by 1.
However, when oblique matching is used instead of orthogonal matching, this is no longer valid.
The spectral norm of the oblique projection is the reciprocal of
the cosine of the angle between the two subspaces $\mathcal{R}(\Psi_{\widehat{J}})$ and $\mathcal{R}(\widetilde{\Psi}_{\widehat{J}})$.
This result is consistent with the known analysis of oblique projections.\footnote{
In a general context, unrelated to CS, it has been shown \cite{Eld03}
that oblique projectors are suboptimal in terms of minimizing the projection residual,
which is however bounded within factor $\frac{1}{\cos\theta}$ of the optimal error.}

For the noiseless case ($z = 0$), the sufficient condition in (\ref{eq:thm:doblmp:cond}) reduces to
\begin{equation}
\theta_{s+1}(\widetilde{\Psi}^* \Psi) < \min_{J \subset J^\star, J \neq \emptyset} \frac{\norm{\Pi_J x^\star}_\infty}{2\norm{\Pi_J x^\star}_2}.
\label{eq:thm:doblmp:cond:noiseless}
\end{equation}
Compared to the sufficient condition for ObThres in (\ref{eq:thm:dthres:cond:noiseless}),
where depending on the dynamic range of $x^\star$, the upper bound on the RBOC can be arbitrary small,
the right-hand side in (\ref{eq:thm:doblmp:cond:noiseless}) is no smaller than $\frac{1}{2\sqrt{s}}$ for any $x^\star$.
Although ObMP is guaranteed under a milder RBOP condition than ObThres,
the corresponding sufficient condition is still demanding compared to those of iterative greedy pursuit algorithms.

However, ObThres and ObMP are important,
since they provide basic building blocks for the iterative greedy pursuit algorithms.
The thresholding and OMP algorithms have been modified to ObThres and ObMP
by replacing two basic blocks, ``$\Psi^*$ followed by hardthresholding'', and ``orthogonal matching'',
to ``$\widetilde{\Psi}^*$ followed by hardthresholding'', and ``oblique matching'', respectively.
The modifications of these two basic blocks will similarly alter the other greedy pursuit algorithms and their RIP-based guarantees.

In the next section, we present the oblique versions of some iterative greedy pursuit algorithms (CoSaMP, SP, IHT, and HTP).
However, the conversion to the oblique version of both algorithm and guarantee is not restricted to these examples.
It applies to any other greedy pursuit algorithm that builds on these basic blocks (e.g., Fast Nesterov's Iterative Hard Thresholding (FNIHT) \cite{CevJaf10}).

\subsection{Iterative Oblique Greedy Pursuit Algorithms}
\label{subsec:iterative_algs}

Compressive Sampling Matching Pursuit (CoSaMP) \cite{NeeTro09} and Subspace Pursuit (SP) \cite{DaiMil09}
are more sophisticated greedy pursuit algorithms that iteratively update the $s$-sparse estimate of $x^\star$.
At a high level, both CoSaMP and SP update the estimate of the true support using the following procedure:
\begin{enumerate}
\item Augment the estimated set by adding more indices that might include the missing elements of the true support.
\item Refine the augmented set to a subset with $s$ elements.
\end{enumerate}
The two algorithms differ in the size of the increment in the augmentation.
More important, SP completes each iteration by updating the residual using an orthogonal projection, which is similar to that of OMP.
CoSaMP and SP provide RIP-based guarantees, which are comparable to those of $\ell_1$-based solutions such as BP.

Both algorithms use the basic building blocks of correlation maximization by hard thresholding and least squares problems.
Therefore, following the same approach we used to modify thresholding and OMP to ObThres and ObMP,
we modify CoSaMP and SP to their oblique versions called \textit{Oblique CoSaMP (ObCoSaMP)} and \textit{Oblique SP (ObSP)}, respectively.
ObCoSaMP and ObSP are summarized in Algorithm~\ref{alg:obcosamp} and Algorithm~\ref{alg:obsp}.

\begin{algorithm}
\LinesNumbered
\SetAlgoNoLine
\caption{Oblique Compressive Matching Pursuit (ObCoSaMP)}
\label{alg:obcosamp}

\BlankLine
\While{stop condition not satisfied}{
    \BlankLine
    $\widetilde{J}_{t+1} \leftarrow \supp{x_t} \cup \supp{H_{2s}\big(\widetilde{\Psi}^*(y - \Psi x_t)\big)}$\;
    \BlankLine
    $\displaystyle \widetilde{x} \leftarrow \arg\min_x \left\{ \big\|\widetilde{\Psi}_{\widetilde{J}_{t+1}}^*(y - \Psi x)\big\|_2 : \supp{x} \subset \widetilde{J}_{t+1} \right\}$\;
    \BlankLine
    $x_{t+1} \leftarrow H_s(\widetilde{x})$\;
    \BlankLine
    $t \leftarrow t+1$\;
    \BlankLine
}
\BlankLine
\end{algorithm}

\begin{algorithm}
\LinesNumbered
\SetAlgoNoLine
\caption{Oblique Subspace Pursuit (ObSP)}
\label{alg:obsp}

\BlankLine
\While{stop condition not satisfied}{
    \BlankLine
    $\widetilde{J}_{t+1} \leftarrow \supp{x_t} \cup \supp{H_s\big(\widetilde{\Psi}^*(y - \Psi x_t)\big)}$\;
    \BlankLine
    $\displaystyle \widetilde{x} \leftarrow \arg\min_x \left\{ \big\|\widetilde{\Psi}_{\widetilde{J}_{t+1}}^*(y - \Psi x)\big\|_2 : \supp{x} \subset \widetilde{J}_{t+1} \right\}$\;
    \BlankLine
    $J_{t+1} \leftarrow \supp{H_s(\widetilde{x})}$\;
    \BlankLine
    $\displaystyle x_{t+1} \leftarrow \arg\min_x \left\{ \big\|\widetilde{\Psi}_{J_{t+1}}^*(y - \Psi x)\big\|_2 : \supp{x} \subset J_{t+1} \right\}$\;
    \BlankLine
    $t \leftarrow t+1$\;
    \BlankLine
}
\BlankLine
\end{algorithm}

\textit{Iterative Hard Thresholding (IHT)} \cite{BluDav09} and \textit{Hard Threshold Pursuit (HTP)} \cite{Fou11HTP}
are two other greedy pursuit algorithms with RIP-based guarantees.
HTP is a modified version of IHT, which updates the residual using orthogonal projection like SP.
Since both IHT and HTP use the same basic building blocks used in the other greedy pursuit algorithms, they too admit the oblique versions.
We name these modified versions \textit{Oblique IHT (ObIHT)} and \textit{Oblique HTP (ObHTP)}.
ObIHT and ObHTP are summarized in Algorithm~\ref{alg:obiht} and Algorithm~\ref{alg:obhtp}.
Note that these iterative oblique greedy pursuit algorithms reduce to their conventional counterparts when $\widetilde{\Psi} = \Psi$.

\begin{algorithm}
\LinesNumbered
\SetAlgoNoLine
\caption{Oblique Iterative Hard Thresholding (ObIHT)}
\label{alg:obiht}

\BlankLine
\While{stop condition not satisfied}{
    \BlankLine
    $x_{t+1} \leftarrow H_s\big(x_t + \widetilde{\Psi}^*(y - \Psi x_t)\big)$\;
    \BlankLine
    $t \leftarrow t+1$\;
    \BlankLine
}
\BlankLine
\end{algorithm}

\begin{algorithm}
\LinesNumbered
\SetAlgoNoLine
\caption{Oblique Hard Thresholding Pursuit (ObHTP)}
\label{alg:obhtp}

\BlankLine
\While{stop condition not satisfied}{
    \BlankLine
    $J_{t+1} \leftarrow \supp{H_s\big(x_t + \widetilde{\Psi}^*(y - \Psi x_t)\big)}$\;
    \BlankLine
    $\displaystyle x_{t+1} \leftarrow \arg\min_x \left\{ \big\|\widetilde{\Psi}_{J_{t+1}}^*(y - \Psi x)\big\|_2 : \supp{x} \subset J_{t+1} \right\}$\;
    \BlankLine
    $t \leftarrow t+1$\;
    \BlankLine
}
\BlankLine
\end{algorithm}

We briefly review the currently available RIP-based guarantees of the original algorithms.
The guarantees of the iterative greedy pursuit algorithms were provided in their original papers \cite{NeeTro09,DaiMil09,BluDav09,Fou11HTP}.
In particular, Needell and Tropp, in their technical report on CoSaMP \cite{NeeTro08},
showed that CoSaMP (with exact arithmetic) converges within a finite number of iterations,
which is at most $O(s)$ for the worst case and can be as small as $O(\ln s)$.
We will show that the same analysis applies to SP, HTP, and their oblique versions.

The guarantees of the iterative greedy pursuit algorithms are provided by sufficient conditions given in a common form
$\delta_{ks}(\Psi) < c$, where the condition becomes more demanding for larger $k$ and smaller $c$.
Recently, Foucart \cite{Fou10} refined the guarantees of CoSaMP and IHT by increasing required $c$.
We will show that the guarantee of SP is similarly improved
using similar techniques and replacing triangle inequalities by the Pythagorean theorem when applicable.\footnote{
As an aside, inspired by the existing RIP analysis
that $\delta_{ks}(\Psi) < c$ holds with $m = O(k s c^{-2} \ln^4 n)$,
Foucart \cite{Fou10} proposed to compare sufficient conditions by comparing the values of $k c^{-2}$.
Nevertheless, this comparison is heuristic and only relies on sufficient conditions for the worst case guarantee.
Therefore, it is not necessarily true that an algorithm with smaller $k c^{-2}$ performs better.}

Next, we show that the RIP-based guarantees of the iterative greedy pursuit algorithms
are replaced by similar guarantees of the corresponding oblique pursuit greedy algorithms, in terms of the RBOP.
In fact, the modification of the guarantees is rather straightforward, as was the modification of the algorithms.
We only provide the full derivation for the RBOP-based guarantee of ObSP.
Replacing $\widetilde{\Psi}$ by $\Psi$ in the result and the derivation will provide an RIP-based guarantee for SP.
The guarantees of the other iterative oblique pursuit algorithms (ObCoSaMP, ObIHT, and ObHTP)
are obtained by similarly modifying the corresponding results \cite{Fou10,Fou11HTP}.
Therefore, we do not repeat the derivations but only state the results.

\begin{theorem}
Let $\texttt{Alg} \in \{\text{ObSP}, \text{ObCoSaMP}, \text{ObIHT}, \text{ObHTP}\}$.
Let $(x_t)_{t \in \mathbb{N}}$ be the sequence generated by algorithm \texttt{Alg}. Then
\begin{equation}
\norm{x_{t+1} - x^\star}_2 \leq \rho \norm{x_t - x^\star}_2 + \tau \norm{z}_2
\label{eq:idgp:recursion}
\end{equation}
where $\rho$ and $\tau$ are positive constants depending on \texttt{Alg}, given as explicit functions of
$\theta_{ks}(\widetilde{\Psi}^* \Psi)$, $\delta_{ks}(\Psi)$, and $\delta_{ks}(\widetilde{\Psi})$.
Moreover, $\rho$, which only depends on $\theta_{ks}(\widetilde{\Psi}^* \Psi)$, is less than 1,
provided that the condition in Table~\ref{tab:rboptbl} specified by \texttt{Alg} is satisfied.
\label{thm:pgoblp}
\end{theorem}
\begin{IEEEproof}[Proof of Theorem~\ref{thm:pgoblp}]
We only provide the proof for ObSP in Appendix~\ref{subsec:proof:thm:pgoblp}.
The formulae for $\rho$ and $\tau$ are provided for all listed algorithms.
\end{IEEEproof}

\begin{table*}
\caption{The RBOP condition required for linear convergence in Theorem~\ref{thm:pgoblp}.}
\label{tab:rboptbl}
\begin{center}
\begin{tabular}{c||*{3}{c|}c}
\hline
\texttt{Alg} & ObCoSaMP & ObSP & ObIHT & ObHTP \\\hline\hline
$\theta_{ks}(\widetilde{\Psi}^* \Psi) < c$
& $\theta_{4s}(\widetilde{\Psi}^* \Psi) < 0.384$
& $\theta_{3s}(\widetilde{\Psi}^* \Psi) < 0.325$
& $\theta_{3s}(\widetilde{\Psi}^* \Psi) < 0.5$
& $\theta_{3s}(\widetilde{\Psi}^* \Psi) < 0.577$ \\\hline
\end{tabular}
\end{center}
\end{table*}

For $\rho < 1$, (\ref{eq:idgp:recursion}) implies that in the noiseless case the iteration converges linearly at rate $\rho$ to the true solution,
whereas in the noisy case the error at convergence is $\norm{x_\infty - x^\star}_2 = \tau/(1 - \rho) \norm{z}_2$.

Unlike ObIHT, the other algorithms (ObCoSaMP, ObSP, and ObHTP) involve the step of updating the estimate by solving a least squares problem.
This additional step provides the property in Lemma~\ref{lemma:fcoblp}, which enables the finite convergence of the algorithms.

\begin{lemma}
Let $\texttt{Alg} \in \{\text{ObSP}, \text{ObCoSaMP}, \text{ObHTP}\}$.
Let $(x_t)_{t \in \mathbb{N}}$ be the sequence generated by \texttt{Alg}.
Then, the approximation error $\norm{x_t - x^\star}_2$ is less than the $\ell_2$ norm of the missed components of $x^\star$ to within a constant factor $\bar{\rho}$ plus the noise term, i.e.,
\begin{equation}
\norm{x_{t+1} - x^\star}_2 \leq \bar{\rho} \norm{\Pi_{J_{t+1}}^\perp x^\star}_2 + \bar{\tau} \norm{z}_2
\label{eq:idgp:uncapenergy}
\end{equation}
where $\bar{\rho}$ and $\bar{\tau}$ are positive constants given as explicit functions (depending on \texttt{Alg}) of
$\theta_{ks}(\widetilde{\Psi}^* \Psi)$, $\delta_{ks}(\Psi)$, and $\delta_{ks}(\widetilde{\Psi})$.
\label{lemma:fcoblp}
\end{lemma}
\begin{IEEEproof}[Proof of Lemma~\ref{lemma:fcoblp}]
Lemma~\ref{lemma:fcoblp} is an intermediate step for proving Theorem~\ref{thm:pgoblp}.
For example, for ObSP, it corresponds to Lemma~\ref{lemma:obsp_step1} in Appendix~\ref{subsec:proof:thm:pgoblp}.
For the other algorithms, we only provide the formulae for $\bar{\rho}$ and $\bar{\tau}$ in Appendix~\ref{subsec:proof:thm:pgoblp}.
\end{IEEEproof}

Needell and Tropp \cite{NeeTro08} showed finite convergence of CoSaMP.
The same analysis also applies to ObCoSaMP, ObSP, and ObHTP.
To show this, let us recall the relevant definitions from the technical report on CoSaMP \cite{NeeTro08}.
The \textit{component bands} $(\Set{B}_j)$ of $x^\star$ are by
\[
\Set{B}_j \triangleq \{ i : 2^{-(j+1)} \norm{x^\star}_2^2 < |(x^\star)_i| \leq 2^{-j} \norm{x^\star}_2^2 \}, \quad \forall j \in \mathbb{Z} \cup \{0\}.
\]
Then, the \textit{profile} of $x^\star$ is defined as the number of nonempty component bands.
By definition, the profile of $x^\star$ is not greater than the sparsity level of $x^\star$.

\begin{lemma}[{A Paraphrase of \cite[Theorem~B.1]{NeeTro08}}]
Let $p$ be the profile of $x^\star$.
Suppose that $(x_t)_{t \in \mathbb{N}}$ satisfies \cref{eq:idgp:recursion,eq:idgp:uncapenergy}.
Then, for
\begin{align}
t > L + p \ln\left( 1 + 2 \left[ \bar{\rho} + \frac{\tau}{\bar{\tau}} (1-\rho-\eta) \right] \sqrt{\frac{s}{p}} \right)
\left[\ln \left(\frac{1}{1-\eta}\right)\right]^{-1},
\label{eq:thm:iterbnd:thres}
\end{align}
it holds that
\[
\norm{x_t - x^\star}_2 \leq
\left[ \rho^L \left( \frac{\tau\bar{\rho}}{1-\rho-\eta} + \bar{\tau} \right) + \left( \frac{1-\rho^L}{1-\rho} \right) \tau \right] \norm{z}_2.
\]
\label{lemma:iterbnd}
\end{lemma}

The minimal number of iterations for the convergence (the right-hand side of (\ref{eq:thm:iterbnd:thres})) is maximized when $p = s$ \cite{NeeTro08}.
The following theorem is a direct consequence of Theorem~\ref{thm:pgoblp}, Lemma~\ref{lemma:fcoblp}, and Lemma~\ref{lemma:iterbnd}.

\begin{theorem}
Let $\texttt{Alg} \in \{\text{ObSP}, \text{ObCoSaMP}, \text{ObHTP}\}$.
Suppose that $\theta_{ks}(\widetilde{\Psi}^* \Psi) < c$ holds depending on \texttt{Alg} as in Table~\ref{tab:rboptbl}.
After $t_{\max} = C_1 (s+1)$ iterations, \texttt{Alg} provides an estimate $\widehat{x}$ satisfying $\norm{\widehat{x} - x^\star}_2 \leq C_2 \norm{z}_2$.
Here, $k$, $c$, $C_1$, and $C_2$ are constants, specified by \texttt{Alg}.
\label{thm:itercnt}
\end{theorem}

The fast convergence of iterative greedy pursuit algorithms that involve the least square steps is important.
When the problem is large (e.g., in CS imaging, the image size is typically $512 \times 512$ pixels),
solving the least squares problems is the most computationally demanding step of the recovery algorithms.
Empirically, as the theory suggests, the iterative algorithms (ObCoSaMP, ObSP, and ObHTP) converge at most within $O(s)$ iterations,
and are even more computationally efficient than the non-iterative ObMP.

\begin{remark}
The extension of greedy pursuit algorithms and their RIP-based guarantees to those based on the RBOP is not restricted to the aforementioned algorithms.
For example, Fast Nesterov's Iterative Hard Thresholding (FNIHT) \cite{CevJaf10} is another promising algorithm
with an RIP-based guarantee, which will extend likewise.
\end{remark}

\section{Restricted Biorthogonality Property}
\label{sec:rbp}

In this section, we show that the RBOP-based guarantees of oblique pursuits apply to realistic models of compressed sensing systems in practice.
For example, when applied to random frame matrices,
the guarantees remain valid even though the i.i.d. sampling is done according to a nonuniform distribution.
Recall that the guarantees of oblique pursuits in Section~\ref{sec:algorithms} required
$\theta_{ks}(\widetilde{\Psi}^* \Psi) < c$ where $k \in \{2,3,4\}$ and $c \in (0,1)$ are constants specified by the algorithm in question.
The noise amplification in the reconstruction for these guarantees also depend on $\delta_{ks}(\Psi)$ and $\delta_{ks}(\widetilde{\Psi})$.
However, unlike $\theta_{ks}(\widetilde{\Psi}^* \Psi)$,
the RICs $\delta_{ks}(\Psi)$ and $\delta_{ks}(\widetilde{\Psi})$ need not be less than 1 to provide the guarantees.
In fact, as discussed later, reasonable upper bounds on $\delta_{ks}(\Psi)$ and on $\delta_{ks}(\widetilde{\Psi})$
(possibly larger than 1) are obtained with no additional conditions whenever $\theta_{ks}(\widetilde{\Psi}^* \Psi) < c$ is achieved.
Therefore, we may focus on the condition $\theta_{ks}(\widetilde{\Psi}^* \Psi) < c$.
Also recall that the guarantees for the corresponding conventional pursuit algorithms require $\delta_{ks}(\Psi) < c$,
for $k \in \{2,3,4\}$, $c \in (0,1)$, with the same $k$ and $c$ as the corresponding oblique pursuits.
To compare the guarantees of the oblique vs. the conventional pursuit algorithms,
assuming $k \in \{2,3,4\}$ and $c \in (0,1)$ arbitrarily fixed constants, we compare the difficulty in achieving
the respective bounds on $\delta_{ks}(\Psi)$ and $\theta_{ks}(\widetilde{\Psi}^* \Psi)$.
While both properties are guaranteed when $m = O(s \ln^4 n)$,
$\theta_{ks}(\widetilde{\Psi}^* \Psi) < c$ is achieved without additional conditions required for achieving $\delta_{ks}(\Psi) < c$,
which are often violated in practical compressed sensing.

\subsection{General Estimate}
We extend \cite[Theorem~8.4]{Rau10} to the following theorem, so that it provides an upper bound on $\theta_s(\widetilde{\Psi}^* \Psi)$.
\begin{theorem}
Let $\Psi, \widetilde{\Psi} \in \mathbb{K}^{m \times n}$ be random matrices not necessarily mutually independent,
each with i.i.d. rows with elements bounded in magnitude as
\begin{equation}
\max_{k,\ell} |(\Psi)_{k,\ell}| \leq \frac{K}{\sqrt{m}} \quad \text{and} \quad
\max_{k,\ell} |(\widetilde{\Psi})_{k,\ell}| \leq \frac{\widetilde{K}}{\sqrt{m}}
\label{eq:thm:urbp:eq1}
\end{equation}
for $K,\widetilde{K} \geq 1$.
Then, $\theta_s(\widetilde{\Psi}^*\Psi) < \delta + \theta_s(\mathbb{E} \widetilde{\Psi}^* \Psi)$ holds with probability $1 - \eta$ provided that
\begin{align}
m {} & \geq C_1 \delta^{-2} \left( K \sqrt{2 + \theta_s(\mathbb{E} \Psi^* \Psi)} + \widetilde{K} \sqrt{2 + \theta_s(\mathbb{E} \widetilde{\Psi}^* \widetilde{\Psi})} \right)^2 s (\ln s)^2 \ln n \ln m, \label{eq:thm:urbp:eq2} \\
m {} & \geq C_2 \delta^{-2} \widetilde{K} \max(K,\widetilde{K}) s \ln(\eta^{-1}) \label{eq:thm:urbp:eq3}
\end{align}
for universal constants $C_1$ and $C_2$.
\label{thm:urbp}
\end{theorem}

\begin{IEEEproof}[Proof of Theorem~\ref{thm:urbp}]
See Appendix~\ref{subsec:proof:thm:urbp}.
\end{IEEEproof}

Letting $\widetilde{\Psi} = \Psi$ in Theorem~\ref{thm:urbp} provides the following corollary.\footnote{
A direct derivation of Corollary~\ref{cor:urip} might provide better constants,
but we do not attempt to optimize the universal constants.}

\begin{corollary}
Let $\Psi \in \mathbb{K}^{m \times n}$ be a random matrix with i.i.d. rows with elements bounded in magnitude as
$\max_{k,\ell} |(\Psi)_{k,\ell}| \leq \frac{K}{\sqrt{m}}$ for $K \geq 1$.
Then, $\delta_s(\Psi) < \delta + \theta_s(\mathbb{E} \Psi^* \Psi)$ holds with probability $1 - \eta$ provided that
\begin{align}
m {} & \geq C_1 \delta^{-2} K^2 4 \left[2 + \theta_s(\mathbb{E} \Psi^* \Psi) \right] s (\ln s)^2 \ln n \ln m, \label{eq:cor:urip:eq2} \\
m {} & \geq C_2 \delta^{-2} K^2 s \ln(\eta^{-1}) \label{eq:cor:urip:eq3}
\end{align}
for universal constants $C_1$ and $C_2$.
\label{cor:urip}
\end{corollary}

The following corollary is obtained by combining Theorem~\ref{thm:urbp} and Corollary~\ref{cor:urip}
applied to $\Psi$ and to $\widetilde{\Psi}$, respectively.
Corollary~\ref{cor:urbp} aims to provide an upper bound on $\theta_s(\widetilde{\Psi}^* \Psi)$.
It also provides upper bounds on both $\delta_s(\Psi)$ and $\delta_s(\widetilde{\Psi})$.

\begin{corollary}
Let $\Psi, \widetilde{\Psi} \in \mathbb{K}^{m \times n}$ be random matrices with i.i.d. rows with elements bounded in magnitude as
$\max_{k,\ell} |(\Psi)_{k,\ell}| \leq \frac{K}{\sqrt{m}}$ and
$\max_{k,\ell} |(\widetilde{\Psi})_{k,\ell}| \leq \frac{\widetilde{K}}{\sqrt{m}}$ for $K, \widetilde{K} \geq 1$.
Then, $\theta_s(\widetilde{\Psi}^*\Psi) < \delta + \theta_s(\mathbb{E} \widetilde{\Psi}^* \Psi)$,
$\delta_s(\Psi) < \delta + \theta_s(\mathbb{E} \Psi^* \Psi)$,
and $\delta_s(\widetilde{\Psi}) < \delta + \theta_s(\mathbb{E} \widetilde{\Psi}^* \widetilde{\Psi})$
hold with probability $1 - \eta$ provided that
\begin{align}
m {} & \geq C_1 \delta^{-2} \max(K^2,\widetilde{K}^2)
4 \Big[2 + \max\left(\theta_s(\mathbb{E} \Psi^* \Psi), \theta_s(\mathbb{E} \widetilde{\Psi}^* \widetilde{\Psi})\right) \Big]
s (\ln s)^2 \ln n \ln m, \label{eq:cor:urbp:eq2} \\
m {} & \geq C_2 \delta^{-2} \max(K^2,\widetilde{K}^2) s \ln(\eta^{-1}) \label{eq:cor:urbp:eq3}
\end{align}
for universal constants $C_1$ and $C_2$.
\label{cor:urbp}
\end{corollary}

Corollary~\ref{cor:urip} and Corollary~\ref{cor:urbp} have very different implications.
Corollary~\ref{cor:urip} guarantees that $\delta_{ks}(\Psi) < c$ holds with high probability when $m = O(s \ln^4 n)$
if $\max_{k,\ell} |(\Psi)_{k,\ell}| = O(\frac{1}{\sqrt{m}})$ and $\theta_{ks}(\mathbb{E} \Psi^* \Psi) < 0.5 c$.
The former condition implies that the rows of $\Psi$ are incoherent to the standard basis vectors and is called the incoherence property.
As will be discussed in later subsections, the latter condition, $\theta_{ks}(\mathbb{E} \Psi^* \Psi) < 0.5 c$,
is often difficult to satisfy for small $c \in (0,1)$, in particular, in practical settings of compressed sensing.
Although this condition has not been shown to be a necessary condition for $\delta_{ks}(\Psi) < c$,
no alternative analysis is available for random frame matrices.
In contrast, $\theta_{ks}(\mathbb{E} \widetilde{\Psi}^* \Psi)$ can be made small by an appropriate choice of $\widetilde{\Psi}$,
which by Corollary~\ref{cor:urbp} suffices to make $\theta_{ks}(\widetilde{\Psi}^* \Psi) < c$.
In fact, it is often the case that $\widetilde{\Psi}$ can be chosen to make $\theta_{ks}(\mathbb{E} \widetilde{\Psi}^* \Psi)$ much smaller than $\theta_{ks}(\mathbb{E} \Psi^* \Psi)$, or even zero, and to satisfy the incoherence property at the same time.
In this case, $\theta_{ks}(\widetilde{\Psi}^* \Psi) < c$ is guaranteed, whereas $\delta_{ks}(\Psi)$ is not guaranteed so.
This key difference in the guarantees in Corollaries~\ref{cor:urip} and \ref{cor:urbp} establishes the advertised result that
the RBOP-based guarantees of oblique pursuits apply to more general cases,
in which the RIP-based guarantees of the corresponding conventional pursuits fail.

In the next subsections, we elaborate the comparison of the two different approaches:
oblique pursuits with RBOP-based guarantees vs. conventional pursuits with RIP-based guarantees
(per Corollaries~\ref{cor:urip} and \ref{cor:urbp}) in more concrete scenarios
in which $\Psi$ is given as the composition of the sensing matrix $A$ obtained from a frame and the dictionary $D$ with certain properties.

\subsection{Case I: Sampled Frame $A$ and Nonredundant $D$ of Full Rank}

We first consider the case of $\Psi = A D$,
where the sensing matrix $A$ is constructed from a frame $(\phi_\omega)_{\omega \in \Omega}$ by (\ref{eq:contrA})
using a probability measure $\nu$, and the sparsifying dictionary $D$ is nonredundant $(n \leq d)$ with full column rank.

Using the isotropy property, $\mathbb{E} A^* A = I_d$,
conventional RIP analysis \cite[Theorem~8.4]{Rau10} showed that
$\delta_s(\Psi) < \delta$ holds with high probability for $m = O(\delta^{-2} s \ln^4 n)$ under the following ideal assumptions:
\begin{description}
\item[\hspace{-3mm}(AI-1)] $(\phi_\omega)_{\omega \in \Omega}$ is a tight frame, i.e., $\Phi \Phi^* = I_d$ where $\Phi,\Phi^*$ denotes the associated synthesis and the analysis operators.
\item[\hspace{-3mm}(AI-2)] $\nu$ is the uniform measure.
\item[\hspace{-3mm}(AI-3)] $D^* D = I_n$.
\end{description}

Corollary~\ref{cor:urip} generalizes \cite[Theorem~8.4]{Rau10}, so that
the same RIP result continues holds when the ideal assumptions are ``slightly'' violated.
To quantify this statement, we introduce the following metrics that measure the deviation from the ideal assumptions.

\begin{itemize}
\item Nonuniform distribution $\nu$:
We additionally assume that $\nu$ is absolutely continuous with respect to $\mu$.\footnote{If $\Omega$ is a finite set, then $\mu$ is the counting measure and any probability measure $\nu$ is absolutely continuous.}
Define
\begin{equation}
\nu_{\min} \triangleq \text{ess} \inf_{\omega \in \Omega} \frac{d\nu}{d\mu}(\omega)
\quad \text{and} \quad
\nu_{\max} \triangleq \text{ess} \sup_{\omega \in \Omega} \frac{d\nu}{d\mu}(\omega)
\label{eq:def_numinmax}
\end{equation}
where the essential infimum and supremum are w.r.t. to the measure $\nu$.
If $\Omega$ is a finite set, then $\frac{d\nu}{d\mu}(\omega)$ reduces to the probability
that $\omega \in \Omega$ will be chosen, multiplied by the cardinality of $\Omega$.
By their definitions, $\nu_{\min}$ and $\nu_{\max}$ satisfy $\nu_{\min} \leq 1 \leq \nu_{\max}$.
Note that $\nu_{\min}$ and $\nu_{\max}$ measure how different $\nu$ is from the uniform measure $\mu$.
In particular, $\nu_{\min} = \nu_{\max} = 1$ if $\nu$ coincides with $\mu$.

\item Non-tight frame $(\phi_\omega)_{\omega \in \Omega}$:
Multiplying $\Psi$ and $y$ by a common scalar does not modify the inverse problem $\Psi x = y$.
Therefore, replacing $\Phi \Phi^*$ by the same matrix multiplied by an appropriate scalar,
we assume without loss of generality that
\begin{equation}
\lambda_1(\Phi \Phi^*) = 1 + \frac{\kappa(\Phi \Phi^*) - 1}{\kappa(\Phi \Phi^*) + 1}
\label{eq:optscale1}
\end{equation}
and
\begin{equation}
\lambda_d(\Phi \Phi^*) = 1 - \frac{\kappa(\Phi \Phi^*) - 1}{\kappa(\Phi \Phi^*) + 1}
\label{eq:optscale2}
\end{equation}
where $\kappa(\Phi \Phi^*)$ denotes the condition number of $\Phi \Phi^*$.
Equations (\ref{eq:optscale1}) and (\ref{eq:optscale2}) imply
\[
\theta_d(\Phi \Phi^*) = \norm{\Phi \Phi^* - I_d} = \frac{\kappa(\Phi \Phi^*) - 1}{\kappa(\Phi \Phi^*) + 1}
\]
where the first identity follows from the definition of $\theta_d$.
Note that $\theta_d(\Phi \Phi^*) = 0$ if $\Phi \Phi^* = I_d$.

\item Non-orthonormal $D$:
Similarly, for nonredundant $D$, we assume without loss of generality that
\begin{equation}
\lambda_1(D^*D) = 1 + \frac{\kappa(D^*D) - 1}{\kappa(D^*D) + 1}
\label{eq:optscale3}
\end{equation}
and
\begin{equation}
\lambda_n(D^*D) = 1 - \frac{\kappa(D^*D) - 1}{\kappa(D^*D) + 1}
\label{eq:optscale4}
\end{equation}
where $\kappa(D^*D)$ denotes the condition number of $D^*D$.
Equations (\ref{eq:optscale3}) and (\ref{eq:optscale4}) imply
\[
\theta_n(D^* D) = \norm{D^*D - I_n} = \frac{\kappa(D^*D) - 1}{\kappa(D^*D) + 1}.
\]
Note that $\theta_n(D^* D) = 0$ if $D$ corresponds to an orthonormal basis, i.e., $D^* D = I_n$.
\end{itemize}

Now, invoking Corollary~\ref{cor:urip} with the above metrics, we obtain the following Theorem~\ref{thm:urip},
of which Theorem~\ref{thm:uripsimple} is a simplified version.
Under the ideal assumptions, $K_0$ vanishes and Theorem~\ref{thm:urip} reduces to \cite[Theorem~8.4]{Rau10}.
\begin{theorem}
Let $(\phi_\omega)_{\omega \in \Omega}$ and $D = [d_1,\ldots,d_n] \in \mathbb{K}^{d \times n}$ satisfy
$\sup_\omega \max_j |\langle \phi_\omega, d_j \rangle| \leq K$ for $K \geq 1$.
Let $A \in \mathbb{K}^{m \times d}$ be constructed from $(\phi_\omega)_{\omega \in \Omega}$ by (\ref{eq:contrA}) using a probability measure $\nu$,
and let $\Psi = A D$.
Let $\nu_{\min}$ and $\nu_{\max}$ be defined in (\ref{eq:def_numinmax}).
Then, $\delta_s(\Psi) < \delta + K_0$ holds with probability $1 - \eta$ provided that
$m \geq C_1 (1+K_0)^2 K^2 \delta^{-2} s (\ln s)^2 \ln n \ln m$ and $m \geq C_2 K^2 \delta^{-2} s \ln(\eta^{-1})$
for universal constants $C_1$ and $C_2$
where $K_0$ is given in terms of $\nu_{\min}$, $\nu_{\max}$, $\delta_s(D)$, and $\theta_d(\Phi \Phi^*)$ by
\begin{align}
K_0 = \max(1 - \nu_{\min}, \nu_{\max} - 1) + \nu_{\max} [\delta_s(D) + \theta_d(\Phi \Phi^*) + \delta_s(D) \cdot \theta_d(\Phi \Phi^*)].
\label{eq:thm:urip:K0}
\end{align}
\label{thm:urip}
\end{theorem}
\begin{IEEEproof}
See Appendix~\ref{subsec:proof:thm:urip}.
\end{IEEEproof}

Theorem~\ref{thm:urip} shows that the ideal assumptions (AI-1) - (AI-3) for achieving the RIP of $\Psi$ can be relaxed to a certain extent.
However, even the relaxed assumptions are still too demanding to be satisfied in many practical applications of compressed sensing.
When the ideal assumptions are not all satisfied,
each deviation increases $K_0$ and the obtained upper bound on $\delta_s(\Psi)$ also increases.
For example, when $\Phi \Phi^* = I_d$ and $D^* D = I_n$, depending on $\nu$,
the upper bound on $\delta_s(\Psi)$ may turn out to be even larger than 1, which fails to provide an RIP-based guarantee.
As another example, when $\nu = \mu$ and $\Phi \Phi^* = I_d$
(the rows of $A$ are obtained from i.i.d. samples from a tight frame according to the uniform distribution),
$\delta_s(D)$ determines the quality of the upper bound.
Although, in general, computation of $\delta_s(D)$ is NP hard,
an easy upper bound on $\delta_s(D)$ is given as $\delta_n(D) = \norm{D^* D - I_n}$.
Now, note that $\delta_n(D) \geq 0.6$ for $\kappa(D) \geq 2$.
Therefore, considering that the RIP-based guarantee of HTP \cite{Fou11HTP} requires $\delta_{3s}(\Psi) < 0.57$,
which is the largest upper bound on $\delta_{3s}(\Psi)$ among all sufficient conditions for known RIP-based guarantees.
This suggests that even when the other ideal assumptions are satisfied, $D$ needs to be near ideally conditioned.
This strong requirement on $D$ is often too restrictive, in particular, for learning a data-adaptive dictionary $D$.

Next, we show that $\theta_s(\widetilde{\Psi}^* \Psi) < c$ is achieved more easily, without the aforementioned restriction on $\Phi$, $\nu$, or $D$.
To this end, we would like to use Corollary~\ref{cor:urbp}; however, the $\widetilde{K}$ parameter in Corollary~\ref{cor:urbp} requires further attention.
While the incoherence parameter $K$ is determined by the inverse problem,
the other incoherence parameter $\widetilde{K}$ is determined by our own choice of $\widetilde{A}$ and $\widetilde{D}$.
Recall the construction of $\widetilde{\Psi} = \widetilde{A} \widetilde{D}$:
matrix $\widetilde{A} \in \mathbb{K}^{m \times d}$ is constructed from the dual frame $(\widetilde{\phi}_\omega)_{\omega \in \Omega}$ by (\ref{eq:contrtildeA}) using the same probability measure $\nu$ used to construct $A$ per (\ref{eq:contrA}),
whereas $\widetilde{D}$ is given as $\widetilde{D} = D (D^*D)^{-1}$, so that $\widetilde{D}^* D = I_n$.
It follows that $\widetilde{K}$ is related to $\Phi$ and $D$, and thus to $K$.
By deriving an upper bound on $\widetilde{K}$ in terms of $K$ and using it in Corollary~\ref{cor:urbp}, we obtain the following theorem.

\begin{theorem}
Let $(\phi_\omega)_{\omega \in \Omega}$ and $D = [d_1,\ldots,d_n] \in \mathbb{K}^{d \times n}$ satisfy
$\sup_\omega \max_j |\langle \phi_\omega, d_j \rangle| \leq K$ for $K \geq 1$.
Let $\nu$ be a probability measure on $\Omega$ such that its derivative is strictly positive.
Let $A,\widetilde{A} \in \mathbb{K}^{m \times n}$ be random matrices constructed from a biorthogonal frame $(\phi_\omega,\widetilde{\phi}_\omega)_{\omega \in \Omega}$ by (\ref{eq:contrA}) and (\ref{eq:contrtildeA}), respectively using $\nu$.
Let $\Psi = A D$ and $\widetilde{\Psi} = \widetilde{A} \widetilde{D}$ where $\widetilde{D} = D (D^*D)^{-1}$.
Let $\nu_{\min}$ and $\nu_{\max}$ be defined in (\ref{eq:def_numinmax}).
Then, $\theta_s(\widetilde{\Psi}^*\Psi) < \delta$,
$\delta_s(\Psi) < \delta + K_1$, and $\delta_s(\widetilde{\Psi}) < \delta + K_1$
hold with probability $1 - \eta$ provided that
\begin{align}
m {} & \geq C_1 (1 + K_1)^2 K_2^2 \delta^{-2} s (\ln s)^2 \ln n \ln m, \\
m {} & \geq C_2 K_2^2 \delta^{-2} s \ln(\eta^{-1})
\end{align}
for universal constants $C_1$ and $C_2$,
where $K_1$ and $K_2$ are given in terms of $K$, $\nu_{\min}$, $\nu_{\max}$, $\delta_n(D)$, and $\theta_d(\Phi \Phi^*)$ by
{\allowdisplaybreaks
\begin{align}
K_1 {} & = \max(1 - \nu_{\max}^{-1}, \nu_{\min}^{-1} - 1) + \max(\nu_{\max}, \nu_{\min}^{-1}) \nonumber \\
{} & \quad \cdot \Bigg\{1 + \frac{\delta_n(D)}{1 - \delta_n(D)}
+ \frac{\theta_d(\Phi \Phi^*)}{1 - \theta_d(\Phi \Phi^*)} + \frac{\delta_n(D) \theta_d(\Phi \Phi^*)}{[1 - \delta_n(D)][1 - \theta_d(\Phi \Phi^*)]}\Bigg\}
\label{eq:thm:urbpBOK1}
\end{align}
and
\begin{align}
K_2 = \frac{\norm{(D^*D)^{-1}}_{\ell_1^n \to \ell_1^n}}{\nu_{\min}^2}
\left[ K + \left(\sup_{\omega \in \Omega} \norm{\phi_\omega}_{\ell_2^d}\right) \cdot \frac{\theta_d(\Phi \Phi^*)}{1 - \theta_d(\Phi \Phi^*)} \cdot \left( \max_{j \in [n]} \norm{d_j}_{\ell_2^d} \right) \right].
\label{eq:thm:urbpBOK2}
\end{align}} %
\label{thm:urbpBO}
\end{theorem}
\begin{IEEEproof}
See Appendix~\ref{subsec:proof:thm:urbpBO}.
\end{IEEEproof}

With any significant violation of the ideal assumptions (AI-1) -- (AI-3),
Theorem~\ref{thm:urip} fails to provide $\delta_{ks}(\Psi) < c$,
whereas Theorem~\ref{thm:urbpBO} still provides $\theta_{ks}(\widetilde{\Psi}^* \Psi) < c$.
Therefore, the RBOP-based guarantee of recovery by oblique pursuits is a significant improvement over the conventional RIP-based guarantees,
in the sense that the former applies to a practical setup (subset selection with a nonuniform distribution, non-tight frame, and non-orthonormal dictionary) while the latter does not.
This is because violation of the ideal assumptions does not affect the upper bound on $\theta_s(\widetilde{\Psi}^* \Psi)$ in Theorem~\ref{thm:urbpBO}.
Instead, it increases the upper bounds on $\delta_s(\Psi)$ and $\delta_s(\widetilde{\Psi})$.
However, in the guarantees of oblique pursuits, unlike $\theta_s(\widetilde{\Psi}^* \Psi)$,
the restricted isometry constants $\delta_s(\Psi)$ and $\delta_s(\widetilde{\Psi})$ need not be bounded from above by a certain threshold.

\begin{example}
We show the implication of Theorem~\ref{thm:urbpBO} in a 2D Fourier imaging example.
The corresponding numerical results for this scenario can be found in Section~\ref{sec:numres}.
The measurements are taken over random frequencies sampled i.i.d. from the uniform 2D lattice grid $\Omega$ with a nonuniform measure $\nu$.
The signal of interest is sparse over a data-adaptive dictionary $D$, which is invertible ($n = d$) and has block diagonal structure.

More specifically, $D$ in this example is constructed as follows.
Recently, Ravishankar and Bresler \cite{RavBre12} proposed an efficient algorithm that learns a data-adaptive square transform $T$
with a regularizer on its condition number.
When the condition number of $T$ is reasonably small, $D$ given by $D = T^{-1}$ serves as a good dictionary for sparse representation.
In particular, they designed a patch-based transform $T$ that applies to each patch of the image.
When the patches are nonoverlapping, $T$ and $D$ have block diagonal structure; hence, applying $D$ and $D^*$ is computationally efficient.
Furthermore, when the patches are much smaller than the image,
each atom in $D$ is sparse and has low mutual coherence to the Fourier transform that applies to the entire image.
For example, $D \in \mathbb{C}^{512 \times 512}$ used in the numerical experiment in Section~\ref{sec:numres}
was designed so that it applies to $8 \times 8$ pixel patches.
It has condition number 1.99, which implies $\delta_n(D) = 0.60$.
We also observed that $D$ satisfies $\norm{(D^* D)^{-1}}_{\ell_1^d \to \ell_1^d} = 2.13$.

Since $(\phi_\omega)_{\omega \in \Omega}$ corresponding to the 2D DFT is tight,
it follows that $\theta_d(\Phi \Phi^*) = 0$.
Therefore, the expressions for $K_1$ and $K_2$ in \cref{eq:thm:urbpBOK1,eq:thm:urbpBOK2} reduce to
\begin{equation}
K_1 = \max(1 - \nu_{\max}^{-1}, \nu_{\min}^{-1} - 1)  + 2.5 \max(\nu_{\max}, \nu_{\min}^{-1})
\label{eq:thm:urbpBOK1simple}
\end{equation}
and
\begin{equation}
K_2 = \frac{2.13}{\nu_{\min}^2} K.
\label{eq:thm:urbpBOK2simple}
\end{equation}
Recall that $\nu_{\min}$ and $\nu_{\max}$ in this scenario correspond to the minimum and maximum probability that
a measurement is taken at a certain frequency component.
The simplified expressions of $K_1$ and $K_2$ in (\ref{eq:thm:urbpBOK1simple}) and (\ref{eq:thm:urbpBOK2simple}) show quantitatively
how the use of nonuniform distribution for the i.i.d. sampling in the construction of a random frame matrix increases the required number of measurements.
\label{example:patchD}
\end{example}

\subsection{Case II: Sampled Frame $A$ and Overcomplete $D$ with the RIP}
\label{subsec:rbopcsi}

The analysis in the previous section focused on the case where the dictionary $D$ is not redundant.
In fact though, the analysis extends to certain cases of redundant/overcomplete $D$.
One such case is when $D$ is, like $A$, a random frame matrix.
Then, using a construction similar to our construction of $\widetilde{A}$ will produce a matrix $\widetilde{D}$ with $\mathbb{E} \widetilde{D}^* D = I_n$, which combined with $\mathbb{E} \widetilde{A}^* A = I_d$ provides $\mathbb{E} \widetilde{\Psi}^* \Psi = I_n$.
However, usually, $D$ is given as a deterministic matrix (e.g., concatenation of analytic bases, analytic frame, data-adaptive dictionary, etc).
Therefore, in the general redundant $D$ case, using the biorthogonal dual of $D$ as $\widetilde{D}$ is not a promising approach.
Instead, we focus in the remainder of this subsection on the case where $D$ satisfies the RIP with small $\delta_s(D)$.
Using $\widetilde{\Psi} = \widetilde{A} D$, we show the RBOP of $(\Psi,\widetilde{\Psi})$ in this case.

{\allowdisplaybreaks
\begin{theorem}
Let $(\phi_\omega)_{\omega \in \Omega}$ and $D = [d_1,\ldots,d_n] \in \mathbb{K}^{d \times n}$ satisfy
$\sup_\omega \max_j |\langle \phi_\omega, d_j \rangle| \leq K$ for $K \geq 1$.
Let $A,\widetilde{A} \in \mathbb{K}^{m \times n}$ be random matrices constructed from a biorthogonal frame $(\phi_\omega,\widetilde{\phi}_\omega)_{\omega \in \Omega}$ by (\ref{eq:contrA}) and (\ref{eq:contrtildeA}), respectively using a probability measure $\nu$.
Suppose that $\delta_s(D) < 1$.
Let $\Psi = A D$, and $\widetilde{\Psi} = \widetilde{A} D$.
Let $\nu_{\min}$ and $\nu_{\max}$ be defined in (\ref{eq:def_numinmax}).
Then, $\theta_s(\widetilde{\Psi}^*\Psi) < \delta + \delta_s(D)$, $\delta_s(\Psi) < \delta + K_1$, and $\delta_s(\widetilde{\Psi}) < \delta + K_1$
hold with probability $1 - \eta$ provided that
\begin{align}
m {} & \geq C_1 (1 + K_1)^2 K_2^2 \delta^{-2} s (\ln s)^2 \ln n \ln m, \\
m {} & \geq C_2 K_2^2 \delta^{-2} s \ln(\eta^{-1})
\end{align}
for universal constants $C_1$ and $C_2$,
where $K_1$ and $K_2$ are given in terms of $K$, $\nu_{\min}$, $\nu_{\max}$, $\delta_s(D)$, and $\theta_d(\Phi \Phi^*)$ by
\begin{align*}
K_1 {} & = \max(1 - \nu_{\max}^{-1}, \nu_{\min}^{-1} - 1) + \max(\nu_{\max}, \nu_{\min}^{-1}) \\
{} & \qquad \cdot \left(1 + \delta_s(D) + \frac{\theta_d(\Phi \Phi^*)}{1 - \theta_d(\Phi \Phi^*)} + \frac{\delta_s(D) \theta_d(\Phi \Phi^*)}{1 - \theta_d(\Phi \Phi^*)}\right).
\end{align*}
and
\begin{align*}
K_2 {} & = \frac{1}{\nu_{\min}}
\Bigg[ K + \left(\sup_{\omega \in \Omega} \norm{\phi_\omega}_{\ell_2^d}\right)
\frac{\theta_d(\Phi \Phi^*)}{1 - \theta_d(\Phi \Phi^*)} \cdot \left( \max_{j \in [n]} \norm{d_j}_{\ell_2^d} \right) \Bigg].
\end{align*}
\label{thm:urbpOC}
\end{theorem}}
\begin{IEEEproof}
See Appendix~\ref{subsec:proof:thm:urbpOC}.
\end{IEEEproof}

\subsection{Case III: Sampled Tight Frame $A$ and Orthonormal Basis $D$ / RIP Matrix $D$}
\label{subsec:orthocase}

In the special case where the use of a nonuniform distribution for the i.i.d. sampling in the construction of $A$ is
the only cause for the resulting failure of the exact/near isotropy property,
the failure of the conventional RIP analysis can be fixed differently.
Recall that the construction of $\widetilde{A}$ in (\ref{eq:contrtildeA})
only involves the weighting of rows of a matrix obtained from the biorthogonal dual frame $(\widetilde{\phi}_{\omega})_{\omega \in \Omega}$,
with sampling at the same indices as used for the construction of $A$ from the frame $(\phi_\omega)_{\omega \in \Omega}$.
Therefore, for the special case when $(\phi_\omega)_{\omega \in \Omega}$ is a tight frame
and $D^* D = I_n$, it is possible to derive the RIP of a preconditioned version of $\Psi$.

We construct a preconditioned sensing matrix $\widehat{A}$ as
\begin{align}
(\widehat{A})_{k,\ell} {} & = \frac{1}{\sqrt{m}} \left[ \frac{d\nu}{d\mu}(\omega_k) \right]^{-1/2} \overline{(\phi_{\omega_k})_\ell}, \quad \forall k \in [m],~ \ell \in [d] \label{eq:contrhatA}
\end{align}
where $(\omega_k)_{k=1}^m$ are the same sampling points used in the construction of $A$ in (\ref{eq:contrA}).
Then, by construction, $\widehat{A}$ satisfies the isotropy property $\mathbb{E} \widehat{A}^* \widehat{A} = I_d$.
Furthermore, if $\sup_\omega \max_j |\langle \phi_\omega, d_j \rangle| \leq K$,
then $\max_{k,\ell} |(\widehat{\Psi})_{k,\ell}| \leq \frac{\nu_{\min}^{-1/2} K}{\sqrt{m}}$ holds.

In this case, it suffices to invoke \cite[Theorem~8.4]{Rau10} to show the RIP of $\widehat{\Psi}$.
Invoking instead Theorem~\ref{thm:urip}, this approach extends in a straightforward way to the case where $D$ satisfies the RIP.
In the case of tight frame $A$ and $D$ that is an orthobasis or an RIP matrix,
these results provide an alternative (and equivalent) approach to obtain guaranteed algorithms, without invoking RBOP.
In particular,
defining $\Lambda$ as the diagonal matrix given by $(\Lambda)_{j,j} = [(d\nu / d\mu)(\omega_k)]^{-1/2}$ for $j \in [m]$,
conventional recovery algorithms with an RIP-based guarantee can be used to solve the modified inverse problem $\Lambda \Psi = \Lambda y$.

As discussed earlier, non-tight frame and/or non-orthonormal or non-RIP dictionaries arise in applications of compressed sensing,
and in these instances too the conventional RIP analysis fails.
We are currently investigating whether, and if so how, the above approach to ``preconditioned'' $\widehat{\Psi}$
may be extended in general beyond the aforementioned cases.

\section{Numerical Results}
\label{sec:numres}

We performed two experiments to compare the oblique pursuits to their conventional counterparts and to other methods.

In the first experiment, we tested the algorithms on a generic data set.
Synthesis operators $\Phi$ and $\widetilde{\Phi}$ for a random biorthogonal frame $(\phi_\omega,\widetilde{\phi}_\omega)_{\omega \in \Omega}$
were generated using random unitary matrices $U,V \in \mathbb{R}^{n \times n}$ and a fixed diagonal matrix $\Sigma$
as $\Phi = U \Sigma V^*$ and $\widetilde{\Phi} = U \Sigma^{-1} V^*$.
The diagonal entries of $\Sigma$ increase linearly from $\sqrt{\frac{2}{3}}$ to $\sqrt{\frac{4}{3}}$.
Sensing matrix $A \in \mathbb{R}^{m \times n}$ was formed by $m$ random rows of $\Phi$ scaled by $\frac{1}{\sqrt{m}}$,
where the row selection was done with respect to the uniform distribution.
Then, the condition number of $\mathbb{E} A^* A$ is 2 and the isotropy property is not satisfied.
In this setting the oblique pursuit algorithms are different from their conventional counterparts.
Signal $x^\star \in \mathbb{K}^n$ is exactly $s$-sparse in the standard basis vectors ($D = I_n$)
and the nonzero elements have unit magnitude and random signs.
The success of each algorithm is defined as the exact recovery of the support.

\begin{figure*}[htb]
\begin{center}
\leavevmode
\makebox[0in]{
\begin{tabular}{cccccc}
\psfig{figure=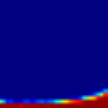,scale=2.0}
&
\psfig{figure=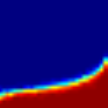,scale=2.0}
&
\psfig{figure=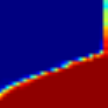,scale=2.0}
&
\psfig{figure=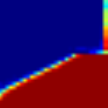,scale=2.0}
&
\psfig{figure=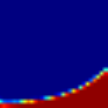,scale=2.0}
&
\psfig{figure=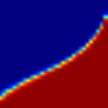,scale=2.0}
\\
Thres & OMP & CoSaMP & SP & IHT & HTP
\\
\psfig{figure=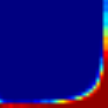,scale=2.0}
&
\psfig{figure=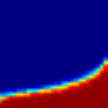,scale=2.0}
&
\psfig{figure=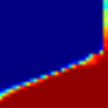,scale=2.0}
&
\psfig{figure=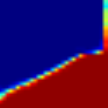,scale=2.0}
&
\psfig{figure=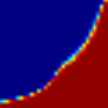,scale=2.0}
&
\psfig{figure=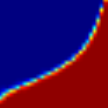,scale=2.0}
\\
ObThres & ObMP & ObCoSaMP & ObSP & ObIHT & ObHTP
\end{tabular}
}
\end{center}
\caption{Phase transition of support recovery by various greedy pursuit algorithms (the horizontal and vertical axes denote
the ratio $m/n$ of number of measurements to number of unknowns and ratio $s/m$ of sparsity level to number of measurements, respectively):
signal $x^\star$ is exactly $s$-sparse with nonzero entries that are $\pm 1$ with random sign.
$n = 1024$, $\text{SNR} = 30 \text{dB}$, $\kappa = 2$.
}
\label{fig:PhaseTransition}
\end{figure*}

Figure~\ref{fig:PhaseTransition} shows the empirical phase transition of each algorithm as a function of $m/n$ and $s/n$.
The results were averaged over 100 repetitions.
Oblique versions of thresholding and IHT showed dramatic improvement in performance
while the performance of the other algorithms is almost the same.
While the oblique pursuit algorithms can be guaranteed without $A$ satisfying the isotropy property,
the modification of the algorithms at least do not result in the degradation of the performance.

In the second experiment, we tested the algorithms on a CS Fourier imaging system.
The partial DFT sensing matrix $A$ used in this experiment
was constructed using the variable density suggested by Lustig \textit{et al.} \cite{LusDP07}.
We used a data-adaptive square dictionary $D$ that applies to non-overlapping patches.
Dictionary $D$ was learned from the fully sampled complex valued brain image using the algorithm proposed by Ravishankar and Bresler \cite{RavBre12}
(See Example~\ref{example:patchD} for more detail).
The resulting $D$ was well conditioned with condition number $\kappa(D) = 1.99$.
The Oblique pursuit algorithms use $\widetilde{\Psi} = \widetilde{A} \widetilde{D}$,
where $\widetilde{D}$ is given as the biorthogonal dual $\widetilde{D} = (D^{-1})^*$.
Since the patches are non-overlapping,
applying $D$, $\widetilde{D}$, and their adjoint operators are patch-wise operations, and are computed efficiently.

The input image was a phantom image obtained by $s$-sparse approximation over the dictionary $D$
of an original brain image with sparsity ratio $s/n = 0.125$.
Our goal in this experiment is not to compete with the state of the art of recovery algorithms in CS imaging system;
rather, we want to check whether the oblique pursuit algorithms perform competitively with their conventional counterparts
in a setting where the RBOP of $(\Psi,\widetilde{\Psi})$ is guaranteed.
This motivates our choice of a simplified test scenario.
We also compare the oblique pursuit algorithms to simple zero filling,
and to NESTA \cite{BBC11nesta} that solves the $\ell_1$ analysis formulation \cite{CENR11}.
In fact, when the original brain image is used as the input image,
all sparsity-based reconstruction algorithms, including NESTA, performed worse than zero filling.\footnote{
To achieve good performance on the original image requires a more sophisticated recovery algorithm with overlapping patches,
and adaptive sparsity level \cite{RavBre11dlmri}.}
To get a meaningful result in this setting, we replaced the input image by an exactly $s$-sparse phantom
obtained by the $s$-sparse approximation of the original brain image.

\begin{table*}[htb]
\caption{Quality (PSNR in decibels) of images reconstructed from noisy variable density Fourier samples with measurement SNR = 30 decibels.
Results averaged over 100 random sampling patterns.}
\label{tab:psnr_noisy}
\setlength{\tabcolsep}{10pt}
\begin{center}
\begin{tabular}{c}
\begin{tabular}{c|*{6}{c|}|c}
\hline
{} & Thres & CoSaMP & SP & IHT & HTP & $\ell_1$-Analysis & Zero Filling \\\hline\hline
conventional & 14.48 & 42.93 & 45.24 & 9.06 & 40.02 & \multirow{2}{*}{34.04} & \multirow{2}{*}{34.73} \\\cline{1-6}
oblique & 37.59 & 43.30 & 44.79 & 44.96 & 45.53 & & \\\hline
\end{tabular}
\\
(a) Downsample by 2
\\
\begin{tabular}{c|*{6}{c|}|c}
\hline
{} & Thres & CoSaMP & SP & IHT & HTP & $\ell_1$-Analysis & Zero Filling \\\hline\hline
conventional & 9.34 & 29.46 & 34.74 & 9.34 & 31.58 & \multirow{2}{*}{30.96} & \multirow{2}{*}{31.55} \\\cline{1-6}
oblique & 31.13 & 32.21 & 36.17 & 31.10 & 36.26 & & \\\hline
\end{tabular}
\\
(b) Downsample by 3
\end{tabular}
\end{center}
\end{table*}

\begin{figure*}[htb]
\begin{center}
\leavevmode
\makebox[0in]{
\begin{tabular}{cccccc}
\psfig{figure=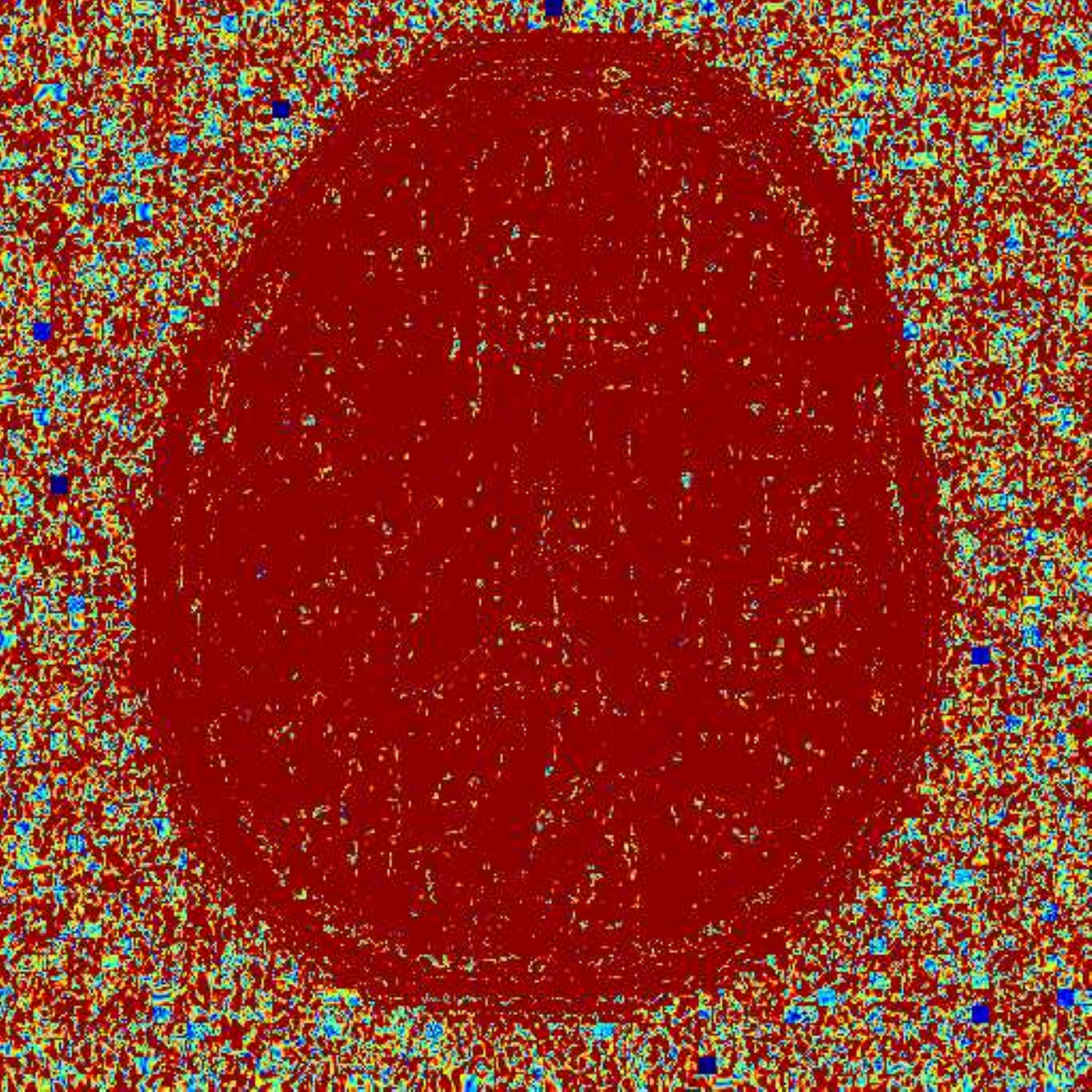,scale=.20}
&
\psfig{figure=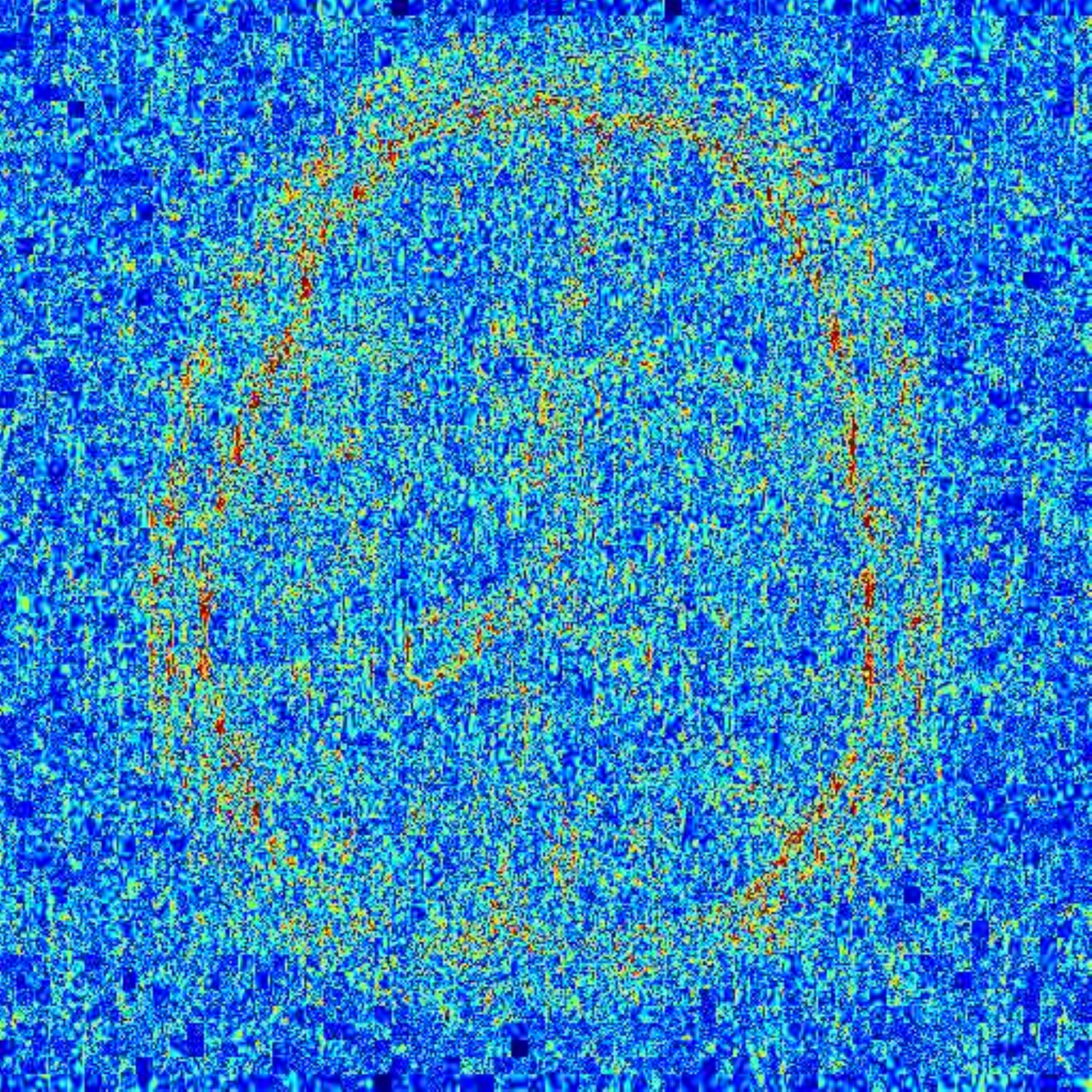,scale=.20}
&
\psfig{figure=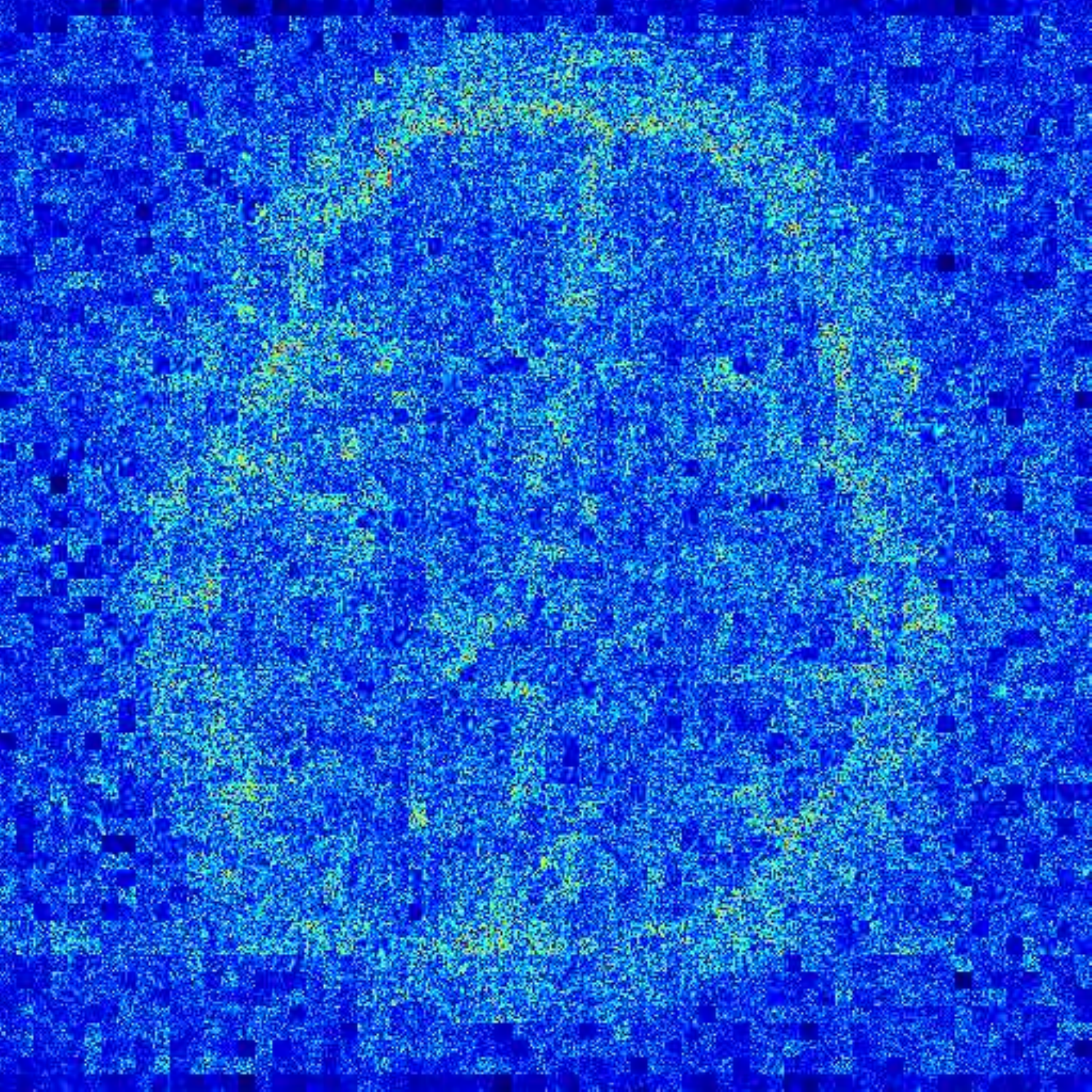,scale=.20}
&
\psfig{figure=./fig/colorbar.pdf,height=38mm} \hspace{-10mm}
\\
Thres & CoSaMP & SP \\
9.34 dB & 29.38 dB & 34.58 dB \\
\psfig{figure=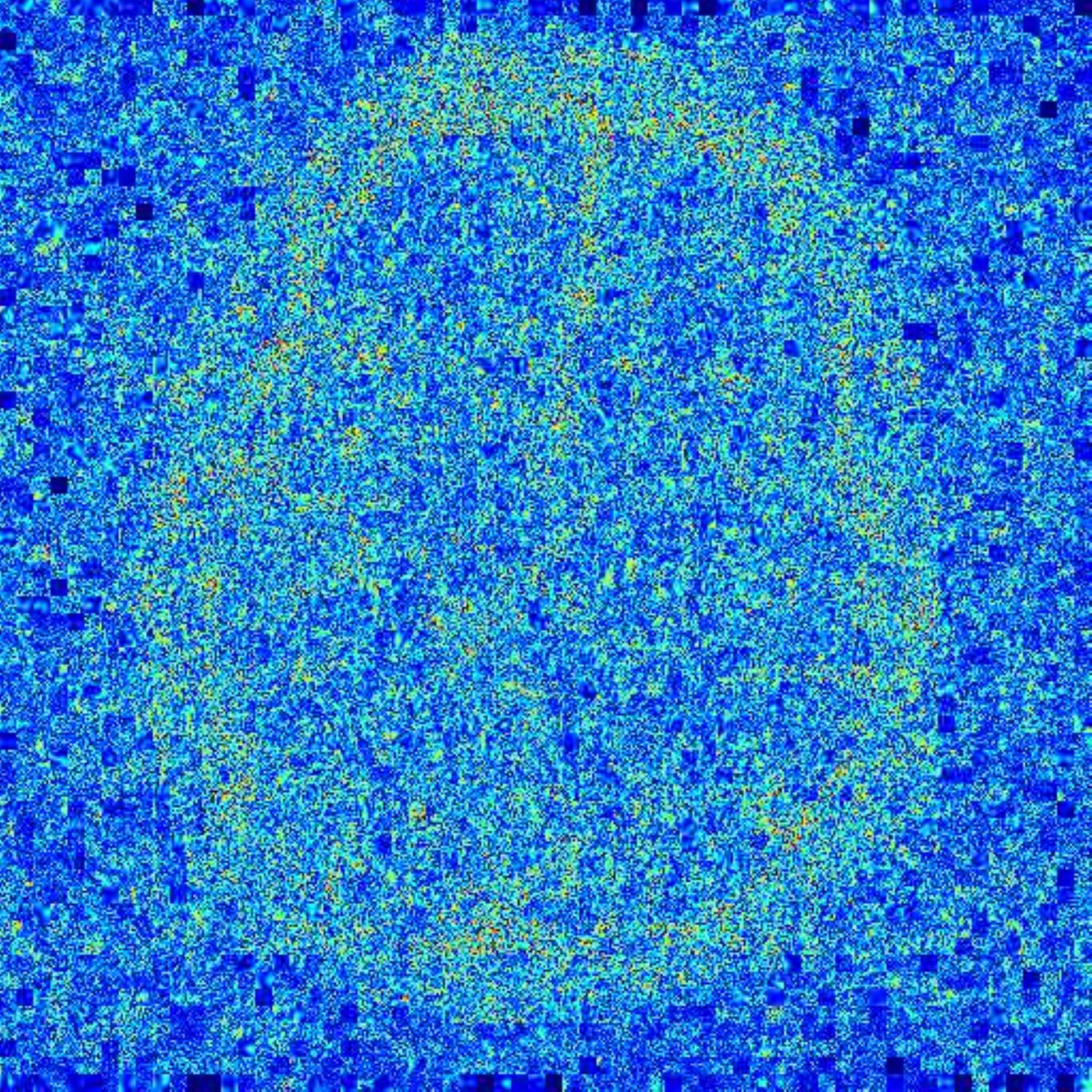,scale=.20}
&
\psfig{figure=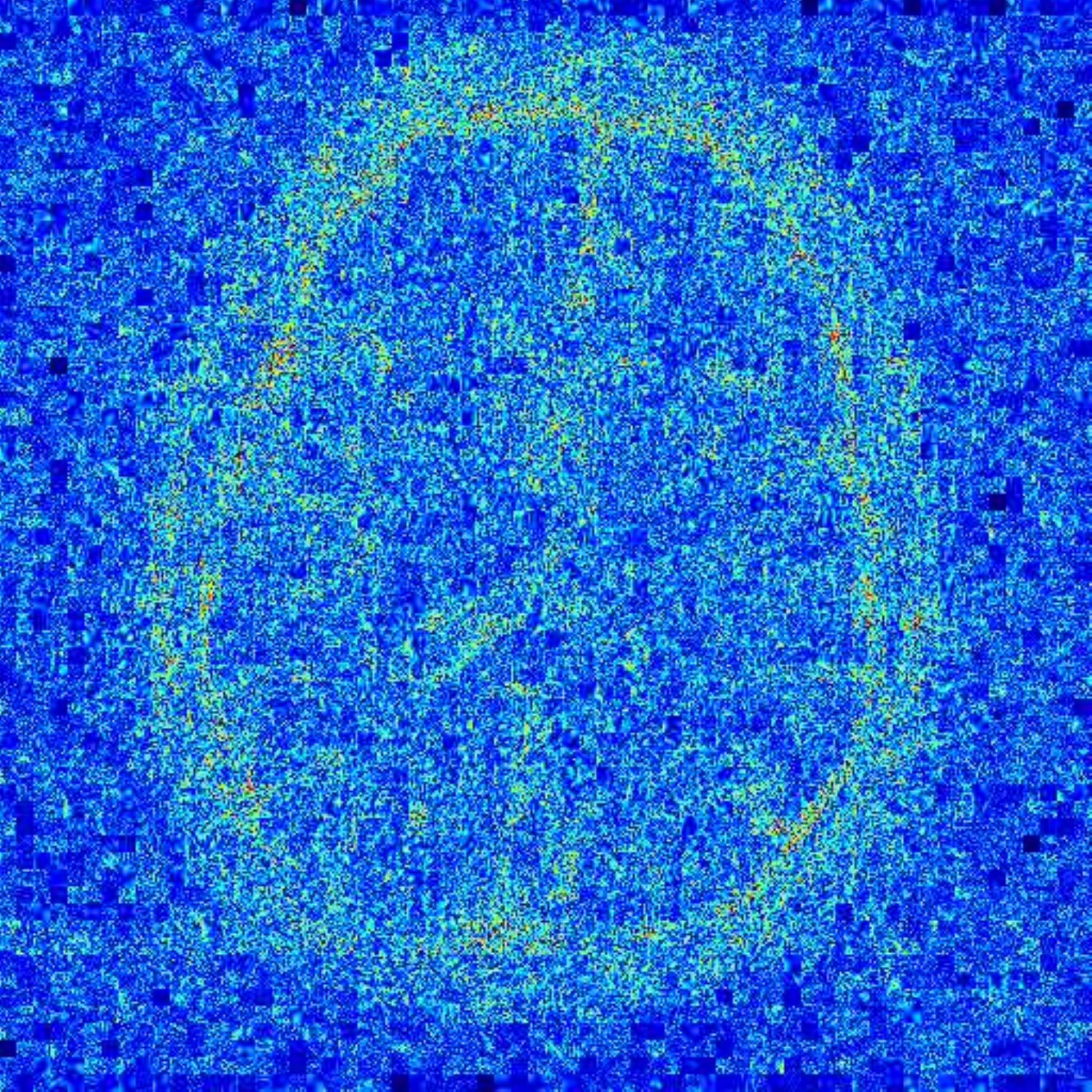,scale=.20}
&
\psfig{figure=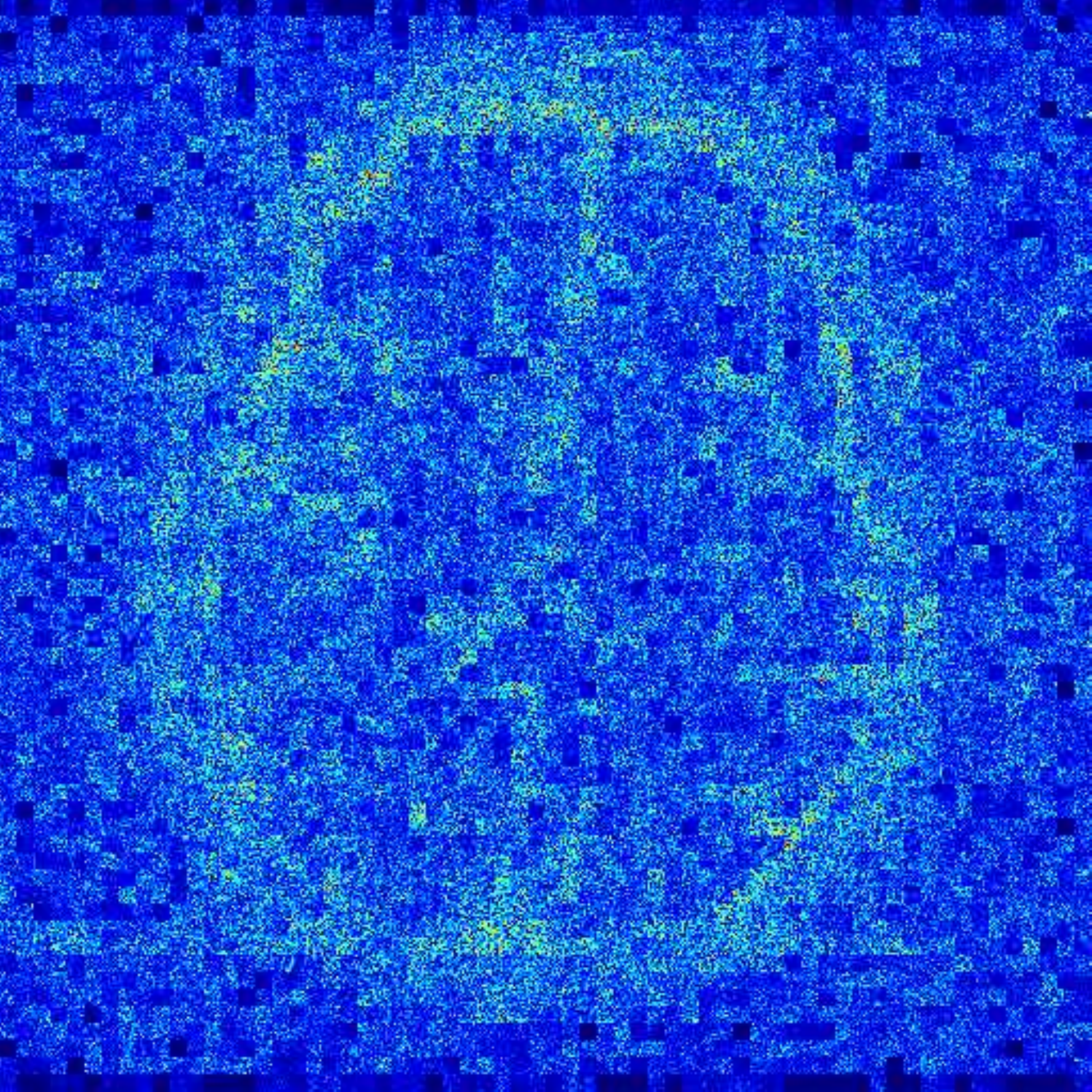,scale=.20}
\\
ObThres & ObCoSaMP & ObSP \\
31.01 dB & 32.27 dB & 36.26 dB \\
\psfig{figure=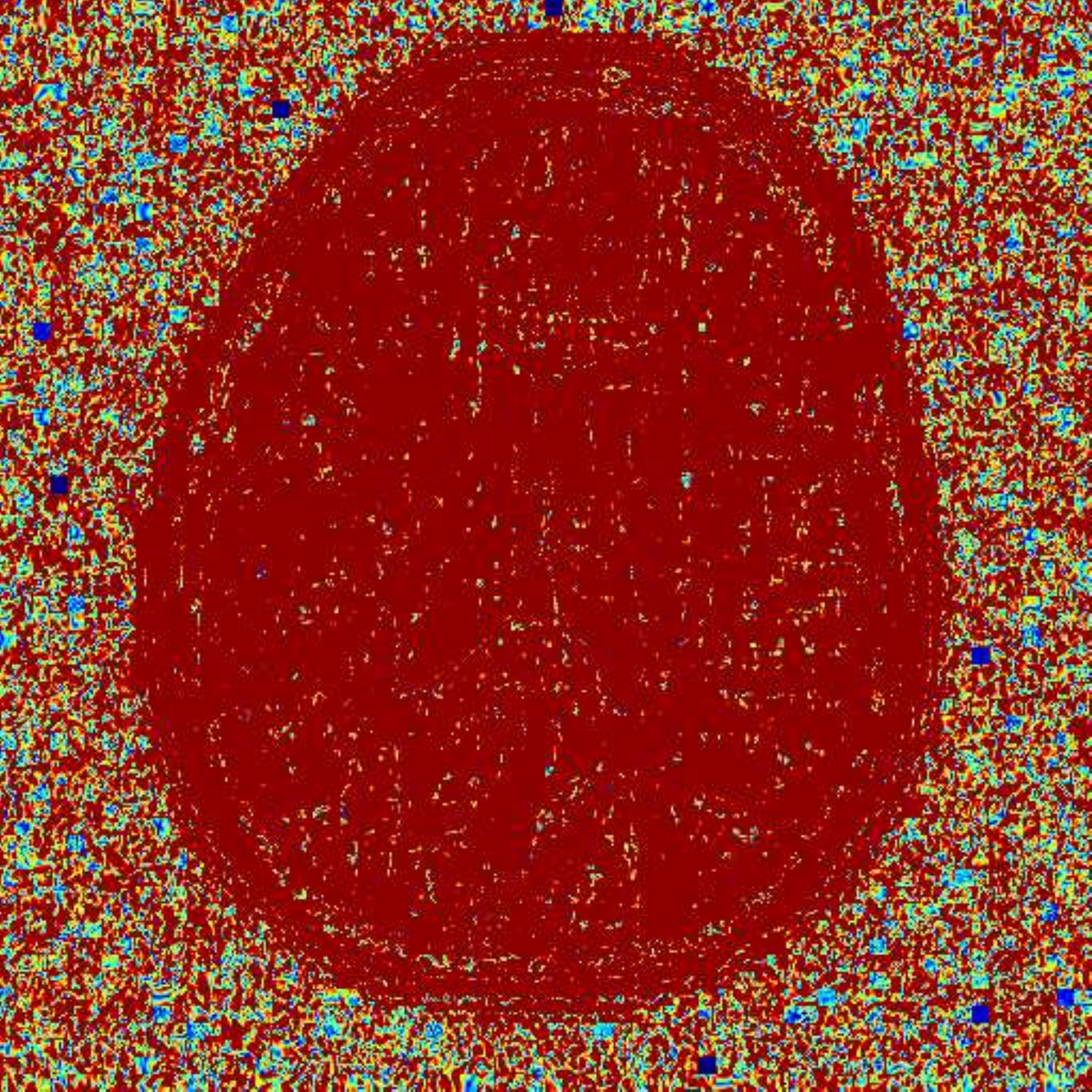,scale=.20}
&
\psfig{figure=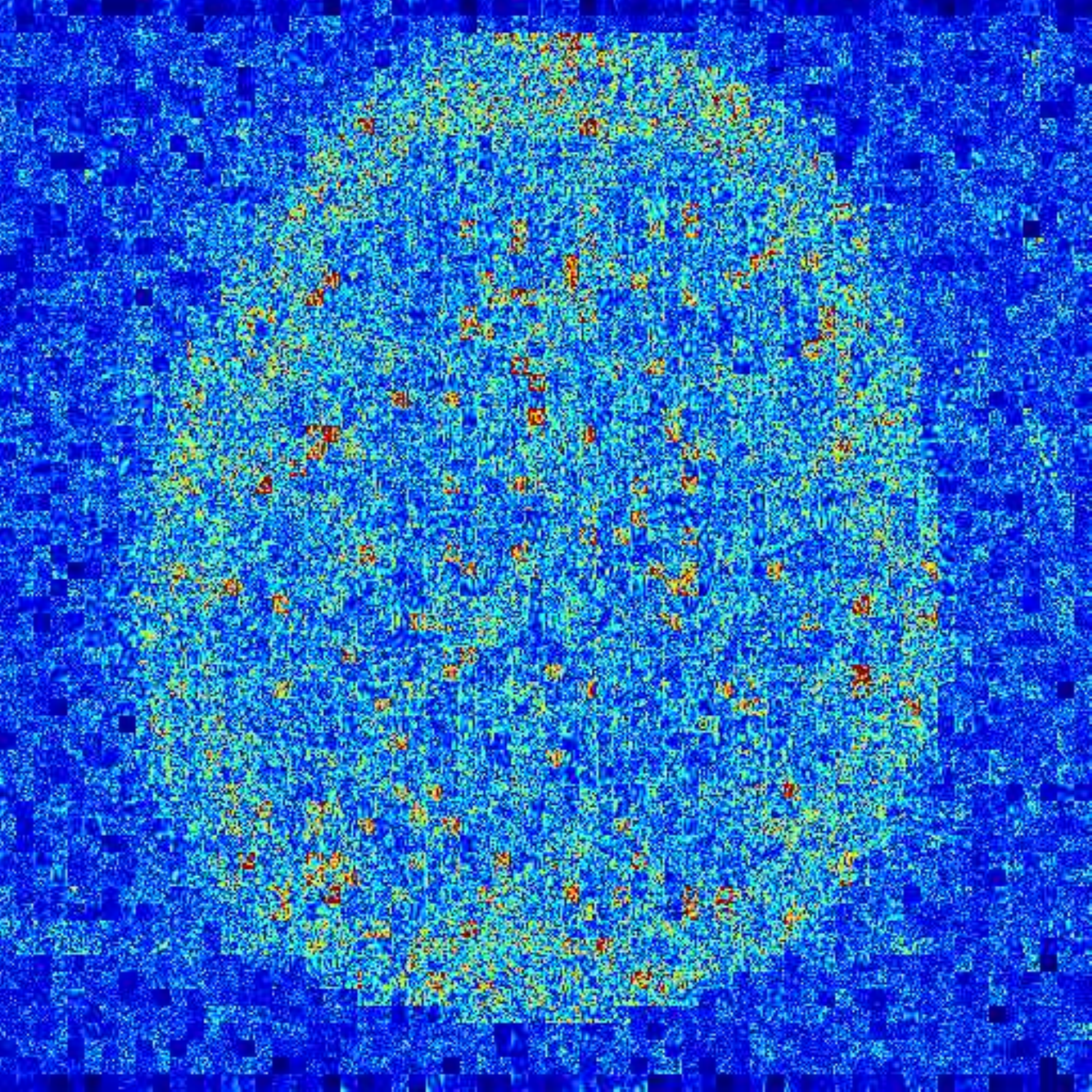,scale=.20}
&
\psfig{figure=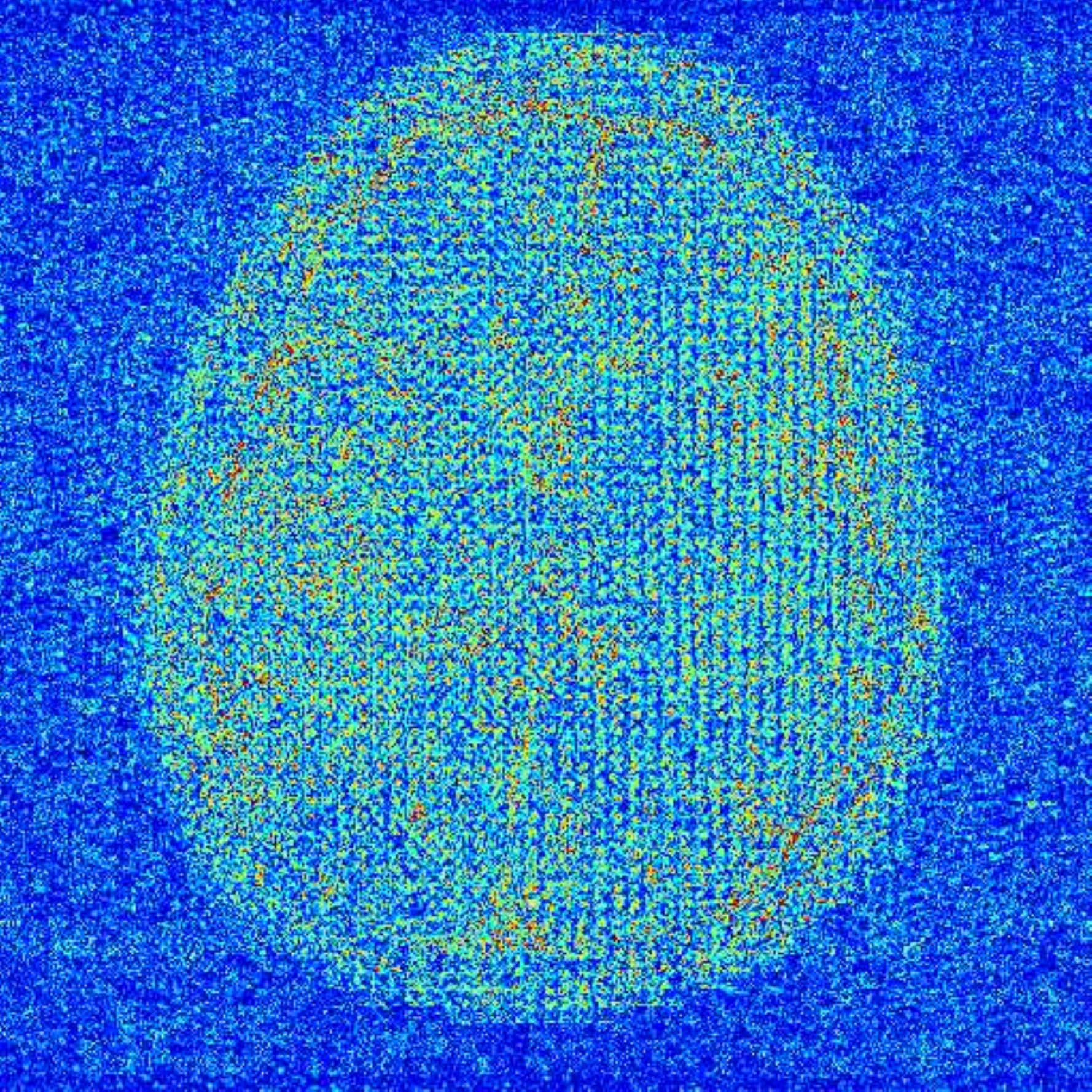,scale=.20}
\\
IHT & HTP & $\ell_1$ Analysis \\
9.34 dB & 31.02 dB & 29.75 dB \\
\psfig{figure=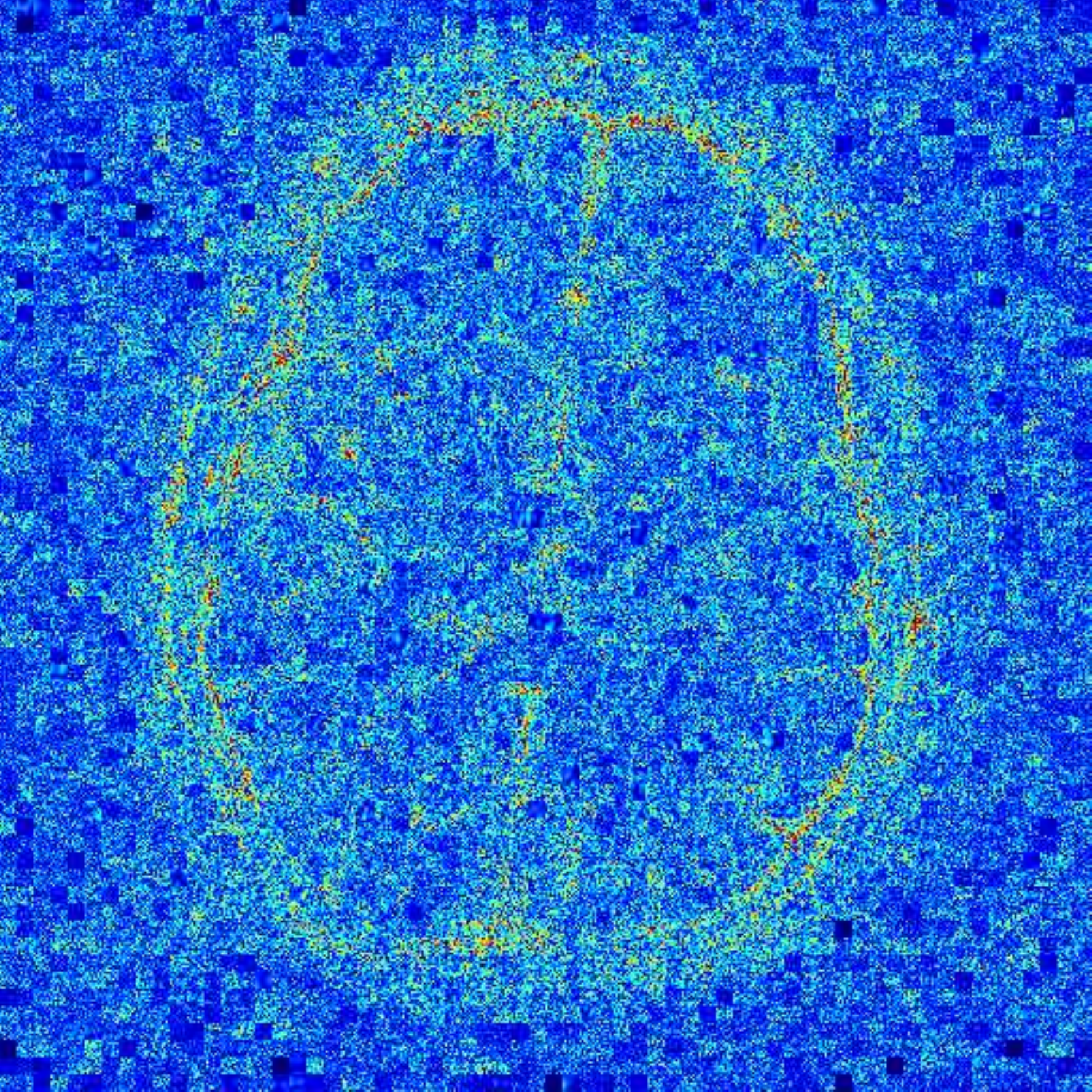,scale=.20}
&
\psfig{figure=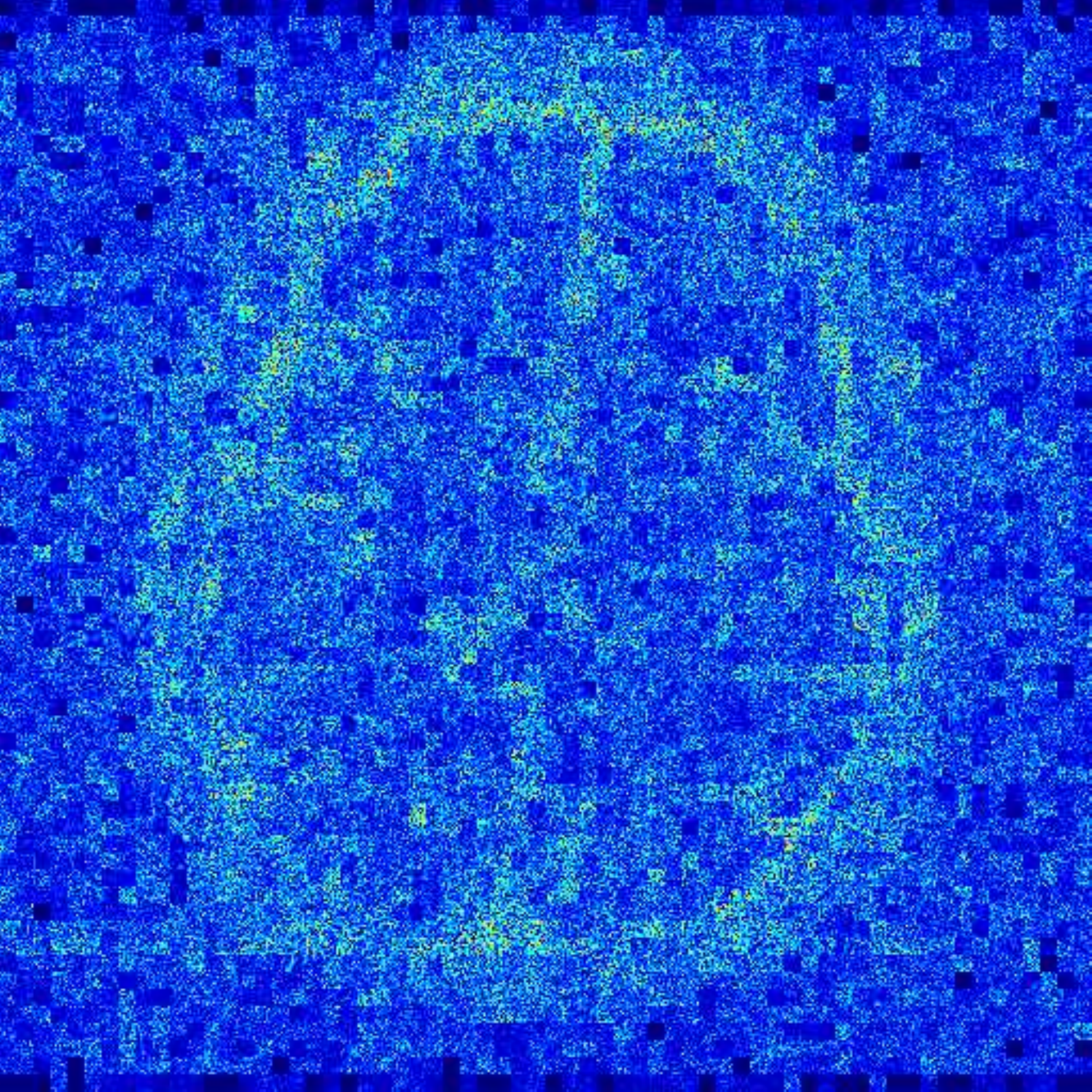,scale=.20}
&
\psfig{figure=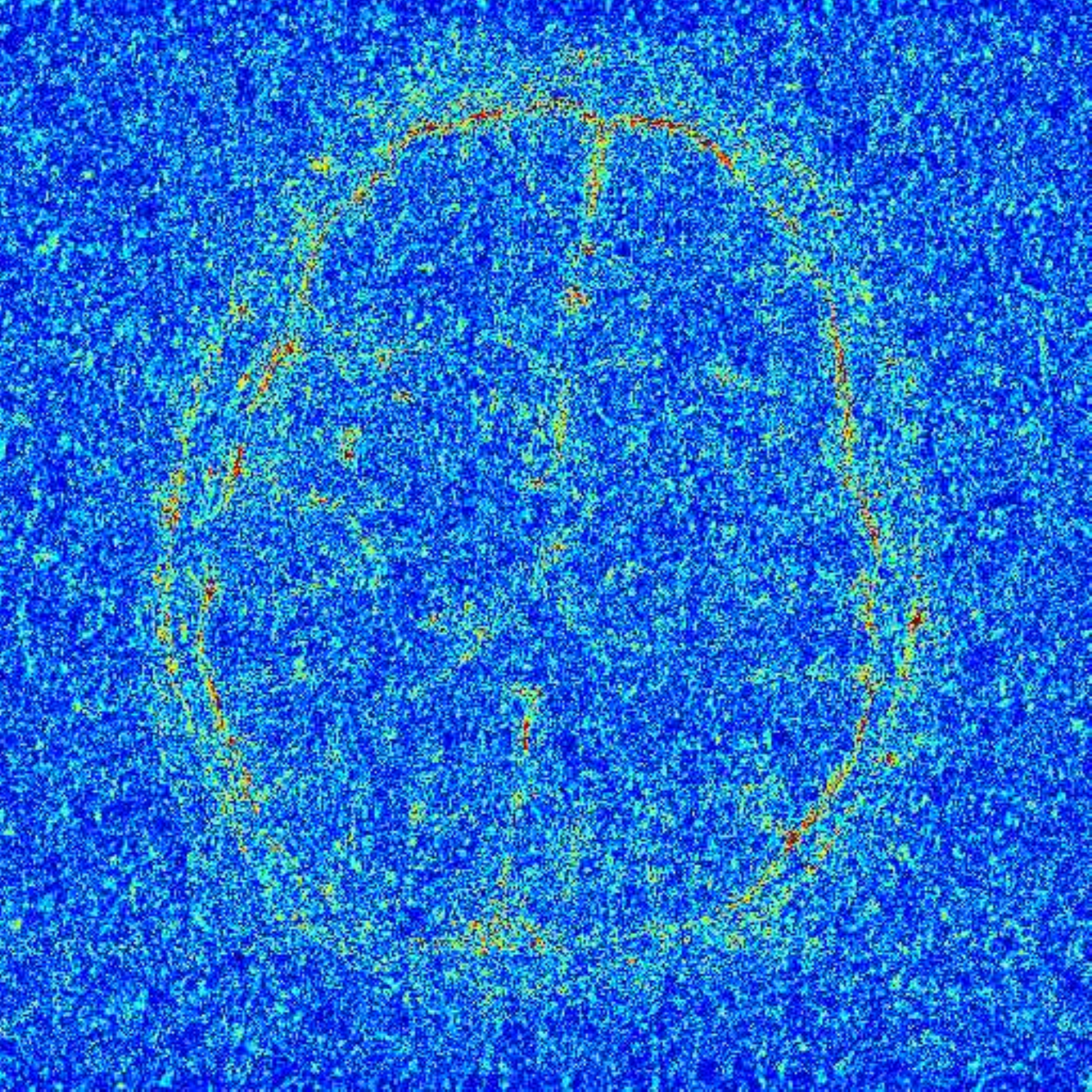,scale=.20}
\\
ObIHT & ObHTP & Zero Filling \\
30.95 dB & 36.40 dB & 31.46 dB
\end{tabular}
}
\end{center}
\caption{
Error images and PSNR for recovery by various algorithms from noisy measurements
(the maximum intensity of the input image is normalized as 1): $\text{SNR} = 30 \text{dB}$, downsample by 3.}
\label{fig:reconimg_noisy}
\end{figure*}

Table~\ref{tab:psnr_noisy} shows the PSNR of the reconstructed images using the various algorithms with different downsampling ratio.
The error images truncated at the maximum magnitude of the input image divided by 10 are shown in Fig.~\ref{fig:reconimg_noisy}.
Downsampling by factors of 2 and 3 is presented, but the results for larger downsampling factor are qualitatively the same.

In most cases, the oblique pursuit algorithms performed better than the conventional counterparts.
In the few exceptions, the difference in performance is not significant.
In particular, ObSP and ObHTP performed significantly better than zero filling.
We observed that thresholding and IHT totally failed in this experiment.
In this experiment, the step sizes of IHT was fixed as 1 for its RIP-based guarantees.
By employing an empirically tuned step size, the performance of IHT might be improved.
In contrast, ObIHT provided a reasonable performance with a fixed step size.

Fig.~\ref{fig:reconimg_noisy} also shows that the error in the reconstruction include blocky artifacts
that are more severe in the reconstruction by the $\ell_1$ analysis formulation.
This issue can be resolved by replacing the non-overlapping patches by overlapping patches.
Furthermore, sparse representation of overlapping patches allows more redundancy,
which helps reduce the sparse approximation error.
In this case, applying the inverse and the biorthogonal dual of the sparsifying transform are no longer patch-wise operations,
but the inverse operation might be still efficiently computed by solving a structured inverse problem.
More generally, the sparsifying dictionary might be replaced by any redundant dictionary.

However, we do not pursue the various possible the improvements of the reconstruction performance in this paper.
As mentioned earlier, the purpose of the numerical results in this section is just to confirm that
the modification made in the oblique pursuit algorithms from the original ones
does not degrade their empirical performance.
It turned out fortuitously that the oblique pursuit algorithms, designed to provide guarantees in terms of the RBOP,
also show significant improvement in empirical performance.

\section{Conclusion}
\label{sec:conclusion}

Previous guarantees for the reconstruction of sparse signals from compressive sensing via random frame matrices
by various practical algorithms were provided in terms of the restricted isometry property (RIP) of the sensing matrix.
Previous works on the RIP focused on scenarios where, to satisfy the isotropy property,
the sensing matrix is constructed from i.i.d. samples from a tight frame according to the uniform distribution.
However, the frame might not be tight due to the physics of the sensing procedure or due to the dictionary that provides a sparse representation.
Furthermore, a non-uniform rather than the uniform distribution is often used
for the i.i.d. sampling in practice in compressed sensing, especially in imaging applications,
due to the signal characteristics or due to the limitation imposed by the physics of the applications.
To derive guarantees without idealized assumptions,
we proposed to exploit the property of biorthogonality that naturally arises in frame theory.
We generalized the RIP to the restricted biorthogonality property (RBOP) that is satisfied without requiring the isotropy property.
To take advantage of the new RBOP, we extended greedy pursuit algorithms with RIP-based guarantees
to new variations -- oblique pursuit algorithms, so that they provide RBOP-based guarantees.
These guarantees apply with relaxed conditions on the sensing matrices and dictionaries, which are satisfied by practical CS imaging schemes.
The extension of greedy pursuit algorithms and their RIP-based guarantees to those based on the RBOP
is not restricted to the specific algorithms studied in this paper.
For example, Fast Nesterov's Iterative Hard Thresholding (FNIHT) \cite{CevJaf10} is another promising algorithm with a RIP-based guarantee,
which will extend similarly.
Finally, we note that although the oblique pursuit algorithms were designed to provide performance guarantees in the worst-case sense,
they also perform competitively with or sometimes significantly better than their conventional counterparts empirically.

\appendix
\allowdisplaybreaks

\numberwithin{equation}{section}
\renewcommand{\theequation}{A.\arabic{equation}}

\subsection{Preliminaries for the Appendix}

\begin{definition}[Dilation \cite{Pau02}]
The dilation of matrix $M$ is defined by
\[
\mathscr{S}(M) \triangleq \begin{bmatrix} 0 & M \\ M^* & 0\end{bmatrix}.
\]
\end{definition}
By definition, $\mathscr{S}(M)$ is a Hermitian matrix and its eigenvalues satisfy
\begin{align*}
\lambda_i(\mathscr{S}(M)) =
\begin{cases}
\sigma_i(M) &  \text{if $i \leq n$} \\
- \sigma_{n-i+1}(M) & \text{if $i > n$}.
\end{cases}
\end{align*}

\begin{definition}[Schur Complement]
Let $M \in \mathbb{K}^{n \times n}$ be a square matrix that can be decomposed as follows:
\[
M = \begin{bmatrix}
M_{11} & M_{12} \\
M_{21} & M_{22}
\end{bmatrix}
\]
where $M_{22} \in \mathbb{K}^{q \times q}$ for $q < n$ is a minor of $M$, which is also a square matrix.
The Schur complement of the block $M_{22}$ of the matrix $M$, denoted by $M/M_{22}$, is the $(n-q) \times (n-q)$ matrix defined by
\[
M/M_{22} \triangleq M_{11} - M_{12} M_{22}^\dagger M_{21}.
\]
\end{definition}

The following lemma extends \cite[Theorem~5]{Smi92} to the non-Hermitian case.
\begin{lemma}
\label{lemma:SINH}
Let $M \in \mathbb{K}^{n \times n}$ be a nonsingular matrix and $M_{22} \in \mathbb{K}^{q \times q}$ for $q < n$ be a minor of $M$.
Then,
\[
\sigma_1(M) \geq \sigma_1(M/M_{22})
\]
and
\[
\sigma_j(M/M_{22}) \geq \sigma_{j+q}(M), \quad \forall j = 1,\ldots,n-q.
\]
\end{lemma}

\begin{remark}
The analogous result for the Hermitian case \cite[Theorem~5]{Smi92} assumed that $M$ is semidefinite
and also showed that
\[
\sigma_j(M) \geq \sigma_j(M/M_{22}), \quad \forall j = 1,\ldots,n-q.
\]

\end{remark}

\begin{IEEEproof}[Proof of Lemma~\ref{lemma:SINH}]
By the Cauchy interlacing theorem, $\sigma_q(M_{22}) \geq \sigma_n(M) > 0$; hence, $M_{22}$ is invertible.
Let
\begin{align*}
M = \begin{bmatrix}
M_{11} & M_{12} \\
M_{21} & M_{22}
\end{bmatrix}
=
\underbrace{
\begin{bmatrix}
M_{11} - M_{12} M_{22}^{-1} M_{21} & 0 \\
0 & 0
\end{bmatrix}
}_{= M_1}
+
\underbrace{
\begin{bmatrix}
M_{12} M_{22}^{-1} M_{21} & M_{12} \\
M_{21} & M_{22}
\end{bmatrix}
}_{= M_2}
.
\end{align*}
Let $M_{22} = U \Sigma V^*$ be the singular value decomposition of $M_{22}$.
Then, $M_2$ is factorized as
\[
M_2
=
\begin{bmatrix}
M_{12} V \Sigma^{-1/2} \\
U \Sigma^{1/2}
\end{bmatrix}
\begin{bmatrix}
\Sigma^{-1/2} U^* M_{21} & \Sigma^{1/2} V^*
\end{bmatrix}
\]
where the left factor has $q$ linearly independent columns and the right factor has $q$ linearly independent rows.
Therefore, $\rank(M_2) = q$.

Now, we use Weyl's inequalities for the eigenvalues of the sum of two Hermitian matrices \cite[Theorem~III.2.1]{Bha97}.
By applying \cite[Theorem~III.2.1]{Bha97} to $\mathscr{S}(M_1)$ and $\mathscr{S}(M_2)$, we obtain
\begin{align*}
{} & \lambda_{j+q}(\mathscr{S}(M_1)+\mathscr{S}(M_2)) \\
{} & \leq \lambda_j(\mathscr{S}(M_1)) + \lambda_{q+1}(\mathscr{S}(M_2)) \\
{} & = \lambda_j(\mathscr{S}(M_1)), \quad \forall j = 1,\ldots,n-q,
\end{align*}
where we used the fact that $\lambda_{q+1}(\mathscr{S}(M_2)) = \sigma_{q+1}(M_2) = 0$ since $\rank(M_2) = q$.
Therefore,
\[
\sigma_{j+q}(M_1 + M_2) \leq \sigma_j(M_1), \quad \forall j = 1,\ldots,n-q.
\]

Since $M$ is invertible, $M/M_{22}$ is also invertible since $\sigma_{n-q}(M/M_{22}) \geq \sigma_n(M) > 0$.
The Schur complement $(M/M_{22})^{-1}$ is a minor of $M^{-1}$; hence,
\begin{align*}
\sigma_1(M)^{-1} {} & = \sigma_n(M^{-1})
\leq \sigma_{n-q}((M/M_{22})^{-1})
= \sigma_1(M/M_{22})^{-1}.
\end{align*}

\end{IEEEproof}

\begin{lemma}
\label{lemma:pertub}
Let $M \in \mathbb{K}^{m \times m}$. Then,
\[
\sigma_1(M - I_m) = \max\left( 1 - \sigma_m(M), \sigma_1(M) - 1 \right).
\]
\end{lemma}

\begin{IEEEproof}[Proof of Lemma~\ref{lemma:pertub}]
If $M$ is a Hermitian matrix, then the proof is straightforward since the eigenvalues of $M - I_m$ are the eigenvalues of $M$ shifted by 1.
Otherwise, by \cite[Theorem~III.2.8]{Bha97}, it follows that
\begin{align*}
{} & \max_{k \in [2m]} |\lambda_k(\mathscr{S}(M)) - \lambda_k(\mathscr{S}(I_m))| \\
{} & \leq \norm{\mathscr{S}(M) - \mathscr{S}(I_m)} \\
{} & \leq \max_{k \in [2m]} |\lambda_k(\mathscr{S}(M)) - \lambda_{2m-k+1}(\mathscr{S}(I_m))|
\end{align*}
where $\mathscr{S}(M)$ and $\mathscr{S}(I_m)$ are the dilations of $M$ and $I_m$, respectively.

Since
\[
\lambda_k(\mathscr{S}(M)) =
\begin{cases}
\sigma_k(M) & k \leq m \\
- \sigma_{m-k+1}(M) & k > m
\end{cases}
\]
and
\[
\lambda_k(\mathscr{S}(I_m)) =
\begin{cases}
1 & k \leq m \\
- 1 & k > m,
\end{cases}
\]
it follows that
\begin{align*}
{} & \max_{k \in [2m]} |\lambda_k(\mathscr{S}(M)) - \lambda_k(\mathscr{S}(I_m))| \\
{} & = \max_{k \in [2m]} |\lambda_k(\mathscr{S}(M)) - \lambda_{2m-k+1}(\mathscr{S}(I_m))| \\
{} & = \max\left( 1 - \sigma_m(M), \sigma_1(M) - 1 \right);
\end{align*}
hence,
\begin{align*}
\norm{M - I_m} {} & = \norm{\mathscr{S}(M - I_m)} \\
{} & = \norm{\mathscr{S}(M) - \mathscr{S}(I_m)} \\
{} & = \max\left( 1 - \sigma_m(M), \sigma_1(M) - 1 \right).
\end{align*}
\end{IEEEproof}

\begin{lemma}
\label{lemma:biorwoblp}
Let $M,\widetilde{M} \in \mathbb{K}^{m \times k}$.
Let $J_1 \subsetneq [k]$ and $J_2 = [k] \setminus J_1$.
Suppose $\widetilde{M}^* M$ has full rank. Then,
\[
\norm{\widetilde{M}_{J_2}^* (I_{|J_1|} - M_{J_1} (\widetilde{M}_{J_1}^* M_{J_1})^\dagger \widetilde{M}_{J_1}^*) M_{J_2} - I_{|J_2|}}
\leq \norm{\widetilde{M}^* M - I_k}.
\]
\end{lemma}

\begin{IEEEproof}[Proof of Lemma~\ref{lemma:biorwoblp}]
To simplify the notation, let $E \triangleq I_{|J_1|} - M_{J_1} (\widetilde{M}_{J_1}^* M_{J_1})^\dagger \widetilde{M}_{J_1}^*$.
By Lemma~\ref{lemma:pertub}, it follows that
\begin{align}
{} & \norm{\widetilde{M}_{J_2}^* E M_{J_2} - I_{|J_2|}} \nonumber \\
{} & = \max\Big\{ 1 - \sigma_{|J_2|}(\widetilde{M}_{J_2}^* E M_{J_2}) ,~ \sigma_1(\widetilde{M}_{J_2}^* E M_{J_2}) - 1 \Big\}.
\label{eq:proof:lemma:biorwoblp:eq1}
\intertext{Furthermore, since $\widetilde{M}^* M$ has full rank, (\ref{eq:proof:lemma:biorwoblp:eq1}) is upper bounded by Lemma~\ref{lemma:SINH} as}
{} & \leq \max\Big\{ 1 - \sigma_k(\widetilde{M}^* M) ,~ \sigma_1(\widetilde{M}^* M) - 1 \Big\} \nonumber \\
{} & = \norm{\widetilde{M}^* M - I_k}
\end{align}
where the last step too follows from Lemma~\ref{lemma:pertub}.
\end{IEEEproof}

\begin{lemma}[{\cite[Corollary~5.2]{IpsMey95}}]
Suppose that $E \in \mathbb{K}^{n \times n}$ is idempotent ($E^2 = E$) and is neither $0$ nor $I_n$.
Then, $\norm{I_n - E} = \norm{E}$.
\label{lemma:ipsen}
\end{lemma}

\begin{lemma}
Let $\Psi,\widetilde{\Psi} \in \mathbb{K}^{m \times n}$.
Let $P \in \mathbb{K}^{n \times n}$ be an orthogonal projector in $\mathbb{K}^n$.
Then, for all $x,y \in \mathbb{K}^n$,
\begin{equation}
\Big| \big|\langle \widetilde{\Psi} P x, \Psi P y \rangle\big| - \big|\langle P x, P y \rangle\big| \Big|
\leq \norm{P \widetilde{\Psi}^* \Psi P - P} \cdot \norm{x}_2 \cdot \norm{y}_2.
\end{equation}
\label{lemma:IPdev}
\end{lemma}

\begin{IEEEproof}[Proof of Lemma~\ref{lemma:IPdev}]
The proof follows from the properties of an inner product:
{\allowdisplaybreaks
\begin{align*}
{} & \Big| \big|\langle \widetilde{\Psi} P x, \Psi P y \rangle\big| - \big|\langle P x, P y \rangle\big| \Big| \\
{} & \overset{\text{(a)}}{\leq} \left| \langle \widetilde{\Psi} P x, \Psi P y \rangle - \langle P x, P y \rangle \right| \\
{} & \overset{\text{(b)}}{=} \left| \langle x, P \widetilde{\Psi}^* \Psi P y \rangle - \langle x, P y \rangle \right| \\
{} & = \left| \langle x, (P \widetilde{\Psi}^* \Psi P - P) y \rangle \right| \\
{} & \leq \norm{P \widetilde{\Psi}^* \Psi P - P} ~ \norm{x}_2 \norm{y}_2
\end{align*}} %
where (a) follows from the triangle inequality, (b) follows since $P^2 = P$ and $P^* = P$.
\end{IEEEproof} 

\subsection{Proof of Theorem~\ref{thm:dthres}}
\label{subsec:thm:dthres}

ObThres is guaranteed to recover $J^\star$ if
\begin{equation}
\min_{j \in J^\star} |(\widetilde{\Psi}^* y)_j| > \max_{j \not\in J^\star} |(\widetilde{\Psi}^* y)_j|.
\label{eq:proof:thm:dthre:eq1}
\end{equation}

The $j$th component of $\widetilde{\Psi}^*y$ is given as
\[
(\widetilde{\Psi}^* y)_j = \widetilde{\psi}_j^* y = e_j^* \Pi_{\{j\}} \widetilde{\Psi}^* \Psi \Pi_J x^\star + \widetilde{\psi}_j^* z
\]
and satisfies
\begin{align}
{} & |(\widetilde{\Psi}^* y)_j - (x^\star)_j| \nonumber \\
{} & \leq |e_j^* \Pi_{\{j\}} \widetilde{\Psi}^* \Psi \Pi_{J^\star} x^\star - (x^\star)_j| + |\widetilde{\psi}_j^* z| \nonumber \\
{} & = |e_j^* (\Pi_{\{j\}} \widetilde{\Psi}^* \Psi \Pi_{J^\star} - \Pi_{\{j\} \cap J^\star}) x^\star| + |\widetilde{\psi}_j^* z| \nonumber \\
{} & \leq \theta_{s+1}(\widetilde{\Psi}^* \Psi) \norm{x^\star}_2 + \max_j \norm{\widetilde{\psi}_j}_2 \norm{z}_2
\label{eq:proof:thm:dthre:eq2}
\end{align}
where the third step follows since
\begin{align*}
{} & \norm{\Pi_{\{j\}} \widetilde{\Psi}^* \Psi \Pi_{J^\star} - \Pi_{\{j\} \cap J^\star}} \\
{} & \leq \norm{\Pi_{\{j\} \cup J^\star} \widetilde{\Psi}^* \Psi \Pi_{\{j\} \cup J^\star} - \Pi_{\{j\} \cup J^\star}} \\
{} & \leq \theta_{|\{j\} \cup J^\star|}(\widetilde{\Psi}^* \Psi) \leq \theta_{s+1}(\widetilde{\Psi}^* \Psi).
\end{align*}
Then, (\ref{eq:thm:dthres:cond}) is obtained by applying (\ref{eq:proof:thm:dthre:eq2}) to (\ref{eq:proof:thm:dthre:eq1}).

\subsection{Proof of Proposition~\ref{prop:doblmp}}
\label{subsec:prop:doblmp}

Given $J \subsetneq J^\star$, the next step of ObMP given $J$ finds an element from $J^\star \setminus J$ if
\begin{align}
{} & \max_{j \in J^\star \setminus J} \big| \widetilde{\psi}_j^* (E_{\mathcal{R}(\widetilde{\Psi}_J)^\perp, \mathcal{R}(\Psi_J)}) y \big| \nonumber \\
{} & > \max_{j \in [n] \setminus J^\star} \big| \widetilde{\psi}_j^* (E_{\mathcal{R}(\widetilde{\Psi}_J)^\perp, \mathcal{R}(\Psi_J)}) y \big|.
\label{eq:proof:prop:doblmp:eq1}
\end{align}

Let $E$ denote $E_{\mathcal{R}(\widetilde{\Psi}_J)^\perp, \mathcal{R}(\Psi_J)}$ to simplify the notation.
Then, $E^* = E_{\mathcal{R}(\Psi_J)^\perp, \mathcal{R}(\widetilde{\Psi}_J)}$ is also an oblique projection.

To derive a sufficient condition for (\ref{eq:proof:prop:doblmp:eq1}),
we first derive a lower bound of the left-hand side of (\ref{eq:proof:prop:doblmp:eq1}) in the following:
{\allowdisplaybreaks
\begin{align}
\max_{j \in J^\star \setminus J} | \widetilde{\psi}_j^* E y |
{} & \geq
\max_{j \in J^\star \setminus J} | \widetilde{\psi}_j^* E \Psi \Pi_{J^\star} x^\star |
- | \widetilde{\psi}_j^* E z | \nonumber \\
{} & \geq
\underbrace{ \max_{j \in J^\star \setminus J} | \widetilde{\psi}_j^* E \Psi \Pi_{J^\star} x^\star | }_{(\star)}
- \norm{\widetilde{\Psi}^*}_{2,\infty} \norm{E} ~ \norm{z}_2.
\label{eq:proof:prop:doblmp:eq2}
\end{align}} %

The term $(\star)$ in (\ref{eq:proof:prop:doblmp:eq2}) is bounded from below by
{\allowdisplaybreaks
\begin{align}
(\star)
{} & = \max_{j \in J^\star \setminus J} \left| \widetilde{\psi}_j^* E \Psi \Pi_{J^\star \setminus J} x^\star \right| \nonumber \\
{} & = \max_{j \in J^\star \setminus J} \left| \langle \widetilde{\psi}_j, ~  E \Psi \Pi_{J^\star \setminus J} x^\star \rangle \right| \nonumber \\
{} & = \max_{j \in J^\star \setminus J} \left| \langle \widetilde{\Psi} \Pi_{J^\star \setminus J} e_j, ~ E \Psi \Pi_{J^\star \setminus J} x^\star \rangle \right| \nonumber \\
{} & \overset{\text(a)}{\geq} \max_{j \in J^\star \setminus J} \left|\langle \Pi_{J^\star \setminus J} e_j, ~ \Pi_{J^\star \setminus J} x^\star \rangle\right| \nonumber \\
{} & \quad - \theta_{s+1}(\widetilde{\Psi}^* \Psi) \norm{\Pi_{J^\star \setminus J} e_j}_2 \norm{\Pi_{J^\star \setminus J} x^\star}_2 \nonumber \\
{} & = \max_{j \in J^\star \setminus J} |(x^\star)_j| - \theta_{s+1}(\widetilde{\Psi}^* \Psi) \norm{\Pi_{J^\star \setminus J} x^\star}_2 \nonumber \\
{} & = \norm{\Pi_{J^\star \setminus J} x^\star}_\infty - \theta_{s+1}(\widetilde{\Psi}^* \Psi) \norm{\Pi_{J^\star \setminus J} x^\star}_2
\label{eq:proof:prop:doblmp:eq3}
\end{align}} %
where (a) holds by Lemma~\ref{lemma:IPdev} since it follows, by Lemma~\ref{lemma:biorwoblp}, that
\begin{align*}
{} & \norm{\Pi_{J^\star \setminus J} \widetilde{\Psi}^* E \Psi \Pi_{J^\star \setminus J} - \Pi_{J^\star \setminus J}} \\
{} & = \norm{\widetilde{\Psi}_{J^\star \setminus J}^* E \Psi_{J^\star \setminus J} - I_{|J^\star \setminus J|}} \\
{} & \leq \norm{\widetilde{\Psi}_J^* \Psi_J - I_s} \leq \theta_{s+1}(\widetilde{\Psi}^* \Psi).
\end{align*}

Next, we derive an upper bound on the right-hand side of (\ref{eq:proof:prop:doblmp:eq1}) in a similar way:
{\allowdisplaybreaks
\begin{align}
\max_{j \in [n] \setminus J^\star} | \widetilde{\psi}_j^* E y |
{} & \leq \max_{j \in [n] \setminus J^\star} | \widetilde{\psi}_j^* E \Psi \Pi_{J^\star} x^\star | + | \widetilde{\psi}_j^* E z | \nonumber \\
{} & \leq \max_{j \in [n] \setminus J^\star} \underbrace{ | \widetilde{\psi}_j^* E \Psi \Pi_{J^\star} x^\star | }_{(\star\star)} + \norm{\widetilde{\Psi}^*}_{2,\infty} \norm{E} ~ \norm{z}_2. \label{eq:proof:prop:doblmp:eq4}
\end{align}} %

The term $(\star\star)$ in (\ref{eq:proof:prop:doblmp:eq4}) is upper bounded by
{\allowdisplaybreaks
\begin{align}
(\star\star)
{} & = \max_{j \in [n] \setminus J^\star} \left| \widetilde{\psi}_j^* E \Psi \Pi_{(J^\star \cup \{j\}) \setminus J} x^\star \right| \nonumber \\
{} & = \max_{j \in [n] \setminus J^\star} \left| \langle \widetilde{\psi}_j, ~  E \Psi \Pi_{(J^\star \cup \{j\}) \setminus J} x^\star \rangle \right| \nonumber \\
{} & = \max_{j \in [n] \setminus J^\star} \left| \langle \widetilde{\Psi} \Pi_{(J^\star \cup \{j\}) \setminus J} e_j, ~ E \Psi \Pi_{(J^\star \cup \{j\}) \setminus J} x^\star \rangle \right| \nonumber \\
{} & \overset{\text(b)}{\leq} \max_{j \in [n] \setminus J^\star} \left|\langle \Pi_{(J^\star \cup \{j\}) \setminus J} e_j, ~ \Pi_{(J^\star \cup \{j\}) \setminus J} x^\star \rangle\right| \nonumber \\
{} & \quad + \theta_{s+1}(\widetilde{\Psi}^* \Psi) \norm{\Pi_{(J^\star \cup \{j\}) \setminus J} e_j}_2 \norm{\Pi_{(J^\star \cup \{j\}) \setminus J} x^\star}_2 \nonumber \\
{} & = \max_{j \in [n] \setminus J^\star} |(x^\star)_j| + \theta_{s+1}(\widetilde{\Psi}^* \Psi) \norm{\Pi_{(J^\star \cup \{j\}) \setminus J} x^\star}_2 \nonumber \\
{} & = \theta_{s+1}(\widetilde{\Psi}^* \Psi) \norm{\Pi_{J^\star \setminus J} x^\star}_2 \label{eq:proof:prop:doblmp:eq5}
\end{align}} %
where (b) follows by Lemma~\ref{lemma:IPdev} since it follows by Lemma~\ref{lemma:biorwoblp} that
{\allowdisplaybreaks
\begin{align*}
{} & \norm{\Pi_{(J^\star \cup \{j\}) \setminus J} \widetilde{\Psi}^* E \Psi \Pi_{(J^\star \cup \{j\}) \setminus J} - \Pi_{(J^\star \cup \{j\}) \setminus J}} \\
{} & \leq \norm{\widetilde{\Psi}_{(J^\star \cup \{j\}) \setminus J}^* E \Psi_{(J^\star \cup \{j\}) \setminus J} - I_{|(J^\star \cup \{j\}) \setminus J|}} \\
{} & \leq \norm{\widetilde{\Psi}_{J \cup \{j\}}^* \Psi_{J \cup \{j\}} - I_{s+1}} \leq \theta_{s+1}(\widetilde{\Psi}^* \Psi).
\end{align*}} %

Applying the bounds in (\ref{eq:proof:prop:doblmp:eq2})~~(\ref{eq:proof:prop:doblmp:eq5}) to (\ref{eq:proof:prop:doblmp:eq1}),
we conclude that, for the success of the next step, it suffices to satisfy
{\allowdisplaybreaks
\begin{align*}
\norm{\Pi_{J^\star \setminus J} x^\star}_\infty - 2 \theta_{s+1}(\widetilde{\Psi}^* \Psi) \norm{\Pi_{J^\star \setminus J} x^\star}_2
> 2 \norm{\widetilde{\Psi}^*}_{2,\infty} \norm{E} ~ \norm{z}_2.
\end{align*}} %
Then, computing an upper bound on $\norm{E}$ will complete the proof.

When $\widetilde{\Psi} = \Psi$, $E$ reduces to an orthogonal projection and satisfies $\norm{E} \leq 1$.
However, since we propose to use $\widetilde{\Psi} \neq \Psi$, $E$ is an oblique projection and $\norm{E}$ is not necessarily bounded by 1.

Since $E$ is idempotent and $E$ is neither $0$ or $I_n$, by Lemma~\ref{lemma:ipsen}, it follows that
\begin{align*}
\norm{E}
{} & = \norm{I_n - E} = \norm{E_{\mathcal{R}(\Psi_J), \mathcal{R}(\widetilde{\Psi}_J)^\perp}} \\
{} & = \norm{\Psi_J (\widetilde{\Psi}_J^* \Psi_J)^{-1} \widetilde{\Psi}_J^*} \\
{} & \leq \frac{\norm{\Psi_J} \norm{\widetilde{\Psi}_J}}{\lambda_s(\widetilde{\Psi}_J^*\Psi_J)}
\leq \frac{\norm{\Psi_{J^\star}} \norm{\widetilde{\Psi}_{J^\star}}}{1-\theta_{s+1}(\widetilde{\Psi}^* \Psi)}.
\end{align*}

\subsection{Proof of Theorem~\ref{thm:pgoblp}}
\label{subsec:proof:thm:pgoblp}

The proof for the ObSP case is done by the following four steps.
To simplify the notations, let
\begin{align*}
\theta = \theta_{3s}(\widetilde{\Psi}^* \Psi), \quad
\delta = \delta_{s}(\Psi), \quad \text{and} \quad
\tilde{\delta} = \delta_{2s}(\widetilde{\Psi}).
\end{align*}

For $J = \{j_1,\ldots,j_\ell\} \subset [n]$, define $R_J: \mathbb{K}^n \to \mathbb{K}^\ell$ by
\[
(R_J x)_k = x_{j_k}, \quad \forall k \in J,~ \forall x \in \mathbb{K}^n,
\]
which is the reduction map to the subvector indexed by $J$.
The adjoint operator $R_J^*: \mathbb{K}^\ell \to \mathbb{K}^n$ satisfies
\[
R_J^* y = \sum_{k=1}^\ell (y)_k e_{j_k}
\]
where $e_k$ is the $k$th column of $I_n$.

\begin{lemma}[Step~1]
Under the assumptions of Theorem~\ref{thm:pgoblp},
\[
\norm{x_{t+1} - x^\star}_2 \leq \rho_1 \norm{\Pi_{J^\star \setminus J_{t+1}} x^\star}_2 + \tau_1 \norm{z}_2
\]
where $\rho_1$ and $\tau_1$ are given by
\[
\rho_1 = \frac{1}{\sqrt{1-\theta^2}}
\quad \text{and} \quad
\tau_1 = \frac{\sqrt{1+\tilde{\delta}}}{1-\theta}.
\]
\label{lemma:obsp_step1}
\end{lemma}

\begin{lemma}[Step~2]
Under the assumptions of Theorem~\ref{thm:pgoblp},
\[
\norm{\Pi_{J^\star \setminus J_{t+1}} x^\star}_2 \leq \rho_2 \norm{\Pi_{J^\star \setminus \widetilde{J}_{t+1}} x^\star}_2 + \tau_2 \norm{z}_2
\]
where $\rho_2$ and $\tau_2$ are given by
\[
\rho_2 = \frac{1+\theta}{1-\theta}
\quad \text{and} \quad
\tau_2 = \frac{2\sqrt{1+\tilde{\delta}}}{1-\theta}.
\]
\label{lemma:obsp_step2}
\end{lemma}

\begin{lemma}[Step~3]
Under the assumptions of Theorem~\ref{thm:pgoblp},
\[
\norm{\Pi_{J^\star \setminus \widetilde{J}_{t+1}} x^\star}_2 \leq \rho_3 \norm{\Pi_{J^\star \setminus J_t} x^\star}_2 + \tau_3 \norm{z}_2
\]
where $\rho_3$ and $\tau_3$ are given by
\[
\rho_3 = \max\left( \frac{\theta}{1-\theta} ,~ \frac{2\theta(1-\theta)}{1+2\theta+2\theta^2} \right)
\]
and
\[
\tau_3 = \max\left( \frac{1}{1-\theta} ,~ \frac{2(1-\theta)}{1+2\theta+2\theta^2} \right) \frac{2\sqrt{1+\delta} (1+\tilde{\delta})}{1-\theta}
\]
\label{lemma:obsp_step3}
\end{lemma}

\textit{(Step~4): } Finally, because $\supp{x_t} = J_t$,
\[
\norm{\Pi_{J^\star \setminus J_t} x^\star}_2
= \norm{\Pi_{J^\star \setminus J_t} (x_t - x^\star)}_2
\leq \norm{x_t - x^\star}_2.
\]

Then, $\rho$ and $\tau$ are given as
\begin{align*}
\rho = \rho_1 \rho_2 \rho_3
\quad \text{and} \quad
\tau = \tau_1 + \rho_1 \tau_2 + \rho_1 \rho_2 \tau_3.
\end{align*}

If we let $\widetilde{\Psi} = \Psi$, ObSP reduces to SP,
and the RBOP-based guarantee for ObSP also reduces to the RIP-based guarantee of SP.
However, compared to the original guarantee \cite{DaiMil09},
the guarantee of SP obtained from Theorem~\ref{thm:pgoblp} requires a less demanding RIP condition.

The results for the other algorithms (ObCoSaMP, ObHTP, and ObIHT) are obtained from
the corresponding results for the conventional algorithms (CoSaMP, HTP, and IHT) \cite{Fou10,Fou11HTP}.
We only need to replace $\Psi^* \Psi$ by $\widetilde{\Psi}^* \Psi$ in the algorithms
and replace $\delta_{ks}(\Psi)$ by $\theta_{ks}(\widetilde{\Psi}^* \Psi)$ in the guarantees.

Constants $\rho$ and $\tau$ are explicitly given as follows:
\begin{itemize}
\item ObCoSaMP
{\allowdisplaybreaks
\begin{align*}
\rho {} & = \sqrt{\frac{4\theta_{4s}(\widetilde{\Psi}^*\Psi)^2(1+3\theta_{4s}(\widetilde{\Psi}^*\Psi)^2)}{1-\theta_{4s}(\widetilde{\Psi}^*\Psi)^2}}, \\
\tau {} & = \left(\sqrt{\frac{2(1+3\theta_{4s}(\widetilde{\Psi}^*\Psi)^2)}{1-\theta_{4s}(\widetilde{\Psi}^*\Psi)^2}} + \frac{\sqrt{1+3\theta_{4s}(\widetilde{\Psi}^*\Psi)^2}}{1-\theta_{4s}(\widetilde{\Psi}^*\Psi)} + \sqrt{3}\right)
\sqrt{1 + \delta_{4s}(\widetilde{\Psi})}.
\end{align*}}

\item ObSP
{\allowdisplaybreaks
\begin{align*}
\rho {} & = \frac{\theta_{3s}(\widetilde{\Psi}^*\Psi)\sqrt{1+\theta_{3s}(\widetilde{\Psi}^*\Psi)}}{\sqrt{1-\theta_{3s}(\widetilde{\Psi}^*\Psi)}} \max\Bigg\{ \frac{1}{(1-\theta_{3s}(\widetilde{\Psi}^*\Psi))^2} ,~ \frac{2}{1+2\theta_{3s}(\widetilde{\Psi}^*\Psi)+2\theta_{3s}(\widetilde{\Psi}^*\Psi)^2} \Bigg\}, \\
\tau {} & = \frac{\sqrt{1 + \delta_{2s}(\widetilde{\Psi})}}{1-\theta_{3s}(\widetilde{\Psi}^*\Psi)}
+ \frac{\sqrt{1 + \delta_{2s}(\widetilde{\Psi})}}{(1-\theta_{3s}(\widetilde{\Psi}^*\Psi)) \sqrt{1 -\theta_{3s}(\widetilde{\Psi}^*\Psi)^2}} \\
{} & \quad + \frac{2\sqrt{1 + \delta_{s}(\Psi)} (1 + \delta_{2s}(\widetilde{\Psi})) \sqrt{1+\theta_{3s}(\widetilde{\Psi}^*\Psi)}}{\sqrt{1-\theta_{3s}(\widetilde{\Psi}^*\Psi)} \cdot (1-\theta_{3s}(\widetilde{\Psi}^*\Psi))} \\
{} & \qquad \cdot \max\Bigg\{ \frac{1}{(1-\theta_{3s}(\widetilde{\Psi}^*\Psi))^2} ,~ \frac{2}{1+2\theta_{3s}(\widetilde{\Psi}^*\Psi)+2\theta_{3s}(\widetilde{\Psi}^*\Psi)^2} \Bigg\}.
\end{align*}}

\item ObHTP
{\allowdisplaybreaks
\begin{align*}
\rho {} & = \sqrt{\frac{2\theta_{3s}(\widetilde{\Psi}^*\Psi)^2}{1-\theta_{3s}(\widetilde{\Psi}^*\Psi)^2}}, \\
\tau {} & = \left(\sqrt{\frac{2}{1-\theta_{3s}(\widetilde{\Psi}^*\Psi)^2}} + \frac{1}{1-\theta_{3s}(\widetilde{\Psi}^*\Psi)}\right)
\sqrt{1 + \delta_{2s}(\widetilde{\Psi})}.
\end{align*}}

\item ObIHT
{\allowdisplaybreaks
\begin{align*}
\rho = 2\theta_{3s}(\widetilde{\Psi}^*\Psi)
\quad \text{and} \quad
\tau = 2 \sqrt{1 + \delta_{2s}(\widetilde{\Psi})}.
\end{align*}}

\end{itemize}

Lemma~\ref{lemma:obsp_step2} is of independent interest to provide the finite convergence in Theorem~\ref{thm:itercnt}.
We stated Lemma~\ref{lemma:obsp_step2} as Lemma~\ref{lemma:fcoblp} in Section~\ref{sec:algorithms}.
For ObCoSaMP and ObHTP, similar lemmata are obtained with a slight modification from the corresponding results \cite{Fou10,Fou11HTP}.
Constants $\bar{\rho}$ and $\bar{\tau}$ in Lemma~\ref{lemma:fcoblp} are explicitly given as follows:
\begin{itemize}
\item ObCoSaMP
{\allowdisplaybreaks
\begin{align*}
\bar{\rho} {} & = \sqrt{\frac{1+3\theta_{4s}(\widetilde{\Psi}^*\Psi)^2}{1-\theta_{4s}(\widetilde{\Psi}^*\Psi)^2}} \\
\bar{\tau} {} & = \left(\frac{\sqrt{1+3\theta_{4s}(\widetilde{\Psi}^*\Psi)^2}}{1-\theta_{4s}(\widetilde{\Psi}^*\Psi)} + \sqrt{3}\right) \sqrt{1 + \delta_{4s}(\widetilde{\Psi})}.
\end{align*}}

\item ObSP
{\allowdisplaybreaks
\begin{align*}
\bar{\rho} {} & = \frac{1}{\sqrt{1-\theta_{3s}(\widetilde{\Psi}^*\Psi)^2}}, \qquad
\bar{\tau} = \frac{\sqrt{1 + \delta_{4s}(\widetilde{\Psi})}}{1-\theta_{3s}(\widetilde{\Psi}^*\Psi)}.
\end{align*}}

\item ObHTP
{\allowdisplaybreaks
\begin{align*}
\bar{\rho} {} & = \frac{1}{\sqrt{1-\theta_{3s}(\widetilde{\Psi}^*\Psi)^2}}, \qquad
\bar{\tau} = \frac{\sqrt{1 + \delta_{4s}(\widetilde{\Psi})}}{1-\theta_{3s}(\widetilde{\Psi}^*\Psi)}.
\end{align*}}

\end{itemize}

\begin{IEEEproof}[Proof of Lemma~\ref{lemma:obsp_step1}]
Lemma~\ref{lemma:obsp_step1} is an extension of the analogous result by Foucart \cite{Fou10} to the biorthogonal case.
The modification is done by replacing some matrices and introducing the RBOP instead of the RIP.
We repeat the proof with appropriate modifications as a guiding example that shows how to modify the derivations using the RBOP.

Recall that $x_{t+1}$ is given as
\[
x_{t+1} = \arg\min_x \{ \norm{\widetilde{\Psi}_{J_{t+1}}^*(y - \Psi x)} : \supp{x} \subset J_{t+1} \}.
\]
Therefore, by the optimality condition of the least square problem, it follows that
\[
(\widetilde{\Psi}_{J_{t+1}}^* \Psi)^* \widetilde{\Psi}_{J_{t+1}}^* (y - \Psi x_{t+1}) = 0,
\]
but, by the RBOP, $\widetilde{\Psi}_{J_{t+1}}^* \Psi$ has full row rank; hence,
\[
\widetilde{\Psi}_{J_{t+1}}^* (y - \Psi x_{t+1})
= \widetilde{\Psi}_{J_{t+1}}^* (\Psi (x^\star - x_{t+1}) + z) = 0,
\]
which implies
\begin{equation}
\Pi_{J_{t+1}} \widetilde{\Psi}^* \Psi (x_{t+1} - x^\star) = \Pi_{J_{t+1}} \widetilde{\Psi}^* z.
\label{eq:proof:lemma:obsp_step1:eq1}
\end{equation}

Now,
{\allowdisplaybreaks
\begin{align}
{} & \norm{\Pi_{J_{t+1}} (x_{t+1} - x^\star)}_2^2 \nonumber \\
{} & = \langle x_{t+1} - x^\star ,~ \Pi_{J_{t+1}} (x_{t+1} - x^\star) \rangle \nonumber \\
{} & = \langle x_{t+1} - x^\star ,~ (I_n - \Psi^* \widetilde{\Psi}) \Pi_{J_{t+1}} (x_{t+1} - x^\star) \rangle \nonumber \\
{} & \quad + \langle x_{t+1} - x^\star ,~ \Psi^* \widetilde{\Psi} \Pi_{J_{t+1}} (x_{t+1} - x^\star) \rangle \nonumber \\
{} & = \langle \Pi_{J^\star \cup J_{t+1}} (x_{t+1} - x^\star) ,~ (I_n - \Psi^* \widetilde{\Psi}) \Pi_{J_{t+1}} (x_{t+1} - x^\star) \rangle \nonumber \\
{} & \quad + \langle \Pi_{J_{t+1}} \widetilde{\Psi}^* \Psi(x_{t+1} - x^\star) ,~ \Pi_{J_{t+1}} (x_{t+1} - x^\star) \rangle \nonumber \\
{} & \overset{\text{(a)}}{=} \langle x_{t+1} - x^\star ,~ \Pi_{J^\star \cup J_{t+1}} (I_n - \Psi^* \widetilde{\Psi}) \Pi_{J_{t+1}} (x_{t+1} - x^\star) \rangle \nonumber \\
{} & \quad + \langle \Pi_{J_{t+1}} \widetilde{\Psi}^* z ,~ \Pi_{J_{t+1}} (x_{t+1} - x^\star) \rangle \nonumber \\
{} & \overset{\text{(b)}}{\leq} \theta \norm{x_{t+1} - x^\star}_2 \norm{\Pi_{J_{t+1}} (x_{t+1} - x^\star)}_2 \nonumber \\
{} & \quad + \sqrt{1+\tilde{\delta}} \norm{z}_2 \norm{\Pi_{J_{t+1}} (x_{t+1} - x^\star)}_2
\label{eq:proof:lemma:obsp_step1:eq2}
\end{align}} %
where (a) follows from (\ref{eq:proof:lemma:obsp_step1:eq1}), and (b) holds since
\begin{align*}
{} & \norm{\Pi_{J^\star \cup J_{t+1}} (I_n - \Psi^* \widetilde{\Psi}) \Pi_{J_{t+1}}} \\
{} & = \norm{\Pi_{J^\star \cup J_{t+1}} (I_n - \Psi^* \widetilde{\Psi}) \Pi_{J^\star \cup J_{t+1}} \Pi_{J_{t+1}}} \\
{} & \leq \norm{I_q - \Psi_{J^\star \cup J_{t+1}}^* \widetilde{\Psi}_{J^* \cup J_{t+1}}} \leq \theta
\end{align*}
where $q = |J^\star \cup J_{t+1}| \leq 2s$.

It follows from (\ref{eq:proof:lemma:obsp_step1:eq2}) that
\[
\norm{\Pi_{J_{t+1}}(x_{t+1} - x^\star)}_2 \leq \theta \norm{x_{t+1} - x^\star}_2 + \sqrt{1+\tilde{\delta}} \norm{z}_2.
\]
Therefore, (\ref{eq:proof:lemma:obsp_step1:eq2}) implies
{\allowdisplaybreaks
\begin{align*}
{} & \norm{x_{t+1} - x^\star}_2^2 \\
{} & = \norm{\Pi_{J_{t+1}} (x_{t+1} - x^\star)}_2^2 + \norm{\Pi_{J^\star \setminus J_{t+1}} (x_{t+1} - x^\star)}_2^2 \\
{} & \leq \left( \theta \norm{x_{t+1} - x^\star}_2 + \sqrt{1+\tilde{\delta}} \norm{z}_2 \right)^2 + \norm{\Pi_{J^\star \setminus J_{t+1}} x^\star}_2^2;
\end{align*}
hence, we have
\begin{align*}
{} & \norm{x_{t+1} - x^\star}_2 \\
{} & \leq \frac{\theta \sqrt{1+\tilde{\delta}} \norm{z}_2 + \sqrt{(1-\theta^2) \norm{\Pi_{J^\star \setminus J_{t+1}} x^\star}_2^2 + (1+\tilde{\delta}) \norm{z}_2^2}}{1-\theta^2} \\
{} & \leq \frac{1}{\sqrt{1-\theta^2}} \norm{\Pi_{J^\star \setminus J_{t+1}} x^\star}_2 + \frac{\sqrt{1+\tilde{\delta}}}{1-\theta} \norm{z}_2.
\end{align*}} %
\end{IEEEproof}

\begin{IEEEproof}[Proof of Lemma~\ref{lemma:obsp_step2}]
Recall that $J_{t+1}$ is chosen as the subset of $\widetilde{J}_{t+1}$
corresponding to the $s$ largest elements of $(\widetilde{\Psi}_{\widetilde{J}_{t+1}}^* \Psi_{\widetilde{J}_{t+1}})^{-1} \widetilde{\Psi}_{\widetilde{J}_{t+1}}^* y$; hence, it satisfies
\begin{align*}
{} & \norm{\Pi_{J_{t+1}} R_{\widetilde{J}_{t+1}}^* (\widetilde{\Psi}_{\widetilde{J}_{t+1}}^* \Psi_{\widetilde{J}_{t+1}})^{-1} \widetilde{\Psi}_{\widetilde{J}_{t+1}}^* y} \\
{} & \geq \norm{\Pi_{J^\star} R_{\widetilde{J}_{t+1}}^* (\widetilde{\Psi}_{\widetilde{J}_{t+1}}^* \Psi_{\widetilde{J}_{t+1}})^{-1} \widetilde{\Psi}_{\widetilde{J}_{t+1}}^* y},
\end{align*}
which implies
\begin{align}
{} & \norm{\Pi_{\widetilde{J}_{t+1} \setminus J_{t+1}} R_{\widetilde{J}_{t+1}}^* (\widetilde{\Psi}_{\widetilde{J}_{t+1}}^* \Psi_{\widetilde{J}_{t+1}})^{-1} \widetilde{\Psi}_{\widetilde{J}_{t+1}}^* y} \nonumber \\
{} & \leq \norm{\Pi_{\widetilde{J}_{t+1} \setminus J^\star} R_{\widetilde{J}_{t+1}}^* (\widetilde{\Psi}_{\widetilde{J}_{t+1}}^* \Psi_{\widetilde{J}_{t+1}})^{-1} \widetilde{\Psi}_{\widetilde{J}_{t+1}}^* y}.
\label{eq:proof:lemma:obsp_step2:eq1}
\end{align}

The left-hand side of (\ref{eq:proof:lemma:obsp_step2:eq1}) is the norm of the following term:
{\allowdisplaybreaks
\begin{align}
{} & \Pi_{\widetilde{J}_{t+1} \setminus J_{t+1}} R_{\widetilde{J}_{t+1}}^* (\widetilde{\Psi}_{\widetilde{J}_{t+1}}^* \Psi_{\widetilde{J}_{t+1}})^{-1} \widetilde{\Psi}_{\widetilde{J}_{t+1}}^* y \nonumber \\
{} & = \Pi_{\widetilde{J}_{t+1} \setminus J_{t+1}} R_{\widetilde{J}_{t+1}}^* (\widetilde{\Psi}_{\widetilde{J}_{t+1}}^* \Psi_{\widetilde{J}_{t+1}})^{-1} \widetilde{\Psi}_{\widetilde{J}_{t+1}}^* (\Psi x^\star + z) \nonumber \\
{} & = \Pi_{\widetilde{J}_{t+1} \setminus J_{t+1}} R_{\widetilde{J}_{t+1}}^* (\widetilde{\Psi}_{\widetilde{J}_{t+1}}^* \Psi_{\widetilde{J}_{t+1}})^{-1} \widetilde{\Psi}_{\widetilde{J}_{t+1}}^* \nonumber \\
{} & \qquad \cdot (\Psi \Pi_{\widetilde{J}_{t+1}} x^\star + \Psi \Pi_{J^\star \setminus \widetilde{J}_{t+1}} x^\star + z).
\label{eq:proof:lemma:obsp_step2:eq2}
\end{align}} %
The first summand in (\ref{eq:proof:lemma:obsp_step2:eq2}) is rewritten as
\begin{align}
{} & \Pi_{\widetilde{J}_{t+1} \setminus J_{t+1}} R_{\widetilde{J}_{t+1}}^* (\widetilde{\Psi}_{\widetilde{J}_{t+1}} \Psi_{\widetilde{J}_{t+1}})^{-1} \widetilde{\Psi}_{\widetilde{J}_{t+1}}^* \Psi \Pi_{\widetilde{J}_{t+1}} x^\star \nonumber \\
{} & = \Pi_{\widetilde{J}_{t+1} \setminus J_{t+1}} R_{\widetilde{J}_{t+1}}^* (\widetilde{\Psi}_{\widetilde{J}_{t+1}}^* \Psi_{\widetilde{J}_{t+1}})^{-1} \widetilde{\Psi}_{\widetilde{J}_{t+1}}^* \Psi_{\widetilde{J}_{t+1}} R_{\widetilde{J}_{t+1}} x^\star \nonumber \\
{} & = \Pi_{\widetilde{J}_{t+1} \setminus J_{t+1}} R_{\widetilde{J}_{t+1}}^* R_{\widetilde{J}_{t+1}} x^\star \nonumber \\
{} & = \Pi_{\widetilde{J}_{t+1} \setminus J_{t+1}} \Pi_{\widetilde{J}_{t+1}} x^\star \nonumber \\
{} & = \Pi_{\widetilde{J}_{t+1} \setminus J_{t+1}} x^\star.
\label{eq:proof:lemma:obsp_step2:eq3}
\end{align}
By the RBOP, the other summands in (\ref{eq:proof:lemma:obsp_step2:eq2}) are bounded from above in the $\ell_2$ norm by
\begin{align}
{} & \norm{\Pi_{\widetilde{J}_{t+1} \setminus J_{t+1}} R_{\widetilde{J}_{t+1}}^* (\widetilde{\Psi}_{\widetilde{J}_{t+1}}^* \Psi_{\widetilde{J}_{t+1}})^{-1} \widetilde{\Psi}_{\widetilde{J}_{t+1}}^* \Psi \Pi_{J^\star \setminus \widetilde{J}_{t+1}} x^\star}_2 \nonumber \\
{} & \leq \norm{(\widetilde{\Psi}_{\widetilde{J}_{t+1}}^* \Psi_{\widetilde{J}_{t+1}})^{-1}} \cdot \norm{\widetilde{\Psi}_{\widetilde{J}_{t+1}}^* \Psi \Pi_{J^\star \setminus \widetilde{J}_{t+1}} x^\star}_2 \nonumber \\
{} & \leq \frac{\theta}{1-\theta} \norm{\Pi_{J^\star \setminus \widetilde{J}_{t+1}} x^\star}_2
\label{eq:proof:lemma:obsp_step2:eq4}
\end{align}
and by
\begin{align}
\norm{\Pi_{\widetilde{J}_{t+1} \setminus J_{t+1}} R_{\widetilde{J}_{t+1}}^* (\widetilde{\Psi}_{\widetilde{J}_{t+1}}^* \Psi_{\widetilde{J}_{t+1}})^{-1} \widetilde{\Psi}_{\widetilde{J}_{t+1}}^* z}_2
\leq \frac{\sqrt{1+\tilde{\delta}}}{1-\theta} \norm{z}_2.
\label{eq:proof:lemma:obsp_step2:eq5}
\end{align}

Combining \cref{eq:proof:lemma:obsp_step2:eq2,eq:proof:lemma:obsp_step2:eq3,eq:proof:lemma:obsp_step2:eq4,eq:proof:lemma:obsp_step2:eq5} implies that the left-hand side of (\ref{eq:proof:lemma:obsp_step2:eq1}) is lower bounded by
\begin{equation}
\norm{\Pi_{\widetilde{J}_{t+1} \setminus J_{t+1}} x^\star}_2
- \frac{\theta}{1-\theta} \norm{\Pi_{J^\star \setminus \widetilde{J}_{t+1}} x^\star}_2
- \frac{\sqrt{1+\tilde{\delta}}}{1-\theta} \norm{z}_2.
\label{eq:proof:lemma:obsp_step2:eq6}
\end{equation}

The right-hand side of (\ref{eq:proof:lemma:obsp_step2:eq1}) is the norm of the following term:
\begin{align}
{} & \Pi_{\widetilde{J}_{t+1} \setminus J^\star} R_{\widetilde{J}_{t+1}}^* (\widetilde{\Psi}_{\widetilde{J}_{t+1}}^* \Psi_{\widetilde{J}_{t+1}})^{-1} \widetilde{\Psi}_{\widetilde{J}_{t+1}}^* y \nonumber \\
{} & = \Pi_{\widetilde{J}_{t+1} \setminus J^\star} R_{\widetilde{J}_{t+1}}^* (\widetilde{\Psi}_{\widetilde{J}_{t+1}}^* \Psi_{\widetilde{J}_{t+1}})^{-1} \widetilde{\Psi}_{\widetilde{J}_{t+1}}^* (\Psi x^\star + z) \nonumber \\
{} & = \Pi_{\widetilde{J}_{t+1} \setminus J^\star} R_{\widetilde{J}_{t+1}}^* (\widetilde{\Psi}_{\widetilde{J}_{t+1}}^* \Psi_{\widetilde{J}_{t+1}})^{-1} \widetilde{\Psi}_{\widetilde{J}_{t+1}}^* \nonumber \\
{} & \quad \cdot (\Psi \Pi_{\widetilde{J}_{t+1}} x^\star + \Psi \Pi_{J^\star \setminus \widetilde{J}_{t+1}} x^\star + z).
\label{eq:proof:lemma:obsp_step2:eq7}
\end{align}
Similarly to (\ref{eq:proof:lemma:obsp_step2:eq3}), the first summand in (\ref{eq:proof:lemma:obsp_step2:eq7}) is rewritten as
\begin{align}
\Pi_{\widetilde{J}_{t+1} \setminus J^\star} R_{\widetilde{J}_{t+1}}^* (\widetilde{\Psi}_{\widetilde{J}_{t+1}}y^* \Psi_{\widetilde{J}_{t+1}})^{-1} \widetilde{\Psi}_{\widetilde{J}_{t+1}}^* \Psi \Pi_{\widetilde{J}_{t+1}} x^\star
= \Pi_{\widetilde{J}_{t+1} \setminus J^\star} x^\star = 0. \label{eq:proof:lemma:obsp_step2:eq8}
\end{align}
In a similar way, the other summands in (\ref{eq:proof:lemma:obsp_step2:eq7}) are bounded from above in the $\ell_2$ norm by
\begin{align}
\norm{\Pi_{\widetilde{J}_{t+1} \setminus J^\star} R_{\widetilde{J}_{t+1}}^* (\widetilde{\Psi}_{\widetilde{J}_{t+1}}^* \Psi_{\widetilde{J}_{t+1}})^{-1} \widetilde{\Psi}_{\widetilde{J}_{t+1}}^* \Psi \Pi_{J^\star \setminus \widetilde{J}_{t+1}} x^\star}_2
\leq \frac{\theta}{1-\theta} \norm{\Pi_{J^\star \setminus \widetilde{J}_{t+1}} x^\star}_2
\label{eq:proof:lemma:obsp_step2:eq9}
\end{align}
and
\begin{align}
\norm{\Pi_{\widetilde{J}_{t+1} \setminus J^\star} R_{\widetilde{J}_{t+1}}^* (\widetilde{\Psi}_{\widetilde{J}_{t+1}}^* \Psi_{\widetilde{J}_{t+1}})^{-1} \widetilde{\Psi}_{\widetilde{J}_{t+1}}^* z}_2
\leq \frac{\sqrt{1+\tilde{\delta}}}{1-\theta} \norm{z}_2.
\label{eq:proof:lemma:obsp_step2:eq10}
\end{align}

Combining \cref{eq:proof:lemma:obsp_step2:eq7,eq:proof:lemma:obsp_step2:eq8,eq:proof:lemma:obsp_step2:eq9,eq:proof:lemma:obsp_step2:eq10} implies that the right-hand side of (\ref{eq:proof:lemma:obsp_step2:eq1}) is upper bounded by
\begin{equation}
\frac{\theta}{1-\theta} \norm{\Pi_{J^\star \setminus \widetilde{J}_{t+1}} x^\star}_2
+ \frac{\sqrt{1+\tilde{\delta}}}{1-\theta} \norm{z}_2.
\label{eq:proof:lemma:obsp_step2:eq11}
\end{equation}
Therefore, by (\ref{eq:proof:lemma:obsp_step2:eq6}) and (\ref{eq:proof:lemma:obsp_step2:eq11}), we have
\begin{equation}
\norm{\Pi_{\widetilde{J}_{t+1} \setminus J_{t+1}} x^\star}_2
\leq \frac{2\theta}{1-\theta} \norm{\Pi_{J^\star \setminus \widetilde{J}_{t+1}} x^\star}_2
+ \frac{2\sqrt{1+\tilde{\delta}}}{1-\theta} \norm{z}_2.
\label{eq:proof:lemma:obsp_step2:eq12}
\end{equation}

Note that $J^\star \setminus J_{t+1} = (J^\star \setminus \widetilde{J}_{t+1}) \cup (J^\star \cap (\widetilde{J}_{t+1} \setminus J_{t+1}))$
and $J^\star \setminus \widetilde{J}_{t+1}$ and $J^\star \cap (\widetilde{J}_{t+1} \setminus J_{t+1})$ are disjoint.
Therefore, since $x^\star$ is supported on $J^\star$, it follows that
\begin{equation}
\norm{\Pi_{J^\star \setminus J_{t+1}} x^\star}_2^2 =
\norm{\Pi_{J^\star \setminus \widetilde{J}_{t+1}} x^\star}_2^2
+ \norm{\Pi_{\widetilde{J}_{t+1} \setminus J_{t+1}} x^\star}_2^2.
\label{eq:proof:lemma:obsp_step2:eq13}
\end{equation}

Applying (\ref{eq:proof:lemma:obsp_step2:eq13}) to (\ref{eq:proof:lemma:obsp_step2:eq12}), we obtain
\begin{align*}
\sqrt{\norm{\Pi_{J^\star \setminus J_{t+1}} x^\star}_2^2 - \norm{\Pi_{J^\star \setminus \widetilde{J}_{t+1}} x^\star}_2^2}
\leq \frac{2\theta}{1-\theta} \norm{\Pi_{J^\star \setminus \widetilde{J}_{t+1}} x^\star}_2
+ \frac{2\sqrt{1+\tilde{\delta}}}{1-\theta} \norm{z}_2,
\end{align*}
which implies the desired inequality after simplification using $\sqrt{a^2+b^2} \leq a + b$ for $a,b \geq 0$.
\end{IEEEproof}

\begin{IEEEproof}[Proof of Lemma~\ref{lemma:obsp_step3}]
The last step in each iteration of ObSP updates $x_t$ by $x_t = R_{J_t}^* (\widetilde{\Psi}_{J_t}^* \Psi_{J_t})^{-1} \widetilde{\Psi}_{J_t}^* y$.
Since $\Psi$ and $\widetilde{\Psi}$ satisfy the RBOP, by Lemma~\ref{lemma:compl},
$\Psi_{J_t} (\widetilde{\Psi}_{J_t}^* \Psi_{J_t})^{-1} \widetilde{\Psi}_{J_t}^*$ is a valid oblique projector onto $\mathcal{R}(\Psi_{J_t})$ along $\mathcal{R}(\widetilde{\Psi}_{J_t})^\perp$.
Then, $I_n - \Psi_{J_t} (\widetilde{\Psi}_{J_t}^* \Psi_{J_t})^{-1} \widetilde{\Psi}_{J_t}^*$
and $\widetilde{\Psi}_{J_t} (\Psi_{J_t}^* \widetilde{\Psi}_{J_t})^{-1} \Psi_{J_t}^*$ are also oblique projectors.
Let $E$ denote the oblique projection $I_n - \Psi_{J_t} (\widetilde{\Psi}_{J_t}^* \Psi_{J_t})^{-1} \widetilde{\Psi}_{J_t}^*$ to simplify the notation.
Then,
\[
\widetilde{\Psi}^*(y - \Psi x_t) = \widetilde{\Psi}^* E y.
\]

Let
\[
\widebar{J} \triangleq \supp{H_s\big(\widetilde{\Psi}^* E y\big)}.
\]
Since $E^* \widetilde{\psi}_j = 0$ for all $j \in J_t$,
it follows that $\widebar{J}$ is disjoint from $J_{t}$.

By definition of $\widebar{J}$, we have
\begin{equation*}
\norm{\widetilde{\Psi}_{\widebar{J}}^* E y}_2
\geq \norm{\widetilde{\Psi}_{J^\star}^* E y}_2.
\end{equation*}
hence, it follows that
\begin{equation}
\norm{\widetilde{\Psi}_{\widebar{J} \setminus J^\star}^* E y}_2
\geq \norm{\widetilde{\Psi}_{J^\star \setminus \widebar{J}}^* E y}_2.
\label{eq:proof:lemma:obsp_step3:eq1}
\end{equation}

Since $E b_j = 0$ for all $j \in J_t$, the left-hand side of (\ref{eq:proof:lemma:obsp_step3:eq1}) is the norm of the following term:
\begin{align}
\widetilde{\Psi}_{\widebar{J} \setminus J^\star}^* E y
= \widetilde{\Psi}_{\widebar{J} \setminus J^\star}^* E (\Psi \Pi_{J^\star \setminus J_t} x^\star + z).
\label{eq:proof:lemma:obsp_step3:eq2}
\end{align}

The first summand in (\ref{eq:proof:lemma:obsp_step3:eq2}) is upper bounded by
\begin{align*}
\norm{\widetilde{\Psi}_{\widebar{J} \setminus J^\star}^* E \Psi \Pi_{J^\star \setminus J_t} x^\star}_2
\leq \underbrace{\norm{\widetilde{\Psi}_{\widebar{J} \setminus J^\star}^* E \Psi_{J^\star \setminus J_t}}}_{(\ast)} \norm{\Pi_{J^\star \setminus J_t} x^\star}_2 \nonumber
\end{align*}
where $(\ast)$ is upper bounded by
\begin{align}
{} & \norm{\Pi_{\widebar{J} \setminus J^\star} \widetilde{\Psi}^* E \Psi \Pi_{J^\star \setminus J_t}} \nonumber \\
{} & = \norm{\Pi_{\widebar{J} \setminus J^\star} (\widetilde{\Psi}^* E \Psi - I_n) \Pi_{J^\star \setminus J_t}} \nonumber \\
{} & \leq \norm{\Pi_{(\widebar{J} \cup J^\star) \setminus J_t} (\widetilde{\Psi}^* E \Psi - I_n) \Pi_{(\widebar{J} \cup J^\star) \setminus J_t}} \nonumber \\
{} & = \norm{\widetilde{\Psi}_{(\widebar{J} \cup J^\star) \setminus J_t}^* E \Psi_{(\widebar{J} \cup J^\star) \setminus J_t} - I_{|(\widebar{J} \cup J^\star) \setminus J_t|}} \nonumber \\
{} & \leq \norm{\widetilde{\Psi}_{\widebar{J} \cup J^\star}^* \Psi_{\widebar{J} \cup J^\star} - I_{|\widebar{J} \cup J^\star|}} \leq \theta \nonumber.
\end{align}
Therefore,
\begin{equation}
\norm{\widetilde{\Psi}_{\widebar{J} \setminus J^\star}^* E \Psi \Pi_{J^\star \setminus J_t} x^\star}_2
\leq \theta \norm{\Pi_{J^\star \setminus J_t} x^\star}_2. \label{eq:proof:lemma:obsp_step3:eq3}
\end{equation}

The first summand in (\ref{eq:proof:lemma:obsp_step3:eq2}) is upper bounded by
\begin{align}
{} & \norm{\widetilde{\Psi}_{\widebar{J} \setminus J^\star}^* E z}_2 \nonumber \\
{} & \leq \norm{\widetilde{\Psi}_{\widebar{J} \setminus J^\star}^*} \norm{E} \norm{z}_2 \nonumber \\
{} & \overset{\text{(a)}}{\leq} \sqrt{1+\tilde{\delta}} \norm{I_n - E} \norm{z}_2 \nonumber \\
{} & = \sqrt{1+\tilde{\delta}} \norm{\Psi_{J_t} (\widetilde{\Psi}_{J_t}^* \Psi_{J_t})^{-1} \widetilde{\Psi}_{J_t}^*} \norm{z}_2 \nonumber \\
{} & \leq \frac{\sqrt{1+\delta} (1+\tilde{\delta})}{1 - \theta} \norm{z}_2
\label{eq:proof:lemma:obsp_step3:eq4}
\end{align}
where (a) follows from Lemma~\ref{lemma:ipsen}.

The right-hand side of (\ref{eq:proof:lemma:obsp_step3:eq1}) is the norm of the following term:
\begin{align}
\widetilde{\Psi}_{J^\star \setminus \widebar{J}}^* E y
{} & = \widetilde{\Psi}_{J^\star \setminus \widetilde{J}_{t+1}}^* E y \nonumber \\
{} & = \widetilde{\Psi}_{J^\star \setminus \widetilde{J}_{t+1}}^* E (\Psi \Pi_{J^\star \setminus J_t} x^\star + z) \nonumber \\
{} & = \widetilde{\Psi}_{J^\star \setminus \widetilde{J}_{t+1}}^* E (\Psi \Pi_{J^\star \setminus \widetilde{J}_{t+1}} x^\star + \Psi \Pi_{J^\star \cap \widebar{J}} x^\star + z) \label{eq:proof:lemma:obsp_step3:eq5}
\end{align}
where the first equality holds since $E^* \widetilde{\psi}_j = 0$ for all $j \in J_t$
and the last equality holds since $J^\star \setminus J_t = (J^\star \setminus \widetilde{J}_{t+1}) \cup (J^\star \cap \widebar{J})$,
and $J^\star \setminus \widetilde{J}_{t+1}$ and $J^\star \cap \widebar{J}$ are disjoint.

The first term in (\ref{eq:proof:lemma:obsp_step3:eq5}) is lower bounded by
\begin{align}
{} & \norm{\widetilde{\Psi}_{J^\star \setminus \widetilde{J}_{t+1}}^* E \Psi \Pi_{J^\star \setminus \widetilde{J}_{t+1}} x^\star}_2 \nonumber \\
{} & \geq \sigma_{|J^\star \setminus \widetilde{J}_{t+1}|}(\widetilde{\Psi}_{J^\star \setminus \widetilde{J}_{t+1}}^* E \Psi_{J^\star \setminus \widetilde{J}_{t+1}}) \norm{\Pi_{J^\star \setminus \widetilde{J}_{t+1}} x^\star}_2 \nonumber \\
{} & \geq \sigma_{|J^\star \setminus \widebar{J}|}(\widetilde{\Psi}_{J^\star \setminus \widebar{J}}^* \Psi_{J^\star \setminus \widebar{J}}) \norm{\Pi_{J^\star \setminus \widetilde{J}_{t+1}} x^\star}_2 \nonumber \\
{} & \geq (1-\theta) \norm{\Pi_{J^\star \setminus \widetilde{J}_{t+1}} x^\star}_2.
\label{eq:proof:lemma:obsp_step3:eq6}
\end{align}

The second term in (\ref{eq:proof:lemma:obsp_step3:eq5}) is lower bounded by
\begin{align*}
\norm{\widetilde{\Psi}_{J^\star \setminus \widetilde{J}_{t+1}}^* E \Psi \Pi_{J^\star \cap \widebar{J}} x^\star}_2
\leq \underbrace{\norm{\widetilde{\Psi}_{J^\star \setminus \widetilde{J}_{t+1}}^* E \Psi \Pi_{J^\star \cap \widebar{J}}}}_{(\ast\ast)}
\norm{\Pi_{J^\star \cap \widebar{J}} x^\star}_2
\end{align*}
where $(\ast\ast)$ is further upper bounded by
\begin{align}
{} & \norm{\widetilde{\Psi}_{J^\star \setminus \widetilde{J}_{t+1}}^* E \Psi \Pi_{J^\star \cap \widebar{J}}} \nonumber \\
{} & = \norm{\Pi_{J^\star \setminus \widetilde{J}_{t+1}} \widetilde{\Psi}^* E \Psi \Pi_{J^\star \cap \widebar{J}}} \nonumber \\
{} & = \norm{\Pi_{J^\star \setminus \widetilde{J}_{t+1}} (\widetilde{\Psi}^* E \Psi - I_n) \Pi_{J^\star \cap \widebar{J}}} \nonumber \\
{} & \leq \norm{\Pi_{(J^\star \setminus \widetilde{J}_{t+1}) \cup (J^\star \cap \widebar{J})} (\widetilde{\Psi}^* E \Psi - I_n) \Pi_{(J^\star \setminus \widetilde{J}_{t+1}) \cup (J^\star \cap \widebar{J})}} \nonumber \\
{} & \leq \norm{\widetilde{\Psi}_{(J^\star \setminus \widetilde{J}_{t+1}) \cup (J^\star \cap \widebar{J})}^* E \Psi_{(J^\star \setminus \widetilde{J}_{t+1}) \cup (J^\star \cap \widebar{J})} - I_{|(J^\star \setminus \widetilde{J}_{t+1}) \cup (J^\star \cap \widebar{J})|}} \nonumber \\
{} & \leq \norm{\widetilde{\Psi}_{(J^\star \setminus \widetilde{J}_{t+1}) \cup (J^\star \cap \widebar{J})}^* E \Psi_{(J^\star \setminus \widetilde{J}_{t+1}) \cup (J^\star \cap \widebar{J})} - I_{|(J^\star \setminus \widetilde{J}_{t+1}) \cup (J^\star \cap \widebar{J})|}} \nonumber \\
{} & \leq \norm{\widetilde{\Psi}_{J^\star \cup J_t}^* \Psi_{J^\star \cup J_t} - I_{|J^\star \cup J_t|}} \leq \theta. \nonumber
\end{align}
Therefore,
\begin{align}
\norm{\widetilde{\Psi}_{J^\star \setminus \widetilde{J}_{t+1}}^* E \Psi \Pi_{J^\star \cap \widebar{J}} x^\star}_2
\leq \theta \norm{\Pi_{J^\star \cap \widebar{J}} x^\star}_2.
\label{eq:proof:lemma:obsp_step3:eq7}
\end{align}

The last term in (\ref{eq:proof:lemma:obsp_step3:eq5}) is upper bounded by
\begin{align}
\norm{\widetilde{\Psi}_{J^\star \setminus J_t}^* E z}_2 {} & \leq \frac{\sqrt{1+\delta} (1+\tilde{\delta})}{1 - \theta} \norm{z}_2.
\label{eq:proof:lemma:obsp_step3:eq8}
\end{align}

Applying \cref{eq:proof:lemma:obsp_step3:eq2,eq:proof:lemma:obsp_step3:eq3,eq:proof:lemma:obsp_step3:eq4,eq:proof:lemma:obsp_step3:eq5,eq:proof:lemma:obsp_step3:eq6,eq:proof:lemma:obsp_step3:eq7,eq:proof:lemma:obsp_step3:eq8} to \cref{eq:proof:lemma:obsp_step3:eq1}, we obtain
\begin{align}
{} & \theta \norm{\Pi_{J^\star \setminus J_t} x^\star}_2
+ \frac{\sqrt{1+\delta} (1+\tilde{\delta})}{1 - \theta} \norm{z}_2 \nonumber \\
{} & \geq (1-\theta) \norm{\Pi_{J^\star \setminus \widetilde{J}_{t+1}} x^\star}
- \theta \norm{\Pi_{J^\star \cap \widebar{J}} x^\star}_2
- \frac{\sqrt{1+\delta} (1+\tilde{\delta})}{1 - \theta} \norm{z}_2.
\label{eq:proof:lemma:obsp_step3:eq9}
\end{align}

Since
\[
\norm{\Pi_{J^\star \setminus J_t} x^\star}_2^2 =
\norm{\Pi_{J^\star \setminus \widetilde{J}_{t+1}} x^\star}_2^2
+ \norm{\Pi_{J^\star \cap \widebar{J}} x^\star}_2^2,
\]
(\ref{eq:proof:lemma:obsp_step3:eq9}) implies
\begin{align}
{} & \theta \norm{\Pi_{J^\star \setminus J_t} x^\star}_2
+ \frac{2 \sqrt{1+\delta} (1+\tilde{\delta})}{1 - \theta} \norm{z}_2 \nonumber \\
{} & \geq (1-\theta) \norm{\Pi_{J^\star \setminus \widetilde{J}_{t+1}} x^\star}_2
- \theta \sqrt{\norm{\Pi_{J^\star \setminus J_t} x^\star}_2^2 - \norm{\Pi_{J^\star \setminus \widetilde{J}_{t+1}} x^\star}_2^2}.
\label{eq:proof:lemma:obsp_step3:eq10}
\end{align}
The final result is obtained by simplifying (\ref{eq:proof:lemma:obsp_step3:eq10}).

To simplify the notation, let
\begin{align*}
a {} & = \norm{\Pi_{J^\star \setminus \widetilde{J}_{t+1}} x^\star}_2 \\
b {} & = \norm{\Pi_{J^\star \setminus J_t} x^\star}_2 \\
c {} & = \frac{2 \sqrt{1+\delta} (1+\tilde{\delta})}{1 - \theta} \norm{z}_2.
\end{align*}
Then, (\ref{eq:proof:lemma:obsp_step3:eq10}) reduces to
\[
\theta b + c \geq (1-\theta) a - \theta \sqrt{b^2 - a^2},
\]
which is equivalent to
\[
\theta \sqrt{b^2 - a^2} \geq (1-\theta) a - (\theta b + c).
\]

If $(1 - \theta) a \leq \theta b + c$, then
\begin{equation}
a \leq \frac{\theta}{1 - \theta} b + \frac{1}{1 - \theta} c.
\label{eq:proof:lemma:obsp_step3:eq11}
\end{equation}

Otherwise, if $(1 - \theta) a > \theta b + c$, we have
\[
\theta^2 (b^2 - a^2) \geq \left((1-\theta) a -\theta b - c\right)^2,
\]
which implies
\[
(2\theta^2 + 2\theta + 1)a^2 - 2(1-\theta)(\theta b + c) a + (\theta b + c)^2 - \theta^2 b^2 \leq 0.
\]
Therefore,
\begin{equation}
a \leq \frac{2\theta(1-\theta)}{2\theta^2 + 2\theta + 1} b
+ \frac{2(1-\theta)}{2\theta^2 + 2\theta + 1} c.
\label{eq:proof:lemma:obsp_step3:eq12}
\end{equation}

Combining \cref{eq:proof:lemma:obsp_step3:eq11,eq:proof:lemma:obsp_step3:eq12} completes the proof.
\end{IEEEproof}

\subsection{Proof of Theorem~\ref{thm:urbp}}
\label{subsec:proof:thm:urbp}

Let $X$ be a random variable defined as
\[
X \triangleq \max_{|J| = s} \Bigg\|\Pi_J \left( \widetilde{\Psi}^* \Psi - \mathbb{E} \widetilde{\Psi}^* \Psi \right) \Pi_J\Bigg\|.
\]

Let $\xi_k$ and $\zeta_k$ be the transposed $k$th row of $\sqrt{m} \Psi$ and $\sqrt{m} \widetilde{\Psi}$, respectively, for all $k \in [n]$.
By the assumption, $(\xi_k)_{k=1}^m$ and $(\zeta_k)_{k=1}^m$ are sequences of independent random vectors such that
\[
\mathbb{E} \xi_k \xi_k^* = \mathbb{E} \Psi^* \Psi \quad \text{and} \quad \mathbb{E} \zeta_k \zeta_k^* = \mathbb{E} \widetilde{\Psi}^* \widetilde{\Psi}
\]
for all $k \in [m]$.
Then, $X$ is rewritten as
\[
X = \max_{|J| = s} \Bigg\|\Pi_J \left(\frac{1}{m} \sum_{k=1}^m (\zeta_k \xi_k^* - \mathbb{E} \zeta_k \xi_k^*) \right) \Pi_J\Bigg\|.
\]

Like the RIP analysis for the Hermitian case \cite{Rau10,RudVer08},
the first step is to show
\[
\mathbb{E} X \leq \frac{8 \delta}{9}.
\]

By symmetrization \cite[Lemma~6.7]{Rau10}, $\mathbb{E} X$ is bounded from above by
{\allowdisplaybreaks
\begin{align}
\mathbb{E} X
\leq 2 \mathbb{E} \max_{|J| = s} \Bigg\|\Pi_J \left(\frac{1}{m} \sum_{k=1}^m \epsilon_k \zeta_k \xi_k^* \right) \Pi_J\Bigg\|
= \frac{2}{m} \mathbb{E} \max_{|J| = s} \Bigg\|\Pi_J \left(\sum_{k=1}^m \epsilon_k \zeta_k \xi_k^* \right) \Pi_J\Bigg\|
\label{eq:bnd_symEX}
\end{align}}
where $(\epsilon_k)_{k=1}^m$ is a Rademacher sequence independent of $(\xi_k)_{k=1}^m$ and $(\zeta_k)_{k=1}^m$.

Define random variables $X_1$ and $X_2$ by
{\allowdisplaybreaks
\begin{align*}
X_1 {} &
\triangleq \max_{|J| = s} \Bigg\|\Pi_J \left( \Psi^* \Psi - \mathbb{E} \Psi^* \Psi \right) \Pi_J\Bigg\| \\
X_2 {} &
\triangleq \max_{|J| = s} \Bigg\|\Pi_J \left( \widetilde{\Psi}^* \widetilde{\Psi} - \mathbb{E} \widetilde{\Psi}^* \widetilde{\Psi} \right) \Pi_J\Bigg\|.
\end{align*}}
Then, $X_1$ and $X_2$ are rewritten as
{\allowdisplaybreaks
\begin{align*}
X_1 {} &
= \max_{|J| = s} \Bigg\|\Pi_J \left(\frac{1}{m} \sum_{k=1}^m \xi_k \xi_k^* - \mathbb{E} \xi_k \xi_k^* \right) \Pi_J\Bigg\| \\
X_2 {} &
= \max_{|J| = s} \Bigg\|\Pi_J \left(\frac{1}{m} \sum_{k=1}^m \zeta_k \zeta_k^* - \mathbb{E} \zeta_k \zeta_k^* \right) \Pi_J\Bigg\|.
\end{align*}}

By symmetrization, $\mathbb{E} X_1$ and $\mathbb{E} X_2$ are bounded from above by
{\allowdisplaybreaks
\begin{align*}
\mathbb{E} X_1 {} & \leq \frac{2}{m} \mathbb{E} \max_{|J| = s} \Bigg\|\Pi_J \left(\sum_{k=1}^m \epsilon_k \xi_k \xi_k^* \right) \Pi_J\Bigg\| \\
\mathbb{E} X_2 {} & \leq \frac{2}{m} \mathbb{E} \max_{|J| = s} \Bigg\|\Pi_J \left(\sum_{k=1}^m \epsilon_k \zeta_k \zeta_k^* \right) \Pi_J\Bigg\|.
\end{align*}}

\begin{lemma}[{\cite[Lemma~3.8]{RudVer08}}]
Let $(h_k)_{k=1}^m$ be vectors in $\mathbb{K}^n$.
Let $K_h \triangleq \max_k \norm{h_k}_{\infty}$.
Then,
{\allowdisplaybreaks
\begin{align*}
{} & \mathbb{E}_\epsilon \max_{|J| = s} \Bigg\|\Pi_J \left(\sum_{k=1}^m \epsilon_k h_k h_k^* \right) \Pi_J\Bigg\| \\
{} & \leq
C_3 \sqrt{s} \ln s \sqrt{\ln n} \sqrt{\ln m}
K_h \max_{|J| = s} \Bigg\|\Pi_J \left(\sum_{k=1}^m h_k h_k^* \right) \Pi_J\Bigg\|^{1/2}.
\end{align*}}
\label{lemma:keyestsym}
\end{lemma}

Since $\max_k \norm{\xi_k}_{\infty} \leq K$, by Lemma~\ref{lemma:keyestsym}, it follows that
\[
\mathbb{E} X_1 \leq 2C_3 \sqrt{\frac{s}{m}} \ln s \sqrt{\ln n} \sqrt{\ln m} K \sqrt{\mathbb{E} X_1 + 1 + \theta_s(\mathbb{E} \Psi^* \Psi)}.
\]
If
\[
2 C_3 \sqrt{\frac{s}{m}} \ln s \sqrt{\ln n} \sqrt{\ln m} K \sqrt{2 + \theta_s(\mathbb{E} \Psi^* \Psi)} \leq \delta_1
\]
for some $\delta_1 < 1$, then $\mathbb{E} X_1 \leq \delta_1$ and it follows that
{\allowdisplaybreaks
\begin{align}
{} & 2C_3 \sqrt{\frac{s}{m}} \ln s \sqrt{\ln n} \sqrt{\ln m}
K \mathbb{E} \max_{|J| = s} \Bigg\|\Pi_J \left(\frac{1}{m} \sum_{k=1}^m \xi_k \xi_k^* \right) \Pi_J\Bigg\|^{1/2} \nonumber \\
{} & \leq 2C_3 \sqrt{\frac{s}{m}} \ln s \sqrt{\ln n} \sqrt{\ln m}
K \sqrt{\mathbb{E} X_1 + 1 + \theta_s(\mathbb{E} \Psi^* \Psi)} \nonumber \\
{} & \leq 2C_3 \sqrt{\frac{s}{m}} \ln s \sqrt{\ln n} \sqrt{\ln m}
K \sqrt{\delta_1 + 1+ \theta_s(\mathbb{E} \Psi^* \Psi)} \nonumber \\
{} & \leq 2C_3 \sqrt{\frac{s}{m}} \ln s \sqrt{\ln n} \sqrt{\ln m}
K \sqrt{2 + \theta_s(\mathbb{E} \Psi^* \Psi)} \leq \delta_1. \label{eq:bnd_delta1}
\end{align}}

Since $\max_k \norm{\zeta_k}_\infty \leq \widetilde{K}$, by Lemma~\ref{lemma:keyestsym}, it follows that
\[
\mathbb{E} X_2 \leq 2C_3 \sqrt{\frac{s}{m}} \ln s \sqrt{\ln n} \sqrt{\ln m} \widetilde{K}
\sqrt{\mathbb{E} X_2 + 1 + \theta_s(\mathbb{E} \widetilde{\Psi}^* \widetilde{\Psi})}.
\]
Similarly, if
\begin{equation*}
2C_3 \sqrt{\frac{s}{m}} \ln s \sqrt{\ln n} \sqrt{\ln m} \widetilde{K} \sqrt{2 + \theta_s(\mathbb{E} \widetilde{\Psi}^* \widetilde{\Psi})} \leq \delta_2
\end{equation*}
for some $\delta_2 < 1$, then $\mathbb{E} X_2 \leq \delta_2$; hence,
{\allowdisplaybreaks
\begin{align}
{} & 2C_3 \sqrt{\frac{s}{m}} \ln s \sqrt{\ln n} \sqrt{\ln m} \widetilde{K} \nonumber \\
{} & \quad \cdot \mathbb{E} \max_{|J| = s} \Bigg\|\Pi_J \left(\frac{1}{m} \sum_{k=1}^m \zeta_k \zeta_k^* \right) \Pi_J\Bigg\|^{1/2}
\leq \delta_2. \label{eq:bnd_delta2}
\end{align}}

Unlike the conventional RIP analyses \cite{RudVer08,Rau10},
matrices $(\zeta_k \xi_k^*)_{k=1}^m$ are not Hermitian symmetric.
The following lemma is modified from Lemma~\ref{lemma:keyestsym}
to get a bound on $\mathbb{E} X$ for the non-Hermitian case.

\begin{lemma}
Let $(h_k)_{k=1}^m$ and $(\widetilde{h}_k)_{k=1}^m$ be vectors in $\mathbb{K}^n$.
Let $K_h \triangleq \max_k \norm{h_k}_{\infty}$ and $K_{\widetilde{h}} \triangleq \max_k \norm{\widetilde{h}_k}_{\infty}$.
Then,
{\allowdisplaybreaks
\begin{align}
{} & \mathbb{E}_\epsilon \max_{|J| = s} \Bigg\|\Pi_J \left(\sum_{k=1}^m \epsilon_k \widetilde{h}_k h_k^* \right) \Pi_J\Bigg\| \nonumber \\
{} & \leq
C_3 \sqrt{s} \ln s \sqrt{\ln n} \sqrt{\ln m} \nonumber \\
{} & \quad \cdot \Bigg[
K_h \max_{|J| = s} \Bigg\|\Pi_J \left(\sum_{k=1}^m h_k h_k^* \right) \Pi_J\Bigg\|^{1/2}
+ K_{\widetilde{h}} \max_{|J| = s} \Bigg\|\Pi_J \left(\sum_{k=1}^m \widetilde{h}_k \widetilde{h}_k^* \right) \Pi_J\Bigg\|^{1/2}
\Bigg]. \label{eq:lemma:keyest}
\end{align}}
\label{lemma:keyest}
\end{lemma}

\begin{IEEEproof}[Proof of Lemma~\ref{lemma:keyest}]
By a comparison principle \cite[inequality (4.8)]{LedTal91}, the left-hand side of (\ref{eq:lemma:keyest}), denoted by $E_1$, is bounded from above by
{\allowdisplaybreaks
\begin{align*}
E_1
\leq \frac{\sqrt{\pi}}{2} \mathbb{E}_g \max_{|J| = s} \Bigg\|\Pi_J \left(\sum_{k=1}^m g_k \widetilde{h}_k h_k^* \right) \Pi_J\Bigg\|
= \frac{\sqrt{\pi}}{2} \mathbb{E}_g \max_{|J| = s} \max_{x,y \in \mathcal{B}_2^J} \left| \sum_{k=1}^m g_k y^* \widetilde{h}_k h_k^* x \right|
\end{align*}}
where $(g_k)_{k=1}^m$ is the standard i.i.d. Gaussian sequence and
$\mathcal{B}_2^J \triangleq \{ x \in \mathbb{K}^n : \norm{x}_2 \leq 1, ~ \supp{x} \subset J \}$.

Define a Gaussian process $G_{x,y}$ indexed by $(x,y) \in \mathbb{K}^n \times \mathbb{K}^n$ as
\[
G_{x,y} \triangleq \sum_{k=1}^m g_k y^* \widetilde{h}_k h_k^* x.
\]

By Dudley's inequality, $E_1$ is bounded from above by
\[
E_1 \leq C_4 \int_0^\infty \Big( \ln N \big( \bigcup_{|J| = s} \big(\mathcal{B}_2^J \times \mathcal{B}_2^J\big), d, u \big) \Big)^{1/2} du
\]
where $N(\Set{B},d,u)$ is the covering number of set $\Set{B}$ with respect to the metric $d$,
induced from the Gaussian process $G_{x,y}$ by
\[
d((x,y),(x',y')) \triangleq \left( \mathbb{E} |G_{x,y} - G_{x',y'}|^2 \right)^{1/2}.
\]

Let
\[
M_1 \triangleq \max_{|J| = s} \Bigg\|\Pi_J \left(\sum_{k=1}^m h_k h_k^* \right) \Pi_J\Bigg\|^{1/2}
\]
and
\[
M_2 \triangleq \max_{|J| = s} \Bigg\|\Pi_J \left(\sum_{k=1}^m \widetilde{h}_k \widetilde{h}_k^* \right) \Pi_J\Bigg\|^{1/2}.
\]

Define
\begin{equation*}
\norm{x}_h \triangleq \max_{k \in [n]} |h_k^*x|
\quad \text{and} \quad
\norm{x}_{\widetilde{h}} \triangleq \max_{k \in [n]} |\widetilde{h}_k^*x|
\end{equation*}
for $x \in \mathbb{K}^n$.
Then, $\norm{\cdot}_h$ and $\norm{\cdot}_{\widetilde{h}}$ are valid norms on $\mathbb{K}^n$
induced by $(h_k)_{k=1}^m$ and $(\widetilde{h}_k)_{k=1}^m$, respectively.

Let $x,x',y',y'$ be arbitrary $s$-sparse vectors in $\mathbb{K}^n$.
Then, $d((x,y),(x',y))$ is upper bounded by
\begin{align}
d((x,y),(x',y))^2
{} & = \mathbb{E} |G_{x,y} - G_{x',y}|^2 \nonumber \\
{} & = \mathbb{E} \big|\sum_{k=1}^m g_k y^* \widetilde{h}_k h_k^* (x - x')\big|^2 \nonumber \\
{} & \leq \max_{k \in [n]} |h_k^*(x - x')|^2 \mathbb{E} \big|\sum_{k=1}^m g_k y^* \widetilde{h}_k\big|^2 \nonumber \\
{} & = \max_{k \in [n]} |h_k^*(x - x')|^2 \sum_{k=1}^m |y^* \widetilde{h}_k|^2 \nonumber \\
{} & = \norm{x-x'}_h^2 |h_k^*(x - x')|^2 \sum_{k=1}^m y^* \widetilde{h}_k \widetilde{h}_k^* y \nonumber \\
{} & \leq M_2^2 \norm{x-x'}_h^2
\label{eq:proof:thm:urbp:eq1}
\end{align}
where the fourth step follows since $(g_k)_{k=1}^m$ is the standard i.i.d. Gaussian sequence.

Similarly, $d((x',y),(x',y'))$ is upper bounded by
\begin{equation}
d((x',y),(x',y')) \leq M_1 \norm{y-y'}_{\widetilde{h}}.
\label{eq:proof:thm:urbp:eq2}
\end{equation}

Then, by the triangle inequality and \cref{eq:proof:thm:urbp:eq1,eq:proof:thm:urbp:eq2}, it follows that
{\allowdisplaybreaks
\begin{align*}
d((x,y),(x',y'))
{} & \leq d((x,y),(x',y)) + d((x',y),(x',y')) \\
{} & \leq M_2 \norm{x-x'}_h + M_1 \norm{y-y'}_{\widetilde{h}};
\end{align*}}
Hence,
{\allowdisplaybreaks
\begin{align*}
{} & \Big( \ln N \big( \bigcup_{|J| = s} \big(\mathcal{B}_2^J \times \mathcal{B}_2^J\big), d, u \big) \Big)^{1/2} \\
{} & \leq \Big( \ln N \big( \bigcup_{|J| = s} \mathcal{B}_2^J, M_2 \norm{\cdot}_h, u/2 \big) \Big)^{1/2} \\
{} & \quad + \Big( \ln N \big( \bigcup_{|J| = s} \mathcal{B}_2^J, M_1 \norm{\cdot}_{\widetilde{h}}, u/2 \big) \Big)^{1/2} \\
{} & \leq 2 M_2 \sqrt{s} \Big( \ln N \big( \bigcup_{|J| = s} \frac{1}{\sqrt{s}} \mathcal{B}_2^J, \norm{\cdot}_h, u \big) \Big)^{1/2} \\
{} & \quad + 2 M_1 \Big( \ln N \big( \bigcup_{|J| = s} \frac{1}{\sqrt{s}} \mathcal{B}_2^J, \norm{\cdot}_{\widetilde{h}}, u \big) \Big)^{1/2}.
\end{align*}}

The remaining steps are identical to the Hermitian case (\cite[Lemma~8.2]{Rau10}, \cite[Lemma~3.8]{RudVer08})
and we do not reproduce the details.
We obtain the desired bound by noting
\begin{align*}
{} & \int_0^\infty \Big( \ln N \big( \bigcup_{|J| = s} \frac{1}{\sqrt{s}} \mathcal{B}_2^J, \norm{\cdot}_h, u \big) \Big)^{1/2} \\
{} & \leq C_4 K_h \ln s \sqrt{\ln n} \sqrt{\ln m}
\end{align*}
and
\begin{align*}
{} & \int_0^\infty \Big( \ln N \big( \bigcup_{|J| = s} \frac{1}{\sqrt{s}} \mathcal{B}_2^J, \norm{\cdot}_{\widetilde{h}}, u \big) \Big)^{1/2} \\
{} & \leq C_4 K_{\widetilde{h}} \ln s \sqrt{\ln n} \sqrt{\ln m},
\end{align*}
which have been shown in the proof of \cite[Lemma~3.8]{RudVer08}.
\end{IEEEproof}

Let
\[
\delta_1 = \frac{K \sqrt{2 + \theta_s(\mathbb{E} \Psi^* \Psi)}}
{K \sqrt{2 + \theta_s(\mathbb{E} \Psi^* \Psi)} + \widetilde{K} \sqrt{2 + \theta_s(\mathbb{E} \widetilde{\Psi}^* \widetilde{\Psi})}}
\cdot \frac{8\delta}{9}
\]
and
\[
\delta_2 = \frac{\widetilde{K} \sqrt{2 + \theta_s(\mathbb{E} \widetilde{\Psi}^* \widetilde{\Psi})}}
{K \sqrt{2 + \theta_s(\mathbb{E} \Psi^* \Psi)} + \widetilde{K} \sqrt{2 + \theta_s(\mathbb{E} \widetilde{\Psi}^* \widetilde{\Psi})}}
\cdot \frac{8\delta}{9}.
\]

Applying Lemma~\ref{lemma:keyest} to (\ref{eq:bnd_symEX}), we obtain the following bound on $\mathbb{E} X$.
If
\begin{align*}
{} & 2C_3 \sqrt{\frac{s}{m}} \ln s \sqrt{\ln n} \sqrt{\ln m} \\
{} & \leq
\frac{(8/9)\delta}{K \sqrt{2 + \theta_s(\mathbb{E} \Psi^* \Psi)} + \widetilde{K} \sqrt{2 + \theta_s(\mathbb{E} \widetilde{\Psi}^* \widetilde{\Psi})}},
\end{align*}
then
{\allowdisplaybreaks
\begin{align*}
\mathbb{E} X
{} & \leq 2 \mathbb{E} \max_{|J| = s} \Bigg\|\Pi_J \left(\sum_{k=1}^m \epsilon_k \zeta_k \xi_k^* \right) \Pi_J\Bigg\| \nonumber \\
{} & \leq 2C_3 \sqrt{\frac{s}{m}} \ln s \sqrt{\ln n} \sqrt{\ln m} \nonumber \\
{} & \quad \cdot \mathbb{E} \Bigg[
K \max_{|J| = s} \Bigg\|\Pi_J \left( \frac{1}{m} \sum_{k=1}^m \xi_k \xi_k^* \right) \Pi_J\Bigg\|^{1/2} \nonumber \\
{} & \qquad + \widetilde{K} \max_{|J| = s} \Bigg\|\Pi_J \left( \frac{1}{m} \sum_{k=1}^m \zeta_k \zeta_k^* \right) \Pi_J\Bigg\|^{1/2}
\Bigg] \\
{} & \leq \delta_1 + \delta_2 = \frac{8\delta}{9}
\end{align*}}
where the last inequality follows from (\ref{eq:bnd_delta1}) and (\ref{eq:bnd_delta2}).

The second step is to show that $X$ is concentrated around $\mathbb{E} X$ with high probability.
The corresponding result for the Hermitian case \cite[Section~8.6]{Rau10} has been derived
using a probabilistic upper bound on a random variable defined as the supremum of an empirical process \cite[Theorem~6.25]{Rau10}.
We show that the derivation for the Hermitian case \cite[Section~8.6]{Rau10} extends to the non-Hermitian case with slight modifications.

Let $\mathcal{B}_2^J \triangleq \{ x \in \mathbb{K}^n : \norm{x}_2 \leq 1, ~ \supp{x} \subset J \}$.
Since $\mathcal{B}_2^J$ is closed under the multiplication with any scalar of unit modulus,
$m X$ is written as
\begin{align*}
m X {} = \max_{|J| = s} \max_{x,y \in \mathcal{B}_2^J} \left| \sum_{k=1}^m y^* (\zeta_k \xi_k^* - \mathbb{E} \zeta_k \xi_k^*) x \right|
= \max_{|J| = s} \max_{x,y \in \mathcal{B}_2^J} \mathrm{Re}\left( \sum_{k=1}^m y^* (\zeta_k \xi_k^* - \mathbb{E} \zeta_k \xi_k^*) x \right).
\end{align*}

Define $f_{x,y} : \mathbb{K}^{n \times n} \to \mathbb{R}$ by
\[
f_{x,y}(Z) \triangleq \mathrm{Re}\left( y^* (Z - \mathbb{E} Z) x \right).
\]
Then, $\mathbb{E} f_{x,y}(\zeta_k \xi_k^*) = 0$ for all $k \in [m]$ and $m X$ is rewritten as
\[
m X = \max_{x,y} \Big\{ \sum_{k=1}^m f_{x,y}(\zeta_k \xi_k^*) : (x,y) \in \bigcup_{|J| = s} \left( \Set{B}_2^J \times \Set{B}_2^J \right)\Big\}.
\]

Let $k \in [m]$ be fixed.
Let $x,y \in \Set{B}_2^J$.
Then,
{\allowdisplaybreaks
\begin{align}
|f_{x,y}(\zeta_k \xi_k^*)|
{} & \leq \left|y^* \Pi_J (\zeta_k \xi_k^* - \mathbb{E} \zeta_k \xi_k^*) \Pi_J x \right| \nonumber \\
{} & \leq \big\| \Pi_J (\zeta_k \xi_k^* - \mathbb{E} \zeta_k \xi_k^*) \Pi_J \big\|_{\ell_2^n \to \ell_2^n} \nonumber \\
{} & \leq \big\| \Pi_J (\zeta_k \xi_k^* - \mathbb{E} \zeta_k \xi_k^*) \Pi_J \big\|_{\ell_1^n \to \ell_1^n}^{1/2} \nonumber \\
{} & \quad \cdot \big\| \Pi_J (\zeta_k \xi_k^* - \mathbb{E} \zeta_k \xi_k^*) \Pi_J \big\|_{\ell_\infty^n \to \ell_\infty^n}^{1/2} \nonumber \\
{} & = \big\| \Pi_J (\zeta_k \xi_k^* - \mathbb{E} \zeta_k \xi_k^*) \Pi_J \big\|_{\ell_1^n \to \ell_1^n}^{1/2} \nonumber \\
{} & \quad \cdot \big\| \Pi_J (\xi_k \zeta_k^* - \mathbb{E} \xi_k \zeta_k^*) \Pi_J \big\|_{\ell_1^n \to \ell_1^n}^{1/2}
\label{eq:proof:thm:urbp:eq3}
\end{align}}
where the third inequality follows from Schur's interpolation theorem \cite{JosMR03}.

We derive an upper bound on $\big\| \Pi_J (\zeta_k \xi_k^* - \mathbb{E} \zeta_k \xi_k^*) \Pi_J \big\|_{\ell_1^n \to \ell_1^n}$ by
\begin{align}
{} & \big\| \Pi_J (\zeta_k \xi_k^* - \mathbb{E} \zeta_k \xi_k^*) \Pi_J \big\|_{\ell_1^n \to \ell_1^n} \nonumber \\
{} & \leq \big\| \Pi_J \zeta_k \xi_k^* \Pi_J \big\|_{\ell_1^n \to \ell_1^n} + \big\| \Pi_J \mathbb{E} \zeta_k \xi_k^* \Pi_J \big\|_{\ell_1^n \to \ell_1^n} \nonumber \\
{} & \leq \big\| \Pi_J \zeta_k \xi_k^* \Pi_J \big\|_{\ell_1^n \to \ell_1^n} + \mathbb{E} \big\| \Pi_J \zeta_k \xi_k^* \Pi_J \big\|_{\ell_1^n \to \ell_1^n}
\nonumber \\
{} & \leq 2 s K \widetilde{K}
\label{eq:proof:thm:urbp:eq4}
\end{align}
where the second inequality follows from Jensen's inequality,
and the last step holds since
\begin{equation*}
\big\| \Pi_J \zeta_k \xi_k^* \Pi_J \big\|_{\ell_1^n \to \ell_1^n}
\leq s \norm{\zeta_k}_\infty \norm{\xi_k}_\infty
\leq s K \widetilde{K}.
\end{equation*}

Similarly, we have
\begin{equation}
\big\| \Pi_J (\xi_k \zeta_k^* - \mathbb{E} \xi_k \zeta_k^*) \Pi_J \big\|_{\ell_1^n \to \ell_1^n} \leq 2 s K \widetilde{K}.
\label{eq:proof:thm:urbp:eq5}
\end{equation}

By applying \cref{eq:proof:thm:urbp:eq4,eq:proof:thm:urbp:eq5} to (\ref{eq:proof:thm:urbp:eq3}), we obtain
\begin{equation}
|f_{x,y}(\zeta_k \xi_k^*)| \leq 2 s K \widetilde{K}.
\label{eq:proof:thm:urbp:eq6}
\end{equation}

Since $k$ was arbitrary, (\ref{eq:proof:thm:urbp:eq6}) implies that $f_{x,y}(\zeta_k \xi_k^*)$ is uniformly bounded
for all $(x,y) \in \bigcup_{|J| = s} \left( \Set{B}_2^J \times \Set{B}_2^J \right)$ and for all $k \in [m]$.

We also verify that the second moment of $f_{x,y}(\zeta_k \xi_k^*)$ is uniformly bounded by
{\allowdisplaybreaks
\begin{align*}
{} & \mathbb{E} |f_{x,y}(\zeta_k \xi_k^*)|^2 \\
{} & = \mathbb{E} \left|y^* \Pi_J (\zeta_k \xi_k^* - \mathbb{E} \zeta_k \xi_k^*) \Pi_J x \right|^2 \\
{} & \leq \mathbb{E} \big\|\Pi_J (\zeta_k \xi_k^* - \mathbb{E} \zeta_k \xi_k^*) \Pi_J x \big\|_2^2 \\
{} & = \mathbb{E} \Big[ x^* \Big( \norm{\Pi_J \zeta_k}_2^2 \Pi_J \xi_k \xi_k^* \Pi_J
- \Pi_J (\mathbb{E} \xi_k \zeta_k^*) \Pi_J \zeta_k \xi_k^* \Pi_J \\
{} & \qquad - \Pi_J \xi_k \zeta_k^* \Pi_J (\mathbb{E} \zeta_k \xi_k^*) \Pi_J
+ \Pi_J (\mathbb{E} \xi_k \zeta_k^*) \Pi_J (\mathbb{E} \zeta_k \xi_k^*) \Pi_J \Big) x \Big] \\
{} & = x^* \Big( \norm{\Pi_J \zeta_k}_2^2 \Pi_J (\mathbb{E} \xi_k \xi_k^*) \Pi_J
- \Pi_J (\mathbb{E} \xi_k \zeta_k^*) \Pi_J (\mathbb{E} \zeta_k \xi_k^*) \Pi_J \Big) x \\
{} & \leq \norm{\Pi_J \zeta_k}_2^2 \norm{\Pi_J (\mathbb{E} \xi_k \xi_k^*) \Pi_J} + \norm{\Pi_J (\mathbb{E} \xi_k \zeta_k^*) \Pi_J}^2 \\
{} & \leq s \widetilde{K}^2 (1 + \theta_s(\mathbb{E} \Psi^* \Psi)) + 1 + \theta_s(\mathbb{E} \widetilde{\Psi}^* \Psi).
\end{align*}}

Then, by \cite[Theorem~6.25]{Rau10}, we obtain (\ref{eq:proof:thm:urbp:eq7}).
\begin{align}
\mathbb{P}(X \geq \delta)
{} & \leq \mathbb{P}\left(m X \geq m \mathbb{E} X + \frac{m \delta}{9}\right) \nonumber \\
{} & \leq \exp\left( - \frac{\left(\frac{\delta}{9} \cdot \frac{m}{2sK\widetilde{K}}\right)^2}
{\frac{2m}{4sK^2} \left(1 + \theta_s(\mathbb{E} \Psi^* \Psi) + \frac{1 + \theta_s(\mathbb{E} \widetilde{\Psi}^* \Psi)}{s\widetilde{K}^2}\right)
+ \frac{32\delta}{9} \cdot \frac{m}{2sK\widetilde{K}}
+ \frac{2\delta}{27} \cdot \frac{m}{2sK\widetilde{K}}} \right) \nonumber \\
{} & = \exp\left( - \frac{m \delta^2}{
6s\widetilde{K} \left( 27 \widetilde{K} \left( 1 + \theta_s(\mathbb{E} \Psi^* \Psi) + \frac{1 + \theta_s(\mathbb{E} \widetilde{\Psi}^* \Psi)}{s\widetilde{K}^2} \right) + 98 \delta K \right)
} \right) \nonumber \\
{} & \leq \exp\left( - \frac{m \delta^2}{1236 s \widetilde{K} \max(K,\widetilde{K})} \right).
\label{eq:proof:thm:urbp:eq7}
\end{align}

Therefore, $\mathbb{P}(\theta_s(\widetilde{\Psi}^* \Psi) \geq \delta) \leq \eta$ holds provided that $m$ satisfies
\begin{align*}
m {} & \geq C_1 \delta^{-2} \left( K \sqrt{2 + \theta_s(\mathbb{E} \Psi^* \Psi)} + \widetilde{K} \sqrt{2 + \theta_s(\mathbb{E} \widetilde{\Psi}^* \widetilde{\Psi})} \right)^2 \\
{} & \quad \cdot s (\ln s)^2 \ln n \ln m
\end{align*}
and
\[
m \geq C_2 \delta^{-2} \widetilde{K} \max(K,\widetilde{K}) s \ln(\eta^{-1})
\]
for universal constants $C_1$ and $C_2$.

\subsection{Proof of Theorem~\ref{thm:urip}}
\label{subsec:proof:thm:urip}

Since $\Psi = A D$, by the construction of $A$ from $(\phi_\omega)_{\omega \in \Omega}$ in (\ref{eq:contrA}), it follows that $\max_{k,\ell} |(\Psi)_{k,\ell}| \leq \sup_\omega \max_j |\langle \phi_\omega, d_j \rangle|$; hence, the incoherence property of $\Psi$ is satisfied by the assumption.
To invoke Corollary~\ref{cor:urip}, it remains to show $\theta_s(\mathbb{E} \Psi^* \Psi) \leq K_0$.
By the definition of $\theta_s$, $\theta_s(\mathbb{E} \Psi^* \Psi)$ is rewritten as
\begin{align}
\theta_s(\mathbb{E} \Psi^* \Psi)
= \max\Bigg[ \max_{|J| = s} \norm{D_J^* \mathbb{E} A^* A D_J} - 1 ,~ 1 - \min_{|J| = s} \lambda_n({D_J^* \mathbb{E} A^* A D_J}) \Bigg].
\label{eq:proof:thm:urip:eq1}
\end{align}
Let $J$ be an arbitrary subset of $[n]$ with $s$ elements.
Then, it follows that
\begin{align}
\norm{D_J^* \mathbb{E} A^* A D_J}
{} & \leq \norm{D_J^*} \norm{\mathbb{E} A^* A} \norm{D_J} \nonumber \\
{} & = \norm{\mathbb{E} A^* A} \norm{D_J}^2 \nonumber \\
{} & \leq \nu_{\max} \norm{\Phi \Phi^*} [1 + \delta_s(D)] \nonumber \\
{} & \leq \nu_{\max} [1 + \theta_d(\Phi \Phi^*)] [1 + \delta_s(D)]
\label{eq:proof:thm:urip:eq2}
\end{align}
and
\begin{align}
\lambda_n(D_J^* \mathbb{E} A^* A D_J)
{} & \geq \sigma_n(D_J^*) \lambda_n(\mathbb{E} A^* A) \sigma_n(D_J) \nonumber \\
{} & \geq \lambda_n(\mathbb{E} A^* A) \lambda_n(D_J^* D_J) \nonumber \\
{} & \geq \nu_{\min} \lambda_n(\Phi \Phi^*) [1 - \delta_s(D)] \nonumber \\
{} & \geq \nu_{\min} [1 - \theta_n(\Phi \Phi^*)] [1 - \delta_s(D)].
\label{eq:proof:thm:urip:eq3}
\end{align}
Applying (\ref{eq:proof:thm:urip:eq2}) and (\ref{eq:proof:thm:urip:eq3}) to (\ref{eq:proof:thm:urip:eq1}),
we verify that $K_0$ given in (\ref{eq:thm:urip:K0}) is a valid upper bound on $\theta_s(\mathbb{E} \Psi^* \Psi)$.
This completes the proof.

\subsection{Proof of Theorem~\ref{thm:urbpBO}}
\label{subsec:proof:thm:urbpBO}

First, we note that the mutual incoherence between $(\phi_\omega)_{\omega \in \Omega}$ and $(d_j)_{j \in [n]}$ is written as an operator norm given by
\begin{align*}
\sup_{\omega \in \Omega} \max_{j \in [n]} |\langle \phi_\omega, d_j \rangle| = \norm{\Phi^* D}_{\ell_1^n \to L_\infty(\Omega,\mu)}.
\end{align*}
Similarly, the mutual incoherence between $(\widetilde{\phi}_\omega)_{\omega \in \Omega}$ and $(\widetilde{d}_j)_{j \in [n]}$ is written as
\begin{align*}
\sup_{\omega \in \Omega} \max_{j \in [n]} |\langle \widetilde{\phi}_\omega, \widetilde{d}_j \rangle| = \norm{\Lambda_\nu^{-1} \widetilde{\Phi}^* \widetilde{D}}_{\ell_1^n \to L_\infty(\Omega,\mu)}
\end{align*}
where $\Lambda_\nu^{-1}: L_2(\Omega,\mu) \to L_2(\Omega,\mu)$ is a diagonal operator defined by
\[
(\Lambda_\nu^{-1} h) (\omega) = \frac{d\mu}{d\nu}(\omega) h(\omega), \quad \forall h \in L_2(\Omega,\mu).
\]

Then, $\norm{\Lambda_\nu^{-1} \widetilde{\Phi}^* \widetilde{D}}_{\ell_1^n \to L_\infty(\Omega,\mu)}$ is upper bounded using $K$ as follows:
\begin{align}
{} & \norm{\Lambda_\nu^{-1} \widetilde{\Phi}^* \widetilde{D}}_{\ell_1^n \to L_\infty(\Omega,\mu)} \nonumber \\
{} & = \norm{\Lambda_\nu^{-1} \Phi^* (\Phi \Phi^*)^{-1} D (D^*D)^{-1}}_{\ell_1^n \to L_\infty(\Omega,\mu)} \nonumber \\
{} & \leq \norm{\Lambda_\nu^{-1} \Phi^* D (D^*D)^{-1}}_{\ell_1^n \to L_\infty(\Omega,\mu)} \nonumber \\
{} & \quad + \norm{\Lambda_\nu^{-1} \Phi^* [(\Phi \Phi^*)^{-1} - I_d] D (D^*D)^{-1}}_{\ell_1^n \to L_\infty(\Omega,\mu)} \nonumber \\
{} & \leq \frac{1}{\nu_{\min}} \norm{\Phi^* D}_{\ell_1^n \to L_\infty(\Omega,\mu)} \norm{(D^*D)^{-1}}_{\ell_1^n \to \ell_1^n} \nonumber \\
{} & \quad + \frac{1}{\nu_{\min}} \norm{\Phi^*}_{\ell_2 \to L_\infty(\Omega,\mu)} \norm{(\Phi \Phi^*)^{-1} - I_d}_{\ell_2^d \to \ell_2^d} \nonumber \\
{} & \qquad \cdot \norm{D}_{\ell_1^n \to \ell_2^d} \norm{(D^*D)^{-1}}_{\ell_1^n \to \ell_1^n} \nonumber \\
{} & \leq \frac{\norm{(D^*D)^{-1}}_{\ell_1^n \to \ell_1^n}}{\nu_{\min}}
\Bigg[ K + \left(\sup_{\omega \in \Omega} \norm{\phi_\omega}_{\ell_2^d}\right) \nonumber \\
{} & \qquad \cdot \frac{\theta_d(\Phi \Phi^*)}{1 - \theta_d(\Phi \Phi^*)} \cdot \left( \max_{j \in [n]} \norm{d_j}_{\ell_2^d} \right) \Bigg].
\label{eq:proof:thm:urbpBO:eq1}
\end{align}
Let $\widetilde{K}$ be the right hand side of (\ref{eq:proof:thm:urbpBO:eq1}).
Then, we apply the incoherence parameters $K$ and $\widetilde{K}$ to Corollary~\ref{cor:urbp}.
Since $\mathbb{E} \widetilde{\Psi}^* \Psi = I_n$, we have $\theta_s(\mathbb{E} \widetilde{\Psi}^* \Psi) = 0$.
Therefore, to obtain a condition on $m$,
it only remains to bound $\theta_s(\mathbb{E} \Psi^* \Psi)$ and $\theta_s(\mathbb{E} \widetilde{\Psi}^* \widetilde{\Psi})$.

In the proof of Theorem~\ref{thm:urip}, we derived an upper bound on $\theta_s(\mathbb{E} \Psi^* \Psi)$ given by
\begin{align}
\theta_s(\mathbb{E} \Psi^* \Psi) {} & \leq \max(1 - \nu_{\min}, \nu_{\max} - 1) \nonumber \\
{} & \quad + \nu_{\max} [\delta_n(D)
+ \theta_d(\Phi \Phi^*) + \delta_n(D) \theta_d(\Phi \Phi^*)].
\label{eq:proof:thm:urbpBO:eq2}
\end{align}
This upper bound is tight in the sense that equality is achieved if $\theta_d(\Phi \Phi^*) = \delta_n(D) = 0$,
which holds, for example, for Fourier compressed sensing with signal sparsity over an orthonormal basis $D$.

Similarly, we derive an upper bound on $\theta_s(\mathbb{E} \widetilde{\Psi}^* \widetilde{\Psi})$.
Recall that $\widetilde{D}$ is written as
\[
\widetilde{D} = D (D^* D)^{-1} = D (D^* D)^{-1/2} (D^* D)^{-1/2}.
\]
Therefore, it follows that
\begin{equation}
\norm{\widetilde{D}} \leq [\lambda_n(D^* D)]^{-1/2} \leq \frac{1}{\sqrt{1 - \delta_n(D)}}
\label{eq:proof:thm:urbpBO:eq3}
\end{equation}
and
\begin{equation}
\sigma_n(\widetilde{D}) \geq [\lambda_1(D^* D)]^{-1/2} \geq \frac{1}{\sqrt{1 + \delta_n(D)}}.
\label{eq:proof:thm:urbpBO:eq4}
\end{equation}

Similarly, since $\widetilde{\Phi}$ is written as
\[
\widetilde{\Phi} = (\Phi \Phi^*)^{-1} \Phi = (\Phi \Phi^*)^{-1/2} (\Phi \Phi^*)^{-1/2} \Phi,
\]
it follows that
\begin{equation}
\norm{\widetilde{\Phi}} \leq [\lambda_d(\Phi \Phi^*)]^{-1/2} \leq \frac{1}{\sqrt{1 - \theta_d(\Phi \Phi^*)}}
\label{eq:proof:thm:urbpBO:eq5}
\end{equation}
and
\begin{equation}
\lambda_d(\widetilde{\Phi} \widetilde{\Phi}^*) \geq [\lambda_1(\Phi \Phi^*)]^{-1/2} \geq \frac{1}{\sqrt{1 + \theta_d(\Phi \Phi^*)}}.
\label{eq:proof:thm:urbpBO:eq6}
\end{equation}

Using \cref{eq:proof:thm:urbpBO:eq3,eq:proof:thm:urbpBO:eq4,eq:proof:thm:urbpBO:eq5,eq:proof:thm:urbpBO:eq6},
we derive upper and lower bounds on the eigenvalues of $\mathbb{E} \widetilde{\Psi}^* \widetilde{\Psi}$ as follows:
\begin{align}
\norm{\mathbb{E} \widetilde{\Psi}^* \widetilde{\Psi}}
{} & \leq \norm{\widetilde{D}^* \mathbb{E} \widetilde{A}^* \widetilde{A} \widetilde{D}} \nonumber \\
{} & \leq \norm{\widetilde{D}^*} \norm{\mathbb{E} \widetilde{A}^* \widetilde{A}} \norm{\widetilde{D}} \nonumber \\
{} & = \norm{\mathbb{E} A^* A} \norm{\widetilde{D}}^2 \nonumber \\
{} & \leq \nu_{\min}^{-1} \norm{\widetilde{\Phi}}^2 [1 - \delta_n(D)]^{-1} \nonumber \\
{} & \leq \nu_{\min}^{-1} [1 - \theta_d(\Phi \Phi^*)]^{-1} [1 - \delta_n(D)]^{-1}
\label{eq:proof:thm:urbpBO:eq7}
\end{align}
and
\begin{align}
\lambda_n(\mathbb{E} \widetilde{\Psi}^* \widetilde{\Psi})
{} & \geq \lambda_n(\widetilde{D}^* \mathbb{E} \widetilde{A}^* \widetilde{A} \widetilde{D}) \nonumber \\
{} & \geq \sigma_n(\widetilde{D}^*) \lambda_d(\mathbb{E} \widetilde{A}^* \widetilde{A}) \sigma_n(\widetilde{D}) \nonumber \\
{} & \geq \lambda_d(\mathbb{E} \widetilde{A}^* \widetilde{A}) [1 + \delta_n(D)]^{-1} \nonumber \\
{} & \geq \nu_{\max}^{-1} \lambda_d(\widetilde{\Phi} \widetilde{\Phi}^*) [1 + \delta_n(D)]^{-1} \nonumber \\
{} & \geq \nu_{\max}^{-1} [1 + \theta_d(\Phi \Phi^*)]^{-1} [1 + \delta_n(D)]^{-1}.
\label{eq:proof:thm:urbpBO:eq8}
\end{align}

{\allowdisplaybreaks
Then, we derive an upper bound on $\theta_s(\mathbb{E} \widetilde{\Psi}^* \widetilde{\Psi})$ using \cref{eq:proof:thm:urbpBO:eq7,eq:proof:thm:urbpBO:eq8}
as follows:
\begin{align}
\theta_s(\mathbb{E} \widetilde{\Psi}^* \widetilde{\Psi})
{} & \leq \theta_n(\mathbb{E} \widetilde{\Psi}^* \widetilde{\Psi}) \nonumber \\
{} & = \max[ 1 - \lambda_n(\mathbb{E} \widetilde{\Psi}^* \widetilde{\Psi}) , \norm{\mathbb{E} \widetilde{\Psi}^* \widetilde{\Psi}} - 1] \nonumber \\
{} & \leq \max\Big\{1 - \nu_{\max}^{-1} [1 + \delta_n(D)]^{-1} [1 + \theta_d(\Phi \Phi^*)]^{-1}, \nonumber \\
{} & \qquad \nu_{\min}^{-1} [1 - \delta_n(D)]^{-1} [1 - \theta_d(\Phi \Phi^*)]^{-1} - 1 \Big\} \nonumber \\
{} & \leq \max(1 - \nu_{\max}^{-1}, \nu_{\min}^{-1} - 1) \nonumber \\
{} & \quad + \nu_{\min}^{-1} \Bigg\{\frac{\delta_n(D)}{1 - \delta_n(D)} + \frac{\theta_d(\Phi \Phi^*)}{1 - \theta_d(\Phi \Phi^*)} \nonumber \\
{} & \qquad + \frac{\delta_n(D) \theta_d(\Phi \Phi^*)}{[1 - \delta_n(D)][1 - \theta_d(\Phi \Phi^*)]}\Bigg\}. \label{eq:proof:thm:urbpBO:eq9}
\end{align}

Combining (\ref{eq:proof:thm:urbpBO:eq2}) and (\ref{eq:proof:thm:urbpBO:eq9}), we obtain
\begin{align*}
{} & \max(\theta_s(\mathbb{E} \Psi^* \Psi), \theta_s(\mathbb{E} \widetilde{\Psi}^* \widetilde{\Psi})) \\
{} & \leq \max(1 - \nu_{\max}^{-1}, \nu_{\min}^{-1} - 1) \\
{} & \quad + \max(\nu_{\max}, \nu_{\min}^{-1}) \Bigg\{1 + \frac{\delta_n(D)}{1 - \delta_n(D)} \\
{} & \qquad + \frac{\theta_d(\Phi \Phi^*)}{1 - \theta_d(\Phi \Phi^*)} + \frac{\delta_n(D) \theta_d(\Phi \Phi^*)}{[1 - \delta_n(D)][1 - \theta_d(\Phi \Phi^*)]}\Bigg\}
\end{align*}
Applying this to Corollary~\ref{cor:urbp} completes the proof.}

\subsection{Proof of Theorem~\ref{thm:urbpOC}}
\label{subsec:proof:thm:urbpOC}

The proof of Theorem~\ref{thm:urbpOC} is almost identical to that of Theorem~\ref{thm:urbpBO}.
The mutual incoherence between $(\widetilde{\phi}_\omega)_{\omega \in \Omega}$ and $(d_j)_{j \in [n]}$ is bounded in terms of $K$ by
\begin{align}
{} & \sup_{\omega \in \Omega} \max_{j \in [n]} |\langle \widetilde{\phi}_\omega, d_j \rangle| \nonumber \\
{} & = \norm{\Lambda_\nu^{-1} \Phi^* (\Phi \Phi^*)^{-1} D}_{\ell_1^n \to L_\infty(\Omega,\mu)} \nonumber \\
{} & \leq \norm{\Lambda_\nu^{-1} \Phi^* D}_{\ell_1^n \to L_\infty(\Omega,\mu)} \nonumber \\
{} & \quad + \norm{\Lambda_\nu^{-1} \Phi^* [(\Phi \Phi^*)^{-1} - I_d] D}_{\ell_1^n \to L_\infty(\Omega,\mu)} \nonumber \\
{} & \leq \frac{1}{\nu_{\min}} \norm{\Phi^* D}_{\ell_1^n \to L_\infty(\Omega,\mu)} \nonumber \\
{} & \quad + \frac{1}{\nu_{\min}} \norm{\Phi^*}_{\ell_2^d \to L_\infty(\Omega,\mu)} \norm{(\Phi \Phi^*)^{-1} - I_d}_{\ell_2^d \to \ell_2^d} \nonumber \\
{} & \qquad \cdot \norm{D}_{\ell_1^n \to \ell_2^d} \nonumber \\
{} & \leq \frac{1}{\nu_{\min}}
\Bigg[ K + \left(\sup_{\omega \in \Omega} \norm{\phi_\omega}_{\ell_2^d}\right)
\frac{\theta_d(\Phi \Phi^*)}{1 - \theta_d(\Phi \Phi^*)} \cdot \left( \max_{j \in [n]} \norm{d_j}_{\ell_2^d} \right) \Bigg].
\label{eq:proof:thm:urbpOC:eq1}
\end{align}
Let $\widetilde{K}$ be the right hand side of (\ref{eq:proof:thm:urbpOC:eq1}).
Then, we apply the incoherence parameters $K$ and $\widetilde{K}$ to Corollary~\ref{cor:urbp}.
It remains to bound $\theta_s(\mathbb{E} \Psi^* \Psi)$ and $\theta_s(\mathbb{E} \widetilde{\Psi}^* \widetilde{\Psi})$.

In the proof of Theorem~\ref{thm:urip}, we derived an upper bound on $\theta_s(\mathbb{E} \Psi^* \Psi)$ given by
\begin{align}
\theta_s(\mathbb{E} \Psi^* \Psi)
{} & \leq \max(1 - \nu_{\min}, \nu_{\max} - 1) \nonumber \\
{} & \quad + \nu_{\max} (\delta_s(D) + \theta_d(\Phi \Phi^*) + \delta_s(D) \theta_d(\Phi \Phi^*)).
\label{eq:proof:thm:urbpOC:eq2}
\end{align}

Similarly to the proof of Theorem~\ref{thm:urbpsimpleBO}, $\theta_s(\mathbb{E} \widetilde{\Psi}^* \widetilde{\Psi})$ is bounded by
\begin{align}
\theta_s(\mathbb{E} \widetilde{\Psi}^* \widetilde{\Psi})
{} & \leq \max\Big\{1 - \nu_{\max}^{-1} [1 - \delta_s(D)] [1 + \theta_d(\Phi \Phi^*)]^{-1}, \nonumber \\
{} & \qquad \nu_{\min}^{-1} [1 + \delta_s(D)] [1 - \theta_d(\Phi \Phi^*)]^{-1} - 1 \Big\} \nonumber \\
{} & \leq \max(1 - \nu_{\max}^{-1}, \nu_{\min}^{-1} - 1) + \nu_{\min}^{-1} \nonumber \\
{} & \quad \cdot \Bigg[\delta_s(D) + \frac{\theta_d(\Phi \Phi^*)}{1 - \theta_d(\Phi \Phi^*)} + \frac{\delta_s(D) \theta_d(\Phi \Phi^*)}{1 - \theta_d(\Phi \Phi^*)}\Bigg].
\label{eq:proof:thm:urbpOC:eq3}
\end{align}
Then, (\ref{eq:proof:thm:urbpOC:eq2}) and (\ref{eq:proof:thm:urbpOC:eq3}) imply
\begin{align}
{} & \max(\theta_s(\mathbb{E} \Psi^* \Psi), \theta_s(\mathbb{E} \widetilde{\Psi}^* \widetilde{\Psi})) \nonumber \\
{} & \leq \max(1 - \nu_{\max}^{-1}, \nu_{\min}^{-1} - 1) + \max(\nu_{\max}, \nu_{\min}^{-1}) \nonumber \\
{} & \quad \cdot \Bigg[1 + \delta_s(D) + \frac{\theta_d(\Phi \Phi^*)}{1 - \theta_d(\Phi \Phi^*)} + \frac{\delta_s(D) \theta_d(\Phi \Phi^*)}{1 - \theta_d(\Phi \Phi^*)}\Bigg].
\label{eq:proof:thm:urbpOC:eq4}
\end{align}

Applying (\ref{eq:proof:thm:urbpOC:eq4}) to Corollary~\ref{cor:urbp} completes the proof.

\section*{Acknowledgements}
The authors thank Saiprasad Ravishankar for providing a sparsifying transform learned using his algorithm \cite{RavBre12},
which was used in the simulations of this paper.


\end{document}